\documentclass[aps,pre]{revtex4-1}
\usepackage{amsmath}
\usepackage{amsthm}
\usepackage{amssymb}
\usepackage{amsfonts}
\usepackage{graphicx}
\usepackage{bm}
\usepackage{epsfig}
\usepackage{hyperref} 

\newcommand{\diag}{{\rm diag\,}}
\newcommand{\sign}{{\rm sign\,}}
\newcommand{\tr}{{\rm tr\,}}
\newcommand{\Str}{{\rm Str\,}}
\newcommand{\Sdet}{{\rm Sdet\,}}

\newcommand{\U}{{\rm U\,}}
\newcommand{\Or}{{\rm O\,}}

\newcommand{\W}{{\rm W\,}}
\newcommand{\rt}{{\rm r\,}}
\newcommand{\lt}{{\rm l\,}}

\newcommand{\RE}{{\rm Re\,}}
\newcommand{\IM}{{\rm Im\,}}

\newcommand{\eins}{\leavevmode\hbox{\small1\kern-3.8pt\normalsize1}}
\newcommand{\be}{\begin{eqnarray}}
\newcommand{\ee}{\end{eqnarray}}

\begin{document}

\newcommand{\hama}{\widehat a_6}
\newcommand{\hala}{\widehat a_7}
\newcommand{\hx}{\widehat x}
\newcommand{\hy}{\widehat y}
\newcommand{\hz}{\widehat z}
\newcommand{\ha}{\widehat a}
\newcommand{\hma}{\widehat m_6}
\renewcommand{\hm}{\widehat m}
\newcommand{\hla}{\widehat \lambda_7}
\newcommand{\nn}{\nonumber} 

\title{Spectral Properties of the Wilson Dirac Operator and random matrix theory}
\author{Mario Kieburg$^{1,2}$}
\email{mkieburg@physik.uni-bielefeld.de}
\author{Jacobus J. M. Verbaarschot$^{1}$}
\email{jacobus.verbaarschot@stonybrook.edu}
\author{Savvas Zafeiropoulos$^{1,3}$}
\email{zafeiropoulos@clermont.in2p3.fr}
\affiliation{$^{1}$ Department of Physics and Astronomy, State University of New York at Stony Brook, NY 11794-3800, USA}
\affiliation{$^{2}$ Fakult\"at f\"ur Physik, Postfach 100131, 33501 Bielefeld, Germany}
\affiliation{$^{3}$ Laboratoire de Physique Corpusculaire, Universit\'e Blaise Pascal, CNRS/IN2P3, 63177 Aubi\`ere Cedex, France}

\date{\today}

\begin{abstract}

Random Matrix Theory has been 
successfully applied to lattice Quantum Chromodynamics. 
In particular, a great deal of progress has been made
on the understanding, numerically as well as analytically, of
the spectral properties of the Wilson Dirac operator.
In this paper, we study the infra-red spectrum of the Wilson Dirac operator 
via Random Matrix Theory including the three leading order $a^2$ correction
terms that appear in the corresponding chiral Lagrangian. 
A derivation of  the joint probability density of the eigenvalues 
is presented. This result is used to calculate the density of the complex
eigenvalues, the density of the real eigenvalues and the distribution of
the chiralities over the real eigenvalues. 
A detailed discussion  of these quantities shows how each 
low energy constant affects the spectrum. Especially we consider the limit 
of small and large (which is almost the mean field limit) lattice spacing. 
Comparisons with Monte Carlo simulations of the Random Matrix Theory show 
a perfect agreement with the analytical predictions. Furthermore we 
present some quantities which can be easily used for comparison of  
lattice data and  the analytical results.
\end{abstract}

\maketitle

 {\bf Keywords:} Wilson Dirac operator, epsilon regime, low energy constants, chiral symmetry, random matrix theory\\
 {\bf PACS:} 12.38.Gc, 05.50.+q, 02.10.Yn, 11.15.Ha

\section{Introduction}\label{sec1}

The drastically increasing computational power as well as algorithmic 
improvements over the last decades provide
us with deep insights in non-perturbative effects of 
Quantum Chromodynamics (QCD). However, the artefacts of the 
discretization, i.e. a finite lattice spacing, are not yet
completely under control. In particular, in the past few years 
a large numerical 
\cite{Michael,757044,Baron,DWW,DHS,Damgaard:2012gy,BBS} 
and analytical \cite{NS,Damgaard:2010cz,Akemann:2010zp,HS,Splittorff:2011np,Kieburg:2012fw, Herdoiza:2013sla} effort 
was undertaken to determine  the low energy constants of the terms
in the chiral Lagrangian that describe the discretization errors. 
It is well known that new phase structures arise such as the Aoki phase 
\cite{Aokiclassic} and the Sharpe-Singleton scenario \cite{SharpeSingleton}. 
A direct analytical understanding of lattice QCD seems to be out of reach.
Fortunately, as was already realized two decades ago, the low lying spectrum of 
the continuum QCD Dirac operator can be described in terms 
of Random Matrix Theories (RMTs) \cite{Shuryak:1992pi,V}.

Recently, RMTs were formulated to describe
discretization effects for 
staggered \cite{Osborn:2012bc} as well as Wilson \cite{Damgaard:2010cz,Akemann:2010zp} fermions.
Although these RMTs are more complicated than  the chiral Random Matrix Theory 
formulated in 
\cite{Shuryak:1992pi,V}, in the case of Wilson fermions a complete
analytical solution of the RMT has been achieved
\cite{Damgaard:2010cz,Akemann:2010zp,Splittorff:2011np,AN,Kie12a,Kieburg:2012fw,Kie12b,Kieburg:2011uf}.
Since the Wilson RMT shares the global symmetries of the Wilson Dirac
operator it will be equivalent to the corresponding  (partially quenched)
chiral Lagrangian in the microscopic domain (also known as 
the $\epsilon$-domain)  
\cite{RS,BRS,Aoki-spec,GSS,Shindler,BNSep}. 

Quite recently, there has been  a breakthrough in deriving 
 eigenvalue statistics of the infra-red spectrum of the Hermitian 
\cite{AN} as well as the non-Hermitian 
\cite{Kieburg:2011uf,Kie12a,Kie12b} Wilson Dirac operator. 
These results explain \cite{Kieburg:2012fw}  
why the Sharpe-Singleton scenario is only observed for the case 
of dynamical fermions
 \cite{Baron,BBS,Aoki:1995yf,Aoki:1997fm,Ilgenfritz:2003gw,DGLPT,Aoki:2004iq,Farchioni:2004us,Farchioni:2004fs} 
and not in the quenched theory \cite{AokiGocksch,Jansen:2005cg} 
while the Aoki phase has been seen in both cases. 
First comparisons of the analytical predictions with lattice 
data show a promising agreement \cite{DWW,DHS,Damgaard:2012gy}. 
Good fits of the low energy constants are expected for the 
distributions of individual eigenvalues  \cite{Damgaard:2010cz,Akemann:2010zp,Akemann:2012pn}.
 
Up to now, 
mostly the effects of $W_8$ \cite{Akemann:2010zp,Kieburg:2011uf,AN}, and 
quite recently also of $W_6$ \cite{HS,Kieburg:2012fw,Rasmuss}, on the Dirac spectrum 
were studied in detail. In this article, 
we will discuss the effect of all three low energy constants. 
Thereby we start from the Wilson RMT for 
the non-Hermitian Wilson Dirac operator proposed in 
Refs.~\cite{Damgaard:2010cz}. In Sec.~\ref{sec2}
we recall this Random Matrix Theory and its properties. 
Furthermore we derive the joint probability density of the eigenvalues 
which  so far was only stated without proof in Refs.~\cite{Kieburg:2012fw,Kie12b}.
 We also discuss the approach to the continuum limit
in terms  of the Dirac spectrum.

In Sec.~\ref{sec3}, we derive the level densities 
of $D_\W$ starting from the joint probability density. 
Note that due to its $\gamma_5$-Hermiticity $D_\W$  has
complex eigenvalues as well as exactly real eigenvalues. 
Moreover, the real modes split into those corresponding to eigenvectors 
with positive and negative chirality. In Sec.~\ref{sec4}, we discuss 
the spectrum  of the quenched non-Hermitian Wilson Dirac operator 
in the microscopic limit in detail. 
In particular the asymptotics at small and large lattice spacing is studied. 
The latter limit is equal to a  mean field limit for some quantities 
which can be trivially read off.

In Sec.~\ref{sec5} we summarize our results. 
In particular we present easily measurable quantities which can be 
used for fitting the three low energy constants $W_{6/7/8}$ 
and the chiral condensate $\Sigma$. 
Detailed derivations are given in several appendices.   
The joint probability density is derived in Appendix~\ref{app1}. Some useful
integral identities are given in Appendix~\ref{app2}
and in Appendix~\ref{app3}
 we perform the microscopic limit of the graded partition function that
enters in the distribution of the chiralities over the real eigenvalues
of $D_\W$. 
 Finally, some 
asymptotic results are derived in Appendix~\ref{app4}.

\section{Wilson Random Matrix Theory  and its Joint Probability Density}\label{sec2}

In Sec.~\ref{sec2.1} we introduce the Random
Matrix Theory for the infra-red spectrum of the Wilson Dirac operator and
 recall its most important properties. Its joint probability density is given in Sec.~\ref{sec2.2}, and the continuum limit is derived in Sec.~\ref{sec2.3}.

\subsection{The random matrix ensemble}\label{sec2.1}

We consider the random matrix ensemble \cite{Damgaard:2010cz,Akemann:2010zp}
\begin{eqnarray}\label{2.1}
 D_\W=\left(\begin{array}{cc} A & W \\ -W^\dagger & B \end{array}\right)
\end{eqnarray}
distributed by the probability density
\begin{eqnarray}
 P(D_\W)&=&\left(\frac{n}{2\pi a^2}\right)^{[n^2+(n+\nu)^2]/2}\left(-\frac{n}{2\pi}\right)^{n(n+\nu)}\exp\left[-\frac{a^2}{2}\left(\mu_\rt^2+\frac{n+\nu}{n}\mu_\lt^2\right)\right]\nonumber\\
 &&\times\exp\left[-\frac{n}{2a^2}(\tr A^2+\tr B^2)-n\tr WW^\dagger+\mu_\rt\tr A+\mu_\lt\tr B\right].\label{2.2}
\end{eqnarray}
The Hermitian matrices $A$ and $B$ break chiral symmetry and their dimensions 
are $n\times n$ and $(n+\nu)\times(n+\nu)$, respectively, where 
$\nu$ is the index of the Dirac operator. Both $\mu_{\rm r}$ and $\mu_{\rm l}$ are one dimensional
real variables.
The chiral RMT describing continuum QCD \cite{Shuryak:1992pi} is given by
the ensemble (\ref{2.1}) with $A$ and $B$ replaced by zero.
The $N_{\rm f}$ flavor RMT partition function is defined by
\be
Z^\nu_{N_{\rm f}}(m) = \int D[D_\W] P(D_\W) {\det}^{N_{\rm f}}(D_\W+m).  
\ee
Without loss of generality we can assume $\nu\geq 0$ since the results are 
symmetric under $\nu\to-\nu$  together with $\mu_\rt \leftrightarrow\mu_\lt$.

 The Gaussian integrals over the two variables 
$\mu_\rt$ and $\mu_\lt$ yield the two low energy constants 
$W_6$ and $W_7$ \cite{Damgaard:2010cz,Akemann:2010zp}. The reason is that
 the integrated probability
density
\begin{eqnarray}
  P(D_\W,W_{6/7}\neq0)&=&\label{2.2b}\int_{-\infty}^\infty P(D_\W)\exp\left[-\frac{a^2(\mu_\rt+\mu_\lt)^2}
{16V|W_6|}-\frac{a^2(\mu_\rt-\mu_\lt)^2}{16V|W_7|}\right]\frac{a^2d\mu_\rt d\mu_\lt}{8\pi V\sqrt{W_6W_7}}\nonumber
\end{eqnarray}
generates the terms $(\tr A+\tr B)^2$ and $(\tr A-\tr B)^2$ which 
 correspond to the squares of traces in the chiral Lagrangian 
\cite{RS,BRS,Aoki-spec,GSS}. In the microscopic domain 
the corresponding partition function for $N_{\rm f}$ fermionic 
flavors is then given by
\begin{eqnarray}
 Z^\nu_{N_{\rm f}}(\widetilde{m})&=&\int\limits_{\U(N_{\rm f})}d\mu(U)\exp\left[\frac{\Sigma V}{2}\tr\widetilde{m}(U+U^{-1})-\widetilde{a}^2 VW_6\tr^2( U+U^{-1})\right]\nonumber\\
 &&\hspace*{-0.5cm}\times\exp\left[-\widetilde{a}^2VW_7\tr^2 (U-U^{-1})-\widetilde{a}^2 VW_8\tr (U^2+U^{-2})\right]{\det}^{\nu}U\label{2.2a}
\end{eqnarray}
 with the physical quark masses 
$\widetilde{m}=\diag(\widetilde{m}_1,\ldots,\widetilde{m}_{N_{\rm f}})$, 
the space-time volume $V$, the physical lattice spacing $\widetilde{a}$ 
and the chiral condensate $\Sigma$. The low energy constant $W_8$ is 
generated by the term $\tr A^2+\tr B^2$ in Eq.~\eqref{2.2} and 
is \textit{a priori} positive. We include the lattice spacing $a$ in the standard 
deviation of $A$ and $B$, cf. Eq.~\eqref{2.2}, out of convenience for deriving the joint probability density.
 We employ the sign convention of Refs.~\cite{Damgaard:2010cz,Akemann:2010zp} for the low energy constants.

The microscopic limit ($n\to\infty$) is performed in Sec.~\ref{sec3}. In this limit 
the rescaled lattice spacing $\widehat{a}_8^2=na^2/2=\widetilde{a}^2VW_8$, 
the rescaled parameters $\widehat{m}_{6}=a^2(\mu_{\rt}+\mu_{\lt})$ and
$\widehat{\lambda}_{7}=a^2(\mu_{\rt}-\mu_{\lt})$, and the rescaled eigenvalues 
$\widehat{Z}=2nZ=\diag(2nz_1,\ldots,$ $2nz_{2n+\nu})$ of $D_{\W}$ are kept fixed for $n\to \infty$.  The mass $\widehat{m}_6$ and axial mass $\widehat{\lambda}_7$ 
are 
distributed with respect to Gaussians with  variance $8\widehat{a}_6^2=-8\widetilde{a}^2VW_6$ 
and $8\widehat{a}_7^2=-8\widetilde{a}^2VW_7$, respectively. Note the minus sign in front of 
$W_{6/7}$. As was shown in
Ref.~\cite{Kieburg:2012fw} the opposite sign is inconsistent with the symmetries
of the Wilson Dirac operator.
The notation is slightly different 
from what is used in the literature to get rid of the imaginary unit 
in $\widehat{a}_6$ and $\widehat{a}_7$. 

The joint probability density $p(Z)$ of the eigenvalues 
$Z=\diag(z_1,\ldots,z_{2n+\nu})$ of $D_\W$ 
can be defined by
\begin{eqnarray}\label{2.3}
 I[f]&=&\int\limits_{\mathbb{C}^{(2n+\nu)\times(2n+\nu)}} f(D_\W)P(D_\W)d[D_\W]=\int\limits_{\mathbb{C}^{(2n+\nu)}} f(Z)p(Z)d[Z],
\end{eqnarray}
where $f$ is an arbitrary $\U(n,n+\nu)$ invariant function. 
The random matrix $D_\W$ is $\gamma_5=\diag(\eins_n,-\eins_{n+\nu})$ Hermitian, i.e.
\begin{eqnarray}\label{2.4}
 D_\W^\dagger=\gamma_5D_\W\gamma_5.
\end{eqnarray}
Hence, the eigenvalues $z$ come in complex conjugate pairs or are exactly real. 
The matrix $D_\W$ has $\nu$ generic real modes and $2(n-l)$ additional real eigenvalues ($0\leq l\leq n$). The index $l$  decreases by one when a complex conjugate pair enters the real axis.

\subsection{The joint probability density of $D_\W$}\label{sec2.2}

Let $D_{l}$ be $D_\W$ if it can be quasi-diagonalized by a non-compact unitary rotation $U\in\U(n,n+\nu)$, i.e. $U\gamma_5U^\dagger=\gamma_5$, to
\begin{eqnarray}\label{2.5}
 D_l=UZ_lU^{-1}=U\left(\begin{array}{cc|cc} x_1 & 0 & 0 & 0 \\ 0 & x_2 & y_2 & 0 \\ \hline 0 & -y_2 & x_2 & 0 \\ 0 & 0 & 0 & x_3 \end{array}\right)U^{-1},
\end{eqnarray}
where the real diagonal matrices 
$x_1=\diag(x_1^{(1)},\ldots,x_{n-l}^{(1)})$, $x_2=\diag(x_1^{(2)},\ldots,$ $ x_{l}^{(2)})$, $y_2=\diag(y_1^{(2)},\ldots, y_{l}^{(2)})$ 
and $x_3=\diag(x_1^{(3)},\ldots,x_{n+\nu-l}^{(3)})$ have the dimension 
$n-l$, $l$, $l$ and $n+\nu-l$, respectively. The matrices $x_1$ and $x_3$ 
comprise all real eigenvalues of $D_l$ corresponding to the right-handed 
and left-handed modes, respectively. We refer to an eigenvector 
$\psi$ of $D_\W$ as right-handed if the chirality is positive definite, i.e.
\begin{eqnarray}\label{2.6}
 \langle\psi|\gamma_5|\psi\rangle>0,
\end{eqnarray}
and as left-handed if the chirality is negative definite. The eigenvectors corresponding to complex eigenvalues have vanishing chirality. The complex conjugate pairs are $(z_2=x_2+\imath y_2,z_2^*=x_2-\imath y_2)$. Note that it is not possible to
diagonalize $D_\W$ with a ${\rm U}(n,n+\nu)$ transformation with complex conjugate
eigenvalues. Moreover we emphasize that almost all $\gamma_5$-Hermitian matrices can be brought to the form~\eqref{2.5} excluding a set of measure zero.

The quasi-diagonalization
$D_l= U Z_l U^{-1}$ determines $U$ up to a  $\U^{2n+\nu-l}(1)\times\Or^{l}(1,1)$ transformation
while the set of eigenvalues $Z_l$ can be permuted  in
$l! (n-l)! (n+\nu-l)! 2^l$ different ways. 
The factor $2^l$ is due to the complex conjugation of each single complex pair. 
The Jacobian of the transformation to eigenvalues and the coset
$\mathbb{G}_l=\U(n,n+\nu)/[\U^{2n+\nu-l}(1)\times\Or^{l}(1,1)]$ is given by
\be
|\Delta_{2n+\nu}(Z_l)|^2,
\ee
where the Vandermonde determinant is defined as
\begin{equation}\label{2.3a}
 \Delta_{2n+\nu}(Z)=\prod\limits_{1\leq i<j\leq 2n+\nu}(z_i-z_j)=(-1)^{n+\nu(\nu-1)/2}\det\left[z_i^{j-1}\right]_{1\leq i,j\leq 2n+\nu}.
\end{equation}

The functional $I[f]$ in Eq.~\eqref{2.3} is a sum over $n+1$ integrations on disjoint sets, i.e.
\begin{eqnarray}\label{2.7}
 I[f]&=&\sum\limits_{l=0}^n\frac{1}{2^l(n-l)!l!(n+\nu-l)!}\int\limits_{\mathbb{R}^{\nu+2(n-l)}\times\mathbb{C}^{l}}f(Z_l)\left[\int\limits_{\mathbb{G}_l}P(UZ_lU^{-1})
d\mu_{\mathbb{G}_l}(U)\right]
 |\Delta_{2n+\nu}(Z_l)|^2d[Z_l],
\end{eqnarray}
where we have normalized the terms with respect to the number of
possible permutations of the eigenvalues in $Z_l$.
Thus we have for the joint probability density over all sectors of eigenvalues
\begin{eqnarray}\label{2.8}
 p(Z)d[Z]&=&\sum\limits_{l=0}^np_l(Z_l)d[Z_l]\sum\limits_{l=0}^n\frac{|\Delta_{2n+\nu}(Z_l)|^2d[Z_l]}{2^l(n-l)!l!(n+\nu-l)!}\int\limits_{\mathbb{G}_l}P(UZ_lU^{-1})d\mu_{\mathbb{G}_l}(U).
\end{eqnarray}
Here $p_l(Z_l)$ is the joint probability density for a fixed number 
of complex conjugate eigenvalue pairs, namely $l$. The integration over $U$ is non-trivial and will be worked
out in detail in Appendix~\ref{app1}.

In a more mathematical language the normalization factor in Eq.~(\ref{2.8}) can be understood
as follows.
If 
the permutation group of $N$ elements is denoted by  $\mathbf{S}(N)$ 
while  the group describing the reflection $y\to-y$ is $\mathbb{Z}_2$,  
the factor $2^l(n-l)!l!(n+\nu-l)!$
is the volume of the finite subgroup 
$\mathbf{S}(n-l)\times \mathbf{S}(l)\times \mathbf{S}(n+\nu-l)\times\mathbb{Z}_2^l$ 
of $\U(n,n+\nu)$ which correctly normalizes each summand. 
Originally we had to divide $\U(n,n+\nu)$ by the set 
$\U^{2n+\nu-l}(1)\times\Or^{l}(1,1)\times \mathbf{S}(n-l)\times \mathbf{S}(l)\times 
\mathbf{S}(n+\nu-l)\times\mathbb{Z}_2^l$ because it is the maximal subgroup 
whose image of the adjoint mapping commutes with $Z_l$. The
 reasoning is as follows. 
Let $\Sigma[Z_l]=\{UZ_lU^{-1}|U\in\U(n,n+\nu)\}$ be the orbit of 
$Z_l$ and $\Sigma_{\rm c}[Z_l]=\{\widehat{Z}_l\in\Sigma[Z_l]|[\widehat{Z}_l,Z_l]_-=\widehat{Z}_lZ_l-Z_l\widehat{Z}_l=0\}$ 
a subset of this orbit. Then all orderings 
in each of the three sets of eigenvalues $x_1$, $(z_2,z_2^*)$ and $x_3$ as 
well as the reflections $y_{j}^{(2)}\to-y_{j}^{(2)}$ are in $\Sigma_{\rm c}[Z_l]$. 
This subset $\Sigma_{\rm c}[Z_l]\subset\Sigma[Z_l]$ can be represented by the 
finite group 
$\mathbf{S}(n-l)\times \mathbf{S}(l)\times \mathbf{S}(n+\nu-l)\times\mathbb{Z}_2^l$. 
This group is called the Weyl group in group theory. 
The Lie group  $\U^{2n+\nu-l}(1)\times\Or^{l}(1,1)$ acts on $\Sigma_{\rm c}[Z_l]$ 
as the identity since it commutes with $Z_l$. 
The group $\U^{2n+\nu-l}(1)$ represents $2n+\nu-l$ complex phases along 
the diagonal commuting with the set which consists of 
$Z_l$ with a fixed $l$. Each non-compact orthogonal group 
$\Or(1,1)$ reflects the invariance of a single complex conjugate eigenvalue 
pair under a hyperbolic transformation which is equal to a Lorentz-transformation in 
a  1+1 dimensional space-time.

There are two ways to deal with the invariance under 
$\U^{2n+\nu-l}(1)\times\Or^{l}(1,1)\times \mathbf{S}(n-l)\times \mathbf{S}(l)\times \mathbf{S}(n+\nu-l)\times\mathbb{Z}_2^l$ 
in an integral such that we correctly weigh all points. 
We have either to divide $\U(n,n+\nu)$ by the whole subgroup or we integrate over a 
larger coset and reweight the measure by the volume of the subgroups not excluded. 
The ordering enforced by $\mathbf{S}(n-l)\times \mathbf{S}(l)\times \mathbf{S}(n+\nu-l)\times\mathbb{Z}_2^l$ 
is difficult to handle in calculations. Therefore, we have decided for a reweighting 
of the integration measure by $1/[(n-l)! (n+\nu-l)!l! 2^l]$. However the Lie group $\U^{2n+\nu-l}(1)\times\Or^{l}(1,1)$, in particular the hyperbolic subgroups, 
has to be excluded since its volume is infinite.

In this section as well as in Appendix~\ref{app1}, we use the non-normalized Haar-measures induced by the pseudo metric
\begin{eqnarray}\label{2.9}
 \tr dD_\W^2=\tr dA^2 +\tr dB^2-2\tr dWdW^\dagger.
\end{eqnarray}
Therefore the measures for $D_\W$ and $Z_l$ are
\begin{eqnarray}
 d[D_\W]&=&\prod\limits_{j=1}^ndA_{jj}\hspace*{-0.2cm}\prod\limits_{1\leq i<j\leq n}2\,d\,\RE A_{ij}d\,\IM A_{ij}\prod\limits_{j=1}^{n+\nu}dB_{jj}\label{2.10}\\
  &\times&\prod\limits_{1\leq i<j\leq n+\nu}2\,d\,\RE B_{ij}d\,\IM B_{ij}\underset{1\leq j\leq n+\nu}{\underset{1\leq i\leq n}{\prod}} (-2)\,d\,\RE W_{ij}d\,\IM W_{ij},\nonumber\\
 d[Z_l]&=&\prod\limits_{j=1}^{n-l}dx_{j}^{(1)}\prod\limits_{j=1}^{l}2\imath\,dx_{j}^{(2)}dy_{j}^{(2)}\prod\limits_{j=1}^{n+\nu-l}dx_{j}^{(3)}.\label{2.10b}
\end{eqnarray}
The Haar measure $d\mu_{\mathbb{G}_l}$ for the coset $\mathbb{G}_l$ is also induced by $d[D_\W]$ and results from the pseudo metric, i.e.
\begin{eqnarray}\label{2.11}
 \tr dD_\W^2=\tr dZ_l^2+\tr [U^{-1}dU,Z_l]_-^2.
\end{eqnarray}
The reason for this unconventional definition is the 
non-normalizability of the measure $d\mu_{\mathbb{G}_l}$ 
because $\mathbb{G}_l$ is non-compact for $l>0$. Hence the normalization 
resulting from definition~\eqref{2.11} seems to be the most natural one,
 and it helps in keeping track of the normalizations.

In Appendix~\ref{app1} we solve the coset integral~\eqref{2.8}. The first step
is to linearize the quadratic terms in $UZ_l U^{-1}$
by introducing auxiliary Gaussian integrals 
over additional matrices which is along the idea presented in Ref.~\cite{AN}. 
In this way we split the integrand in a part invariant under 
$\U(n,n+\nu)$ and a non-invariant part resulting from an external source. 
The group integrals appearing in this calculations are 
reminiscent of the Itsykson-Zuber integral. 
However they are over non-compact groups and, thus, much more 
involved than in Ref.~\cite{AN}. 
Because of the $\U(n)\times \U(n+\nu)$ invariance of the probability density
of $D_\W$, the joint eigenvalue distribution is a symmetric function 
 of $n$ eigenvalues which we label by ``r'' and $n+\nu$
eigenvalues labelled by ``l''. The $\gamma_5$-Hermiticity  imposes reality 
constraints on the eigenvalues resulting in Dirac delta-functions
in the joint probability density. 
Similarly to the usual Itzykson-Zuber integral, the symmetric function
of the eigenvalues turns out to be particularly simple (see Appendix~\ref{app1})
\begin{eqnarray}
 p(Z)d[Z]&=&c(1+a^2)^{-n(n+\nu-1/2)}a^{-n-\nu^2}\exp\left[-\frac{a^4}{4(1+a^2)}(\mu_\rt-\mu_\lt)^2\right]\label{2.12}\\
 &&\hspace*{-0cm}\displaystyle\times\Delta_{2n+\nu}(Z)\det\left[\begin{array}{c} \displaystyle \left\{g_2(z_i^{(\rt)},z_j^{(\lt)})d x_i^{(\rt)}d y_i^{(\rt)}d x_j^{(\lt)}d y_j^{(\lt)}\right\}\underset{1\leq j \leq n+\nu}{\underset{1\leq i\leq n}{}} \\ \displaystyle \left\{\left(x_j^{(\lt)}\right)^{i-1}g_1(x_j^{(\lt)})\delta(y_j^{(\lt)})d x_j^{(\lt)}d y_j^{(\lt)}\right\}\underset{1\leq j \leq n+\nu}{\underset{1\leq i\leq \nu}{}} \end{array}\right].\nonumber
\end{eqnarray}
The last $\nu$ rows become zero in the continuum limit resulting in
$\nu$ exact zero modes (see subsection~\ref{sec2.3}). At finite $a$ they
can be interpreted as broadened  ``zero modes''.  The functions in the
determinant are given by
\begin{eqnarray}
 g_2(z_1,z_2)&=&g_r(x_1,x_2)\delta(y_1)\delta(y_2)+g_{\rm c}(z_1)\delta(x_1-x_2)\delta(y_1+y_2),\label{2.12a}\\
 g_{\rm r}(x_1,x_2)&=&\exp\left[-\frac{n}{4a^2}\left(x_1+x_2-\frac{a^2(\mu_\rt+\mu_\lt)}{n}\right)^2+\frac{n}{4}(x_1-x_2)^2\right]\label{2.13}\\
 &&\hspace*{-0cm}\times\left[\sign(x_1-x_2)-{\rm erf}\left[\sqrt{\frac{n(1+a^2)}{4a^2}}(x_1-x_2)-\sqrt{\frac{a^2}{4n(1+a^2)}}(\mu_\rt-\mu_\lt)\right]\right],\nonumber\\
 g_{\rm c}(z)&=&-2\imath\ \sign(y)\exp\left[-\frac{n}{a^2}\left(x-\frac{a^2(\mu_\rt+\mu_\lt)}{2n}\right)^2-ny^2\right],\label{2.14}\\
 g_1(x)&=&\exp\left[-\frac{n}{2a^2}\left(x-\frac{a^2\mu_\lt}{n}\right)^2\right].\label{2.14a}
 \end{eqnarray}
We employ the error function ``${\rm erf}$'' and the function ``${\rm sign}$'' which yields the sign of the argument. The constant is equal to
 \begin{equation}
 \frac{1}{c}=(-1)^{\nu(\nu-1)/2+n(n-1)/2}\left(\frac{16\pi}{n}\right)^{n/2}(2\pi)^{\nu/2}n^{-\nu^2/2-n(n+\nu)}\prod\limits_{j=0}^{n}j!\prod\limits_{j=0}^{n+\nu}j!,
\label{2.14b}
\end{equation}
 and is essentially the volume of the coset $[\U(n)\times\U(n+\nu)]/[\mathbf{S}(n)\times\mathbf{S}(n+\nu)]$.

The two-point weight $g_2$ consists of two parts. The first term, $g_{\rm r}$, 
represents a pair of real modes where one eigenvalue corresponds 
to a right-handed eigenvector and the other one to a 
left-handed one. 
The second term, $g_{\rm c}$, enforces that a complex eigenvalue comes with 
its complex conjugate, only. 
The function $g_1$ is purely Gaussian. As we will see in the next subsection,
in the small $a$ limit this will result in a distribution of the former
zero modes that is broadened to 
the Gaussian Unitary Ensemble (GUE)
\cite{Akemann:2010zp,AN,Kieburg:2011uf,DWW,DHS,Damgaard:2012gy}. 

For $N_{\rm f}$ dynamical quarks with quark mass $m_f$ the joint probability density is simply given by
\cite{Kieburg:2012fw}
\be
p^{(N_{\rm f})}(z) = \prod_{f=1}^{N_{\rm f}} \prod_{k=1}^{2n+\nu} (z_k+m_f)p(Z).
\ee

The expansion in $g_{\rm c}$ yields the joint probability density 
for a fixed number of complex conjugate pairs,
\begin{eqnarray}\label{2.15}
 p_l(Z_l)d[Z_l]&=&\frac{(-1)^{(n-l)l}c(1+a^2)^{-n(n+\nu-1/2)}a^{-n-\nu^2}n!(n+\nu)!}{(n-l)!l!(n+\nu-l)!}\exp\left[-\frac{a^4}{4(1+a^2)}(\mu_\rt-\mu_\lt)^2\right]\\
 &&\hspace*{-0cm}\times\Delta_{2n+\nu}(Z)\det\left[\begin{array}{c} \displaystyle\{g_r(x_i^{(1)},x_j^{(3)})dx_i^{(1)}dx_j^{(3)}\}\underset{1\leq j \leq n+\nu-l}
{\underset{1\leq i\leq n-l}{}} \\ \displaystyle\{(x_j^{(3)})^{i-1}g_1(x_j^{(3)})dx_j^{(3)}\}\underset{1\leq j \leq n+\nu-l}{\underset{1\leq i\leq \nu}{}} \end{array}\right]\prod\limits_{j=1}^lg_c(z_j^{(2)})dx_j^{(2)}dy_j^{(2)}.\nonumber
\end{eqnarray}
The factorials in the prefactor are the combinatorial factor which results 
from the expansion of the  determinant in co-factors with 
 $l$ columns and $l$ rows less. 
Note that they correspond to the coset of finite groups, 
$[\mathbf{S}(n)\times\mathbf{S}(n+\nu)]/[\mathbf{S}(n-l)\times\mathbf{S}(l)\times\mathbf{S}(n+\nu-l)]$,
 which naturally occurs when diagonalizing $D_\W$ in a fixed sector, 
see the discussion after Eq.~\eqref{2.8}.

\subsection{The continuum limit}\label{sec2.3}

In this section, we take the continuum limit of the joint 
probability density $p$, i.e. $a\to0$ at fixed $z$, $\mu_\rt$ and 
$\mu_\lt$. In this limit the probability density~\eqref{2.2} of $D_\W$ 
trivially becomes the one of chiral RMT which is equivalent to continuum QCD 
in the $\epsilon$-regime \cite{Shuryak:1992pi}. We expect that this is 
also the case for the joint probability density.

The small $a$ limit of the two point weight~\eqref{2.12a} is given by
\begin{eqnarray}
 g_2(z_1,z_2)&\overset{a\ll1}{=}&-2\imath\ \sign(y_1)\sqrt{\frac{a^2\pi}{n}}\exp\left[-ny_1^2\right]\delta(x_1)\delta(x_2)\delta(y_1+y_2).
\label{2.16}
\end{eqnarray}
The function $g_{\rm r}$ vanishes due to the error function 
which cancels with the sign function. The expansion of the 
determinant~\eqref{2.12} yields $(n+\nu)!/\nu!$ terms which are 
all the same. Thus, we have
\begin{eqnarray}
 \underset{a\to0}{\lim}\ p(Z) d[Z]&=& c (-1)^{\nu(\nu-1)/2}\frac{(n+\nu)!}{\nu!}\left(-2\imath\sqrt{\frac{\pi}{n}}\right)^n\label{2.17}\\
 &&\hspace*{-0cm}\times\underset{a\to0}{\lim}\ a^{-\nu^2}\Delta_{2n+\nu}(\imath y,-\imath y, x)\Delta_\nu(x)\prod\limits_{j=1}^n\sign(y_j)\exp\left[-ny_j^2\right]dy_j\prod\limits_{j=1}^{\nu}\exp\left[-\frac{n}{2a^2}x_j^2\right]dx_j.\nonumber
\end{eqnarray}
Thereby we have already evaluated the Dirac delta-functions. 
The real part of the complex eigenvalues $z_{j}^{(\rt/\lt)}$, $1\leq j\leq n$, 
and the imaginary part of $z_{j}^{(\lt)}$, $n+1\leq j\leq n+\nu$, vanish and they 
become the variables $\pm\imath y_j$, $1\leq j\leq n$, 
and $x_j$, $1\leq j\leq \nu$, respectively. Note, that 
the random variables $x$ scale with $a$ while $y$ is of order $1$.
 Therefore the distribution of the two sets of eigenvalues
factorizes into a product that can be identified as
the  joint probability density of a $\nu\times\nu$ dimensional 
GUE on the scale of $a$ and the chiral Unitary Ensemble on the scale 1,
\begin{eqnarray}
 \underset{a\to0}{\lim}\ p(Z)d[Z] &=&
 \frac{1}{(2\pi)^{\nu/2}}\left(\frac{n}{a^2}\right)^{\nu^2/2}\prod\limits_{j=0}^\nu\frac{1}{j!}\Delta^2_\nu(x)
\prod\limits_{j=1}^{\nu}\exp\left[-\frac{n}{2a^2}x_j^2\right]dx_j
\nn\\
&&\hspace*{-0cm}\times\frac{n^{n^2+\nu n}}{n!}\prod\limits_{j=0}^{n-1}\frac{1}{(j+\nu)!j!}
\Delta_{n}^2( y^2)\prod\limits_{j=1}^n 2\Theta(y_j)y_j^{2\nu+1}\exp\left[-ny_j^2\right]dy_j,\label{2.18}
\end{eqnarray}
where $\Theta$ is the Heaviside distribution.

\section{From the Joint Probability Density to the Level Densities}\label{sec3}

The level density is  obtained by integrating the joint probability 
density~\eqref{2.12} over all eigenvalues of $D_\W$ except  one. 
We can choose to exclude an eigenvalue of $z^{(\rt)}$ or one of the $z^{(\lt)}$'s.
When we exclude $z_1^{(\rt)}$ we have to expand the determinant (\ref{2.12}) 
with
respect to the first row. All resulting terms are the same and consist of a term for which $z_1^{(\rt)}$ is complex and a term for which $z_1^{(\rt)}$
is real. We thus have
 \cite{Kieburg:2011uf}
\begin{eqnarray}
\int p(Z)\prod\limits_{z_j\neq z_1^{(\rt)}}d[z_j]&=&\rho_\rt(x_1^{(\rt)})\delta(y_1^{(\rt)})+\frac{1}{2}\rho_{\rm c}(z_1^{(\rt)}).\label{3.1}
\ee
When excluding $z_1^{(\lt)}$ and expanding the determinant (\ref{2.12}) with
respect to the first column we notice that the first $n$ terms are the same
while the remaining $\nu$  terms
have to be treated separately.  Again the spectral density is the
sum of the density of the
 real modes, which are left-handed in this case,  and the density of the complex modes 
 \cite{Kieburg:2011uf}
\be
\int p(Z)\prod\limits_{z_j\neq z_1^{(\lt)}}d[z_j]&=&\rho_\lt(x_1^{(\lt)})\delta(y_1^{(\lt)})+\frac{1}{2}\rho_{\rm c}(z_1^{(\lt)}).\label{3.2}
\ee
The level densities $\rho_{\rt}$ and $\rho_{\lt}$ are the densities of 
the real right- and left-handed modes, respectively. 
Interestingly the level density of the complex modes appears 
 symmetrically in both equations. 
The reason is the vanishing chirality of  
eigenvectors corresponding to the  complex eigenvalues.

Let us consider the case when excluding $z_1^{(\rt)}$. The Vandermonde determinant without a factor  
$(z_1^{(\rt)}-z_1^{(\lt)})\prod_{k=2}^n(z_1^{(\rt)}-z_k^{(\rt)})(z_1^{(\lt)}-z_k^{(\rt)})\prod_{j=2}^{n+\nu}(z_1^{(\rt)}-z_j^{(\lt)})(z_1^{(\lt)}-z_j^{(\lt)})$ and 
the cofactor from expanding the first row of the determinant
is equal to  the joint probability density with one pair  
$(z^{(\rt)}, z^{(\lt)})$ less. The $z_1^{(\lt)}$-integral over this distribution  together
with the factor $\prod_{k=2}^n(z_1^{(\rt)}-z_k^{(\rt)})(z_1^{(\lt)}-z_k^{(\rt)})\prod_{j=2}^{n+\nu}(z_1^{(\rt)}-z_j^{(\lt)})(z_1^{(\lt)}-z_j^{(\lt)})$
can be identified as  the partition function with two additional flavors.
We thus find  
\begin{eqnarray}
\rho_\rt(x)&\propto&\int\limits_{-\infty}^\infty g_{\rm r}(x,x^\prime)(x-x^\prime)Z_{N_{\rm f}+2}^{n-1, \nu}(x,x^\prime,m_k)dx^\prime, \label{3.3}\\
\rho_{\rm c}(z)&\propto&g_{\rm c}(z)(z-z^*)Z_{N_{\rm f}+2}^{n-1, \nu}(z,z^*, m_k).\label{3.4}
\end{eqnarray}
The fermionic partition function is given by
\begin{eqnarray}
Z_{N_{\rm f}+2}^{n-1,\nu}(z_1,z_2, m_k)&=&\int \det(D_\W-z_1\eins_{2n+\nu-2})\det(D_\W-z_2\eins_{2n+\nu-2})
\prod_{k=1}^{N_{\rm f}}(D_\W+m_k\eins_{2n+\nu-2}) P(D_\W)d[D_\W],
\label{3.5}
\end{eqnarray}
where $D_\W$ is given as in Eq.~\eqref{2.1} only that $n$ is replaced by $n-1$. In the microscopic limit this is simply a unitary matrix integral which can
be easily numerically  evaluated.
Note that the integral over the variables $\mu_{\rt/\lt}$ which introduces the low energy constants $W_{6/7}$ can already be performed at this step.

Considering the exclusion of $z_{1}^{(\lt)}$ 
we have  to expand the determinant in the joint probability density with respect
to the first column
resulting in a much more complicated expression
\be\label{h1}
\rho_\lt(x) &\propto&  n\int\limits_{\mathbb{C}} d[\widetilde{z}] (x-\widetilde{z})g_2(x,\widetilde{z}) Z^{n-1,\nu}_{N_{\rm f}=2}(x,\widetilde{z})+ \alpha\delta(y)\sum_{p=1}^{\nu} (-1)^{\nu-p} 
\nn \\ && \times\left ( \begin{array}{c} n+\nu-1 \\ \nu-p \end{array} \right )x^{p-1} g_1(x) \nn 
\int\limits_{\mathbb{R}^{\nu-p}} \prod\limits_{j=1}^{\nu-p}dx_jx_j^pg_1(x_j)
 \Delta_{\nu-p}(x_1,\cdots ,x_{\nu-p}) \nn \\ && \times \Delta_{\nu-p+1}(x,x_1,\cdots ,x_{\nu-p})
Z^{n,\, p}_{N_{\rm f}= \nu-p+1}(x,x_1,\cdots ,x_{\nu-p})
\ee
with a certain constant $\alpha$ which we will specify in the microscopic limit. The global proportionality constant is up to a factor $n$ the same as the one in Eqs.~\eqref{3.3} and \eqref{3.4}. Again $g_2(z,z_\rt)$ is the sum of a term comprising the density of the complex
eigenvalues and a term giving the real eigenvalue density.
For the complex eigenvalue density we find the same expression as obtained
by integration over $z_{1}^{(\lt)}$.

For  $\nu =1$, the density of the real eigenvalues  simplifies to
\be
\rho_\lt(x)|_{\nu=1} \propto n\int\limits_{-\infty}^\infty dx_\rt (x_\rt-x)g_r(x_\rt,x) Z^{n-1,1}_{N_{\rm f}=2}(x_\rt,x)
+ \alpha g_1(x) 
Z^{n,\, 0}_{N_{\rm f}= 1}(x)
\ee
since there is no integration in the term proportional to $\alpha$, cf. Eq.~\eqref{h1}. The distribution of chirality over the real modes is the difference
\begin{eqnarray}
\rho_\chi(x)&=&\rho_\lt(x)-\rho_\rt(x),\label{3.6}
\end{eqnarray}
resulting in
\be \rho_\chi|_{\nu=1}  \sim \alpha g_1(x) 
Z^{n,\, 0}_{N_{\rm f}= 1}(x)+ 
n\int\limits_{-\infty}^\infty dx'(x'-x)(g_r(x',x) + g_r(x,x')) Z^{n-1,1}_{N_{\rm f}=2}(x',x),\label{rhochinu1}
\ee
where we used that the two-flavor partition function is symmetric in $x$ and $x'$.
For $\mu_r=\mu_l$, the last two terms cancel resulting in a very simple
expression for $\rho_\chi(x)$.  
Note that the integral over the second term always vanishes such that it does
not contribute to the normalization of the distribution of chirality over the real modes
\be
\int dx \rho_\chi(x) =\nu
\ee
which is $1$ for $\nu=1$. For $\nu = 2$ we find 
\be
\rho_\chi(x)|_{\nu=2}&\sim&\alpha x g_1(x) Z_{N_{\rm f}=1}^{n,1}(x) - \alpha (n+1)g_1(x) \int\limits_{-\infty}^\infty dx' x' (x-x')g_1(x')Z_{N_{\rm f} =2}^{n,0}(x,x') \nn \\
&&+n\int\limits_{-\infty}^\infty dx'(x'-x)(g_r(x',x) +g_r(x,x')) Z^{n-1,2}_{N_{\rm f}=2}(x',x).
\label{rhochinu2}
\ee
In the microscopic limit the two-flavor partition functions can be
replaced by a unitary matrix integral which still can be easily numerically 
evaluated including the integrals over $\widehat{m}_6$ and $\widehat{\lambda}_7$.

For large values of $\nu$ the expression of the distribution of chirality over the real modes obtained
from expanding the determinant
gets increasingly complicated.
However, there is an alternative expression in terms  of a supersymmetric partition function 
\cite{Splittorff:2011bj,Kieburg:2011uf}, 
\begin{eqnarray}
\rho_\chi(x)&\propto&\underset{\varepsilon\to0}{\lim}\IM\left.\frac{\partial}{\partial J}\right|_{J=0}\int \frac{\det(D_\W-(x+J)\eins_{2n+\nu})}{\det(D_\W-x\eins_{2n+\nu}-\imath\varepsilon\gamma_5)}P(D_\W)d[D_\W].\label{3.7}
\end{eqnarray}
In the ensuing sections we will use this expression to calculate the microscopic limit of the distribution of chirality over the real modes.

\subsection{Microscopic Limit of the Eigenvalue Densities}

The goal of  this section is to  derive 
the microscopic limit of  $\rho_{\rm r}$, $\rho_\chi$, and 
$\rho_{\rm c}$ including those terms involving non-zero values of $W_6$ and $W_7$ in the chiral Lagrangian. We only give results for the quenched case.
It is straightforward to include dynamical quarks but this will be
worked out in a forthcoming publication. The result for the distribution of chirality over the real modes with dynamical quarks for $W_6=W_7=0$
was already given in \cite{Splittorff:2011bj}, and an explicit expression for the density of the
complex eigenvalues in the presence of dynamical quarks and non-zero values
$W_6$, $W_7$ and $W_8$
was derived in \cite{Kieburg:2012fw}. 

The microscopic limit of the spectral densities is obtained
from the microscopic limit of the partition functions and the functions appearing in the joint probability density. We remind the reader that
the microscopic parameters which are kept fixed for $V\to \infty$, are defined by
\be
\ha_6^2 &=& - \widetilde a^2 V W_6,\qquad
\ha_7^2 = - \widetilde a^2 V W_7,\qquad
\ha_8^2 =  na^2/2=\widetilde a^2 V W_8,\\
\hma &=&  a^2 (\mu_r+\mu_l),\qquad
\hla =   a^2 (\mu_r-\mu_l),\qquad
\hx = 2n x.\nonumber
\ee 
The microscopic limit of the probability density
of $\hm_6$ and $\hla$ is given by
\be 
 p(\hma, \hla) &=& \frac 1{16\pi \widehat a_6 \widehat a_7} \exp\left[-\frac{\hma^2}{16\ha_6^2} -\frac{\hla^2}{16\ha^2_7 }\right],
\ee
and the functions that appear in the joint probability density simplify to
\be
\widehat g_\rt(\hx,\hx',\hma,\hla) & =&\exp\left[-\frac{(\hx+\hx'-2\hma)^2}{32 \ha^2_8}\right]
\left [ \sign(\hx-\hx') -{\rm erf} \left [
\frac {(\hx-\hx')/2 - \hla} {\sqrt{8\ha_8} }  \right ] \right ],\label{term1}\\
\widehat g_{\rm c}(\hz) &=& -2\imath\, \sign(\hy) \exp\left[-\frac{(\hx-\hma)^2}{8\ha^2_8 }\right],\\
\widehat g_1(\hx) &=&\exp\left[-\frac{(\hx-\hma + \hla)^2}{16 \ha^2_8}\right].
\ee
The microscopic limit of the spectral densities obtained in Eqs.
\eqref{3.3}, \eqref{3.4} and \eqref{3.7} is given by
\be
\rho_\rt(\hx) &=& \frac {1}{32\sqrt{2\pi}\widehat{a}_8}\int_{\mathbb{R}^3} d\hma d\hla d\widehat x' p(\hma,\hla) (\widehat{x}-\widehat{x}^\prime)\widehat g_\rt(\hx,\hx',\hma,\hla) \nn \\ && \times Z_{2/0}^\nu(\hx+\hma,\hx'+\hma,\hla,\ha_8),\label{rhor}\\
\rho_{\rm c}(\hz) &=& \frac {\imath\widehat{y}}{32\sqrt{2\pi}\widehat{a}_8}\int_{\mathbb{R}^2} d\hma d\hla p(\hma,\hla)  \widehat g_c(\hz ,\hz^*,\hm_6 )\nonumber\\
&&\times Z_{2/0}^\nu(\hz+\hma,\hz^*+\hma,\hla,\ha_8),\label{rhoc}\\
\rho_\chi(\hx) &=&\frac 1{\pi} \underset{\varepsilon\to0}{\lim}{\rm Im} \int d\hma d\hla p(\hma,\hla) 
 G_{1/1}(\hx+\hma,\hla+\imath\varepsilon,\ha_8).\label{rhochi}
\ee
The resolvent  $G_{1/1}$ follows from the graded partition function
\begin{eqnarray}\label{Greenfunc}
 G_{1/1}(\hx+\hma,\hla+\imath\varepsilon,\ha_8) &=& \left .  \frac d{d\hx'} Z_{1/1}^\nu(\hx+\hma,\hx'+\hma,\hla+\imath\varepsilon,\ha_8) \right  |_{\hx'=\hx}\\
 &=&\underset{n\to\infty}{\lim}\frac{1}{2n}\int \tr\frac{1}{D_\W-2n\hx\eins_{2n+\nu}-\imath\varepsilon\gamma_5}P(D_\W)d[D_\W].\nonumber
\end{eqnarray}
The two-flavor partition function is up to a constant defined by
\begin{eqnarray}
 Z_{2/0}^\nu(z_1+ m_6,z_2+m_6,\lambda_7,a)&\propto&\int \det(z_1\eins_{2n+\nu-2}-D_\W)\det(z_2\eins_{2n+\nu-2}-D_\W)P(D_\W)d[D_\W].\nonumber\\ \label{twoflav}
\end{eqnarray}
The microscopic limit of the two-flavor partition function follows from the chiral Lagrangian~\eqref{2.2a}. In the diagonal representation of the unitary $2\times 2$ matrix, it
can be simplified by means of an Itzykson-Zuber integral and is given by
\be
Z_{2/0}^\nu(\widehat z_1,\widehat z_2,\widehat \lambda_7,\widehat a_8)& =& \frac {1}{2\pi^2}\int d\varphi_1 d\varphi_2  
\sin^2 ((\varphi_1-\varphi_2)/2)e^{\imath \nu (\varphi_1+\varphi_2)}\exp\left[\imath\widehat \lambda_7(\sin\varphi_1+\sin\varphi_2)-4 \widehat a_8^2(\cos^2 \varphi_1 + \cos^2 \varphi_2)\right]
 \nn\\
&&\times\frac{ \exp\left[\widehat z_1\cos\varphi_1 +\widehat z_2\cos\varphi_2\right]-\exp\left[\widehat z_2\cos\varphi_1 +\widehat z_1\cos\varphi_2\right]}{(\cos\varphi_1-\cos \varphi_2)(\widehat z_1-\widehat z_2)}. \label{z2diag}
\ee
The normalization is chosen such that we find the well known result \cite{Verbaarschot:1995yi},
\be\label{Z2cont}
Z_{2/0}^\nu(\widehat z_1,\widehat z_2,\widehat \lambda_7=0,\widehat a_8=0)& =& \frac{\widehat z_1I_{\nu+1}(\widehat z_1)I_\nu(\widehat z_2)-\widehat z_2I_{\nu+1}(\widehat z_2)I_\nu(\widehat z_1)}{\widehat z_1^2-\widehat z_2^2}, \hspace*{0.5cm}
\ee
 at vanishing lattice spacing, where $I_\nu$ is the modified Bessel function of the first kind.

The microscopic limit of  the graded partition function follows from the
chiral Lagrangian \cite{Damgaard:2010cz}  which can be written as an integral over a $(1/1)\times(1/1)$ supermatrix \cite{Splittorff:2011bj}
\begin{eqnarray}\label{3.2.7}
 U=\left[\begin{array}{cc} e^\vartheta & \eta^* \\ \eta & e^{\imath\varphi} \end{array}\right],\ \vartheta\in\mathbb{R},\ \varphi\in[0,2\pi],
\end{eqnarray}
with $\eta$ and $\eta^*$ two independent Grassmann variables, see Refs.~\cite{Guh06,Som07,LSZ08,KSG09,Guh11} for the supersymmetry method in random matrix theory. Let the normalization of the integration over the Grassmann variables be
\begin{eqnarray}\label{3.2.8}
 \int \eta^*\eta d\eta d\eta^*=\frac{1}{2\pi}.
\end{eqnarray}
Then the graded partition function is
\begin{eqnarray}\label{3.2.9a}
 Z_{1/1}^\nu(\widehat z_1,\widehat z_2,\hla\pm\imath\varepsilon,\ha_8)&=&\int \frac{\imath d e^{\imath\varphi}}{2\pi}d e^{\vartheta}d\eta d\eta^* \Sdet^\nu U\exp[-\widehat{a}^2_8\Str(U^2+U^{-2})]\\
 &&\times\exp\left[\pm\frac{\imath}{2}\Str\widehat Z(U-U^{-1})-\left(\varepsilon\pm\frac{\imath\widehat{\lambda}_7}{2}\right)\Str(U+U^{-1})\right],\nonumber
\end{eqnarray}
where $\widehat Z = {\rm diag}(\widehat z_1,\widehat z_2)$ and the normalization adjusted by the continuum limit \cite{SplVer03,FyoAke03}
\be\label{Z11cont}
Z_{1/1}^\nu(\widehat z_1,\widehat z_2,\hla=0,\ha_8=0)=\widehat z_1K_{\nu+1}(\widehat z_1)I_\nu(\widehat z_2)-\widehat z_2I_{\nu+1}(\widehat z_2)K_\nu(\widehat z_1).
\ee
The function $K_\nu$ is the modified Bessel function of the second kind. 

There are various ways to calculate the integral~\eqref{3.2.9a}. One possibility is a brute force evaluation of the Grassmann integrals
as in \cite{Damgaard:2010cz,Rasmuss}. Then the Gaussian integrals over $\hma$ and $\hla$ can be performed analytically leaving us with a non-singular
two-dimensional integral. A second possibility would be to rewrite the integrals as in \cite{Splittorff:2011bj}. Then we  end
up with a two dimensional singular integral (see Appendix~\ref{app3})
which can be evaluated numerically with some effort. The third way to evaluate
the integral, is a variation
of the method in \cite{Splittorff:2011bj} and results in a one dimensional integral and a sum over Bessel functions that can be  easily numerically evaluated (see section \ref{sec4.3}).

\section{The Eigenvalue Densities and their Properties}\label{sec4}

To illustrate the effect of non-zero $\ha_6$ and $ \ha_7$ we first discuss the case $\ha_8 = 0$.
For the general case, with $\ha_8$ also non-zero,  we will discuss
 the density of the additional real eigenvalues, 
the density of the complex eigenvalues, and finally the distribution of chirality over the real eigenvalues of $D_\W$.

\subsection{Spectrum of $D_\W$ for $\ha_8 =0$}.

The low-energy constants
 $\ha_6$ and $\ha_7$ are introduced 
through the addition of  the Gaussian stochastic variables
$\widehat{m}_6 + \widehat{\lambda}_7 \gamma_5 $ to $D_\W|_{a=0}$ resulting in the massive Dirac operator \cite{footnote}
\be
D= D_\W|_{a=0} +(\widehat{m} + \widehat{m}_6)\eins + \widehat{\lambda}_7 \gamma_5.
\ee
For $\ha_8=0$ the Dirac operator $D_\W|_{a=0}$ is anti-Hermitian, 
and the eigenvalues of $D_\W(\widehat{\lambda}_7,\widehat{m}_6)=D-\widehat{m}$ are given by
\be
\widehat{z}_\pm = \widehat{m}_6\pm \imath\sqrt{\lambda_\W^2-\widehat{\lambda}_7^2},
\label{ev67}
\ee
where $\imath\lambda_\W$ is an eigenvalue of $D_\W|_{a=0}$. 
The density of the eigenvalues of $ D$ is obtained after integrating over the Gaussian distribution of $\widehat{m}_6$ and $\widehat{\lambda}_7$.

As can be seen from Eq. (\ref{ev67}), in case $\ha_6=\ha_8 =0 $ 
and $\ha_7\ne 0$, the eigenvalues of $D$ are either purely 
imaginary or purely real  depending on whether $\widehat{\lambda}_7$ is smaller or larger than $\lambda_\W$, respectively.
Paired imaginary eigenvalues penetrate the real axis only through the origin when varying 
$\widehat{\lambda}_7$, see Fig.~\ref{fig1}. Introducing a non-zero $W_6$, broadens the spectrum
by a Gaussian parallel to the real axis but nothing crucial happens because $\widehat{m}_6$ is just an additive constant to the eigenvalues, cf. Fig.~\ref{fig1}.

In the continuum  the low lying spectral density of the quenched Dirac operator is given by \cite{Shuryak:1992pi}
\begin{eqnarray}
\label{4.0.2}
\rho_{\rm cont.}(\widehat{z})&=&\delta(\widehat{x})\left[\nu\delta(\widehat{y})+\frac{|\widehat{y}|}{2}(J_\nu^2(\widehat{y})-J_{\nu-1}(\widehat{y})J_{\nu+1}(\widehat{y}))\right]=\delta(\widehat{x})\left[\nu\delta(\widehat{y})+\rho_{\rm NZ}(\widehat{y})\right].
\end{eqnarray}
The function $J_\nu$ is the Bessel function of the first kind. The level density $\rho_{\rm NZ}$ describes the density of the generic non-zero eigenvalues, only.

For non-zero $W_{6/7}$ the  
distribution of
the zero modes represented 
by the Dirac delta-functions in Eq.~\eqref{4.0.2} is broadened by a Gaussian,  i.e.
\begin{eqnarray}\label{4.0.3}
 \rho_\chi(\widehat{z},\ha_8=0)&=&\frac{\nu}{\sqrt{16\pi(\widehat{a}_6^2+\widehat{a}_7^2)}}\exp\left[-\frac{\widehat{x}^2}{16(\widehat{a}_6^2+\widehat{a}_7^2)}\right].
\end{eqnarray}
Complex modes have vanishing chirality and do not contribute to the distribution of chirality over the real modes. Additional pairs of real modes
also do not contribute to $ \rho_\chi$. The reason is the symmetric integration of $\hla$ over the real axis. The eigenvalues remain the same under the change $\widehat{\lambda}_7\to-\widehat{\lambda}_7$, see Eq.~\eqref{ev67}. However the corresponding eigenvectors interchange the sign of the chirality which can be seen by the symmetry relation
\begin{eqnarray}\label{ident}
 D_\W(\widehat{\lambda}_7,\widehat{m}_6)=-\gamma_5 D_\W(-\widehat{\lambda}_7,-\widehat{m}_6)\gamma_5.
\end{eqnarray}
Thus the normalized eigenfunctions ($\langle\psi_\pm|\psi_\pm\rangle=1$) corresponding to the eigenvalues $\widehat{z}_\pm$, i.e.
\begin{eqnarray}\label{eveq1}
D_\W(\widehat{\lambda}_7,\widehat{m}_6)|\psi_\pm\rangle=\widehat{z}_\pm|\psi_\pm\rangle,
\end{eqnarray}
 also fulfills the identity
\begin{eqnarray}\label{eveq2}
 D_\W(-\widehat{\lambda}_7,-\widehat{m}_6)\gamma_5|\psi_\pm\rangle=-\widehat{z}_\pm\gamma_5|\psi_\pm\rangle.
\end{eqnarray}
Since the quark mass $\widehat{m}_6$ enters with unity we have also
\begin{eqnarray}\label{eveq3}
 D_\W(-\widehat{\lambda}_7,\widehat{m}_6)\gamma_5|\psi_\pm\rangle=\widehat{z}_\mp\gamma_5|\psi_\pm\rangle.
\end{eqnarray}
The wave-functions $\gamma_5|\psi_\pm\rangle$ share the same chirality with $|\psi_\pm\rangle$. Moreover $|\psi_+\rangle$ and $|\psi_-\rangle$ have opposite chirality because the pair of eigenvalues $\widehat{z}_\pm$ is assumed to be real and their difference $|\widehat{z}_+-\widehat{z}_-|$ non-zero. This can be seen by the eigenvalue equations
\begin{eqnarray}\label{eveq4}
D_\W(\widehat{\lambda}_7,\widehat{m}_6=0)|\psi_\pm\rangle&=&\pm\sqrt{\widehat{\lambda}_7^2-\lambda_\W^2}|\psi_\pm\rangle,\\
\langle\psi_\pm|D_\W(-\widehat{\lambda}_7,\widehat{m}_6=0)&=&\langle\gamma_5D_\W(-\widehat{\lambda}_7,\widehat{m}_6=0)\gamma_5\psi_\pm|=\mp\sqrt{\widehat{\lambda}_7^2-\lambda_\W^2}\langle\psi_\pm|.\nonumber
\end{eqnarray}
In the second equation we used the $\gamma_5$-Hermiticity of $D_\W$. We multiply the first equation with $\langle\psi_\pm|$ and the second with $|\psi_\pm\rangle$ and employ the normalization of the eigenmodes such that we find
\begin{eqnarray}\label{eveq5}
\langle\psi_\pm|D_\W(\widehat{\lambda}_7,\widehat{m}_6=0)|\psi_\pm\rangle&=&\pm\sqrt{\widehat{\lambda}_7^2-\lambda_\W^2},\\
\langle\psi_\pm|D_\W(-\widehat{\lambda}_7,\widehat{m}_6=0)|\psi_\pm\rangle&=&\mp\sqrt{\widehat{\lambda}_7^2-\lambda_\W^2}.\nonumber
\end{eqnarray}
We subtract the second line from the first and use the identity $D_\W(\widehat{\lambda}_7,\widehat{m}_6=0)-D_\W(-\widehat{\lambda}_7,\widehat{m}_6=0)=2\widehat{\lambda}_7\gamma_5$, i.e.
\begin{eqnarray}\label{eveq6}
\widehat{\lambda}_7\langle\psi_\pm|\gamma_5|\psi_\pm\rangle&=&\pm\sqrt{\widehat{\lambda}_7^2-\lambda_\W^2},
\end{eqnarray}
which indeed shows the opposite chirality of $|\psi_+\rangle$ and $|\psi_-\rangle$. Thus $|\psi_+\rangle$ and $\gamma_5|\psi_-\rangle$ have opposite sign of chirality but their corresponding eigenvalues are the same. Therefore the average of their chiralities at a specific eigenvalue vanishes.

The density of the complex eigenvalues can be obtained by integrating over those $\lambda_\W$ fulfilling the condition $|\lambda_\W| > |\widehat{\lambda}_7|$. After averaging 
over $\widehat{m}_6$ and $\widehat{\lambda}_7$ we find
\begin{eqnarray}
 \rho_{\rm c}(\widehat{z}=\hx+i\hy,\ha_8=0)
&=&\frac{\exp\left[-\widehat{x}^2/(16\widehat{a}_6^2)\right]}{16\pi|\widehat{a}_6\widehat{a}_7|}
 \int\limits_{\mathbb{R}^2}\rho_{\rm NZ}(\lambda_\W)\exp\left[-\frac{\widehat{\lambda}_7^2}{16\widehat{a}_7^2}\right]\delta\left(\sqrt{\lambda_\W^2-\widehat{\lambda}_7^2}-|\widehat{y}|\right)
\Theta(|\lambda_\W|-|\widehat{\lambda}_7|)d\lambda_\W d\widehat{\lambda}_7\nonumber\\
 &=&\frac {\exp\left[-\widehat{x}^2/(16\widehat{a}_6^2)\right]} {4\pi|\widehat{a}_6\widehat{a}_7|}
\int\limits_{|\widehat{y}|}^\infty \frac{|\widehat{y}|\rho_{\rm NZ}(\lambda_\W)d\lambda_\W}{\sqrt{\lambda_\W^2-\widehat{y}^2}}
\exp\left[\frac{\lambda_\W^2-\widehat{y}^2}{16\widehat{a}_7^2}\right ] .\label{4.0.4}
\end{eqnarray}
The original continuum result is smoothened by a distribution with a Gaussian tail.
The oscillations in the microscopic spectral density
dampen due to a non-zero  $W_7$ similar to  the effect of 
a non-zero value $W_8$, cf. Ref.~\cite{Kieburg:2011uf}.  
We also expect a loss of the height of the first eigenvalue distributions around the origin. Pairs of eigenvalues are
moving from the imaginary axis into the real axis and thus lowering their probability density on the imaginary axis.
The density $\rho_{\rm c}$ for non-zero $\ha_8$ will be discussed in full detail in Sec.~\ref{sec4.2}.

\begin{figure}
 \includegraphics[width=1\textwidth]{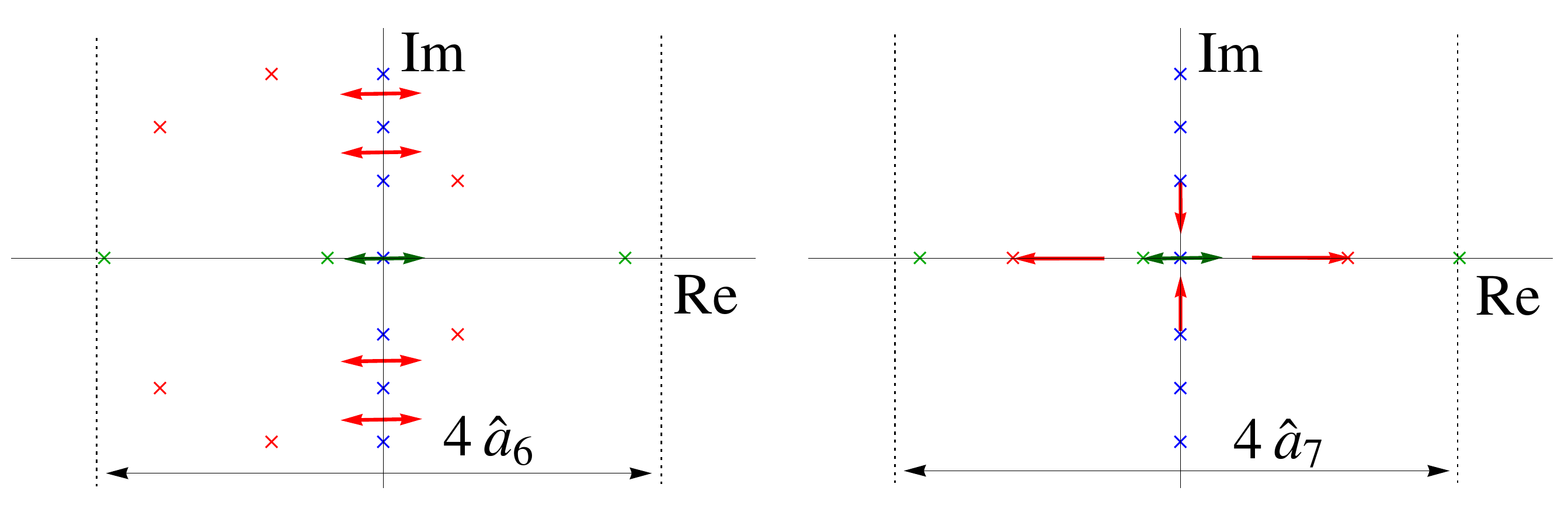}
 \caption{\label{fig1}Schematic plots of the effects of $W_6$ (left plot) and of $W_7$ (right plot). 
The low energy constant $W_6$  broadens the spectrum   
parallel to the real axis according to a Gaussian with
 width $4\ha_6=4\sqrt{-VW_6\tilde a^2}$,
but does not change the continuum spectrum 
in a significant way. 
When $W_7$ is switched on and $W_6=0$
the purely imaginary eigenvalues invade the real axis through the origin
and 
only the real (green crosses along the real axis) are broadened by a Gaussian 
with width  $4\widehat{a}_7=4\sqrt{-V W_7\widetilde{a}^2}$.}
\end{figure}

The density of the additional real modes can be obtained by integrating
the  continuum distribution, $\rho_{\rm NZ}$ over  $ |\lambda_\W| < |\lambda_7 |$
  analogous to the complex case. We find
\begin{eqnarray}
 \rho_{\rm r}(\widehat{x},W_8=0)
&=&\frac{1}{16\pi|\widehat{a}_6\widehat{a}_7|}\int\limits_{\mathbb{R}^3}
\rho_{\rm NZ}(\lambda_\W)\exp\left[-\frac{\widehat{m}_6^2}{16\widehat{a}_6^2}-\frac{\widehat{\lambda}_7^2}{16\widehat{a}_7^2}\right]\delta\left(\sqrt{\widehat{\lambda}_7^2-\lambda_\W^2}-|\widehat{m}_6-\widehat{x}|\right)
\Theta(|\widehat{\lambda}_7|-|\lambda_\W|)d\lambda_\W d\widehat{\lambda}_7d\widehat{m}_6\nonumber\\
 &=&\int\limits_{\mathbb{R}^2}\frac{|\widehat{m}_6|d\widehat{m}_6d \lambda_\W}
{8\pi|\widehat{a}_6\widehat{a}_7|\sqrt{\lambda_\W^2+\widehat{m}_6^2}} 
\rho_{\rm NZ}(\lambda_\W)\exp\left[-\frac{\lambda_\W^2+\widehat{m}_6^2}{16\widehat{a}_7^2}
-\frac{(\widehat{m}_6+\widehat{x})^2}{16\widehat{a}_6^2}\right] .\label{4.0.5}
\end{eqnarray}
The number of additional real modes given by the integral of $\rho_{\rm r}(\widehat x)$
 over $\widehat{x}$ only depends on $ \ha_7 $,  as it should be since 
 $\widehat{m}_6$ is just an additive constant to the eigenvalues.
 Moreover $\rho_{\rm r}$ will inherit the oscillatory behavior of
$\rho_{\rm NZ}$
 although most of it will be damped by the Gaussian cut-off. 
The mixture of this effect with the effect of a non-zero $W_8$ 
is highly non-trivial, but we expect that, at small lattice spacings, 
we can separate both contributions. For a  sufficiently small value of $\widehat{a}_6$
the behavior of $\rho_{\rm r}(\hx)$ 
for $\hx \to 0$ is given by $\rho_\rt(\widehat{x})=\widetilde{c}|\widehat{x}|+\ldots$ with $\widetilde{c}>0$ for vanishing $W_8$ and, thus,
 $\rho_\rt(\widehat{x})=c_0+c_1\widehat{x}^2+\ldots$ with $c_0,c_1>0$ for non-zero $W_8$. 
Hence, we will see a soft repulsion of the additional real eigenvalues from the origin which still allows real eigenvalues to be zero. 

 The discussion of  the real modes for non-zero $\ha_8$ as well
is given in Sec.~\ref{sec4.1}.

\subsection{Eigenvalue densities for non-zero values of $W_6$, $W_7$ and $W_8$}

In this subsection all three low-energy constants are non-zero. As in the previous subsection, we will consider
the density of the real eigenvalues of $D_\W$, the density of the complex eigenvalues of $D_\W$, and the
distribution of the chiralities over the real eigenvalues of $D_\W$. The expressions for these distributions were
already given in section \ref{sec3}, but in this section we further simplify them and calculate the asymptotic expressions
for large and small values of $\ha$.

\subsubsection{Density of the additional real modes}\label{sec4.1}

The quenched eigenvalue density of the additional real modes is given by 
Eq.~\eqref{rhor}. The Gaussian 
 average over the variables $\widehat{m}_6$ and $\widehat{\lambda}_7$ can be worked out analytically. The result is given
by (see Appendix~\ref{app2} for integrals that were used to obtain this result)
\begin{eqnarray}
\rho_{\rm r}(\widehat{x})=\frac{1}{ 16\pi^2}\int\limits_{[0,2\pi]^2}d\varphi_1d\varphi_2\sin^2\left[\frac{\varphi_1-\varphi_2}{2}\right]e^{\imath\nu(\varphi_1+\varphi_2)}\frac{\widetilde{k}(\widehat{x},\varphi_1,\varphi_2)-\widetilde{k}(\widehat{x},\varphi_2,\varphi_1)}{\cos\varphi_2-\cos\varphi_1}\label{4.1.2}
\end{eqnarray}
with
\begin{eqnarray}
 \widetilde{k}(\widehat{x},\varphi_1,\varphi_2)&=&\exp\left[4\widehat{a}_6^2(\cos\varphi_1-\cos\varphi_2)^2-4\widehat{a}_7^2(\sin\varphi_1+\sin\varphi_2)^2+4\widehat{a}_8^2 \left(\cos\varphi_1-\frac{\widehat{x}}{8\widehat{a}_8^2}\right)^2-4\widehat{a}_8^2 \left(\cos\varphi_2-\frac{\widehat{x}}{8\widehat{a}_8^2}\right)^2\right]\nonumber\\
 &&\hspace*{-2.3cm}\times\left[{\rm erf}\left[\frac{\widehat{x}-8(\widehat{a}_6^2+\widehat{a}_8^2)\cos\varphi_1+8\widehat{a}_6^2\cos\varphi_2}{\sqrt{8(\widehat{a}_8^2+2\widehat{a}_6^2)}}\right]+{\rm erf}\left[\frac{8(\widehat{a}_6^2+\widehat{a}_8^2)\cos\varphi_1-8\widehat{a}_6^2\cos\varphi_2-8\imath \widehat{a}_7^2\sin\varphi_1-8\imath \widehat{a}_7^2\sin\varphi_2-\widehat{x}}{\sqrt{16(\widehat{a}_8^2+\widehat{a}_6^2+\widehat{a}_7^2)}}\right]\right] .\nn\\
 &&\label{4.1.3}
\end{eqnarray}
 The effect of each low energy constant on $\rho_\rt$ is shown in Fig.~\ref{fig3}.

\begin{figure}
\center {\includegraphics[width=1\textwidth]{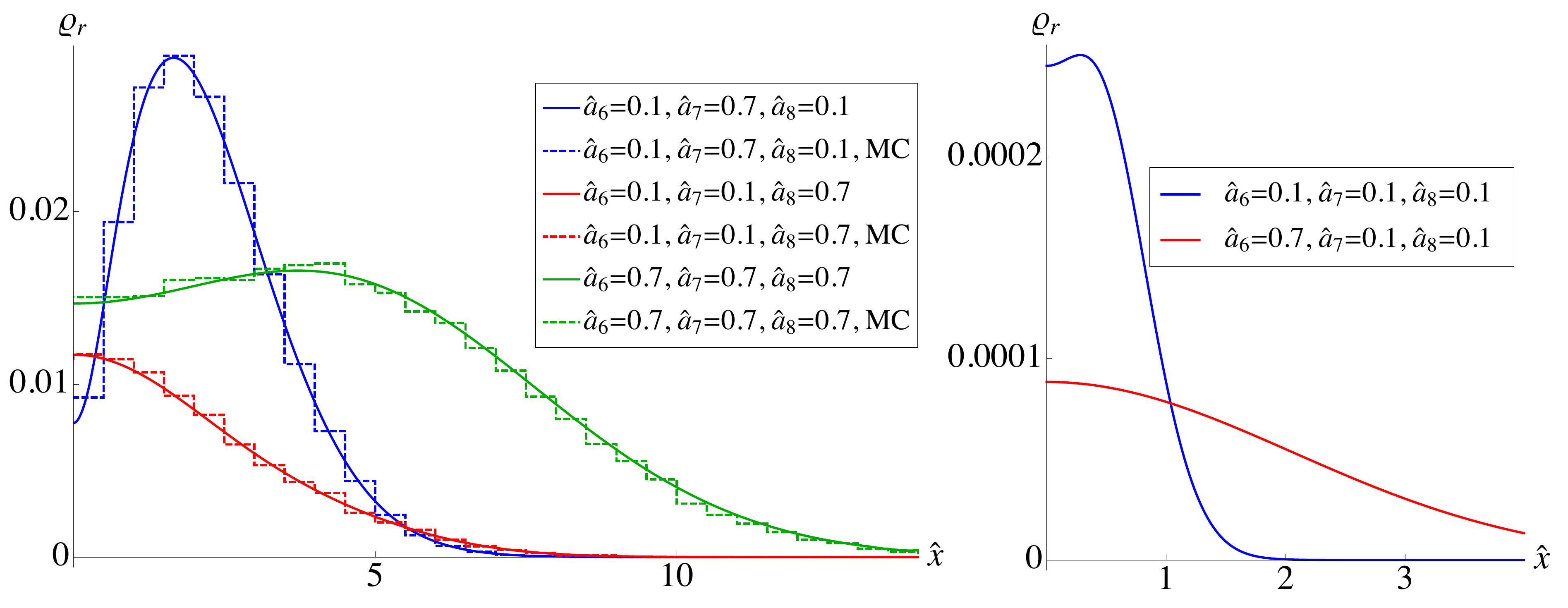}}
 \caption{\label{fig3} The density of additional real modes is shown for various  parameters $\widehat{a}_{6/7/8}$. 
The analytical results (solid curves)  agree with the Monte Carlo simulations of the Random Matrix Theory (histogram [MC] 
with bin size 0.5 and with different ensemble and matrix sizes such that statistics are about 1-5\%) for $\nu=1$. 
We plot only the positive real axis since $\rho_\rt$ 
is symmetric. Notice that the height of the two curves for $\widehat{a}_7=\widehat{a}_8=0.1$ (right plot) are two orders smaller than the height of the other 
curves (left plot) and because of bad statistics we have not performed simulations for this case. 
Notice the soft repulsion of the additional real modes from the origin at large $\widehat{a}_7=\sqrt{-VW_7}\widetilde{a}$ as discussed
in the introductory section. The parameter $\widehat{a}_6=\sqrt{-VW_6}\widetilde{a}$  smoothens the distribution.}
\end{figure}

At small lattice spacing, $\ha\ll1$, the density $\rho_{\rm r}$ has support on the scale of $\ha$. 
In particular it is given by derivatives of a specific function, i.e.
\begin{eqnarray}
\rho_{\rm r}(\widehat{x})&\overset{\widehat{a}\ll1}{=}&\frac{1}{4}\left.\left(\frac{1}{(\nu!)^2}\frac{\partial^{2\nu}}{\partial t_1^\nu\partial t_2^\nu}-\frac{1}{(\nu-1)!(\nu+1)!}\frac{\partial^{2\nu}}{\partial t_1^{\nu-1}\partial t_2^{\nu+1}}\right)\right|_{t_1=t_2=0}\frac{\widehat{k}(\widehat{x},t_1,t_2)-\widehat{k}(\widehat{x},t_2,t_1)}{t_2-t_1},\label{4.1.3a}
\end{eqnarray}
where
\begin{eqnarray}
 \widehat{k}(\widehat{x},t_1,t_2)&=&\exp\left[\widehat{a}_6^2(t_1-t_2)^2+\widehat{a}_7^2(t_1+t_2)^2+\widehat{a}_8^2 \left(t_1-\frac{\widehat{x}}{4\widehat{a}_8^2}\right)^2-\widehat{a}_8^2 \left(t_2-\frac{\widehat{x}}{4\widehat{a}_8^2}\right)^2\right]\nonumber\\
 &&\hspace*{-0cm}\times\left[{\rm erf}\left[\frac{\widehat{x}-4(\widehat{a}_6^2+\widehat{a}_8^2)t_1+4\widehat{a}_6^2t_2}{\sqrt{8(\widehat{a}_8^2+2\widehat{a}_6^2)}}\right]+{\rm erf}\left[\frac{4(\widehat{a}_6^2+\widehat{a}_7^2+\widehat{a}_8^2)t_1-4(\widehat{a}_6^2-\widehat{a}_7^2)t_2-\widehat{x}}{\sqrt{16(\widehat{a}_8^2+\widehat{a}_6^2+\widehat{a}_7^2)}}\right]\right].
\label{4.1.3b}
\end{eqnarray}
The error functions guarantee a Gaussian tail on the scale of  $\widehat{a}$. Furthermore, the height of the density 
is of order $\widehat{a}^{2\nu+1}$. Hence, additional real modes are strongly suppressed for $\nu > 0$ 
and the important contributions only result from $\nu=0$. This behavior  becomes clearer for the expression of
the average number of the additional real modes. This quantity directly follows from the result~\eqref{4.1.3},
\begin{eqnarray}\label{4.1.4}
 N_{\rm add}&=&2\int\limits_{-\infty}^\infty 
\rho_{\rm r}(\widehat{x})d\widehat{x}\\
 &=&\int\limits_0^{2\pi}\frac{d\Phi}{ 4\pi}\cos[2\nu\Phi]\frac{1-\exp\left[-(4\widehat{a}_8^2+8\widehat{a}_7^2) \sin^2\Phi\right]I_0\left[(4\widehat{a}_8^2-8\widehat{a}_7^2)\sin^2\Phi\right]}{\sin^2\Phi}\nonumber\\
 &=&\sum\limits_{n=\nu+1}^\infty\sum\limits_{j=0}^{\lfloor n/2\rfloor}(-1)^{\nu-1+n}\frac{(2n-2)!\left(\widehat{a}_8^2-2\widehat{a}_7^2\right)^{2j}(\widehat{a}_8^2+2\widehat{a}_7^2)^{n-2j}}{2^{2j-1}\Gamma(n-\nu)\Gamma(n+\nu)\Gamma(n-2j+1)(j!)^2},\nonumber
\ee
where the symbol $\lfloor n/2\rfloor$ denotes the largest integer smaller than or equal to $n/2$.

\begin{figure}
\center {\includegraphics[width=1\textwidth]{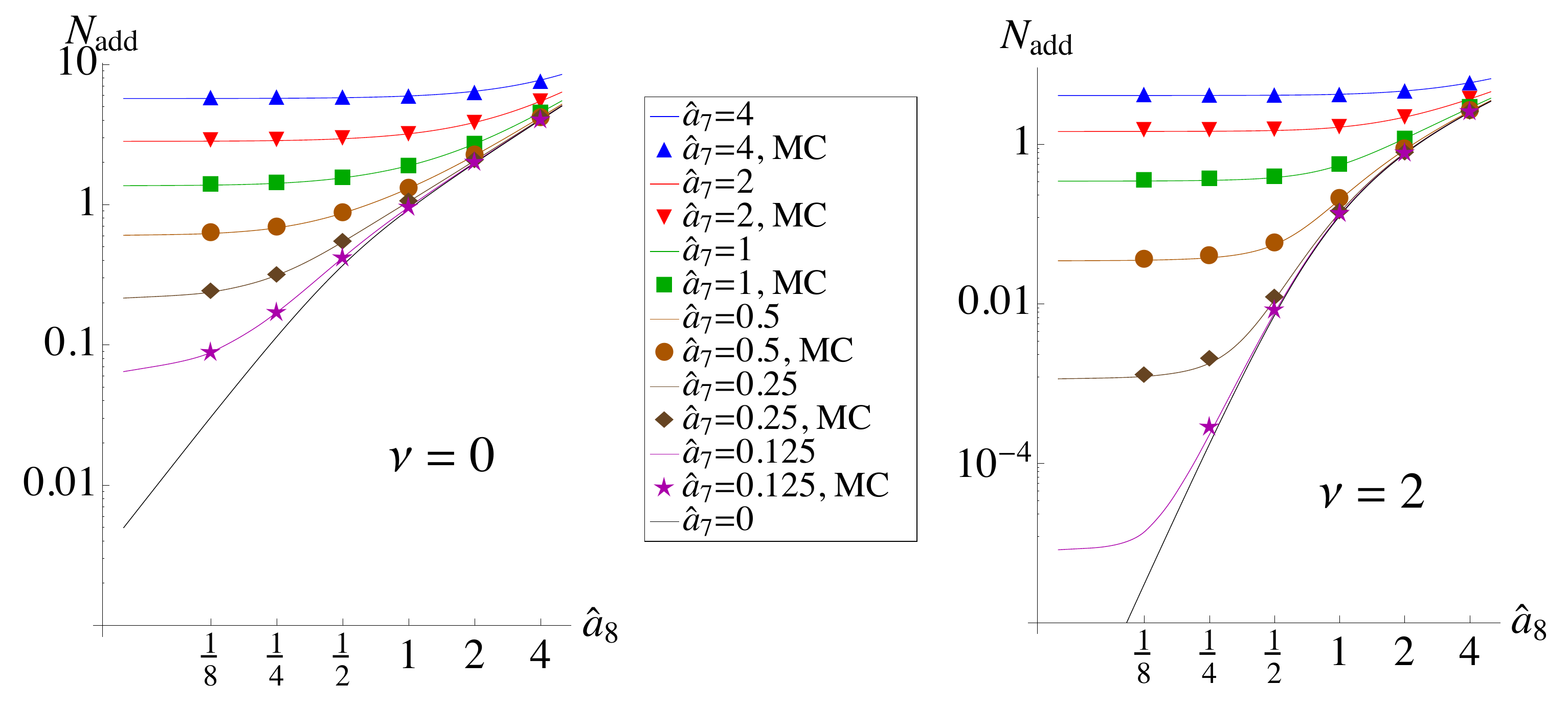}}
 \caption{\label{fig2}Log-log plots of $N_{\rm add}$ as a function 
of $\widehat{a}_8=\sqrt{VW_8\widetilde{a}^2}$ for $\nu=0$ (left plot) and $\nu=2$ (right plot). The analytical results (solid curves)  
are compared to Monte Carlo simulations of RMT
(symbols; ensemble and matrix size varies such that the statical error is about 1-5\%). Notice that $W_6$ has no 
effect on $N_{\rm add}$. The saturation around zero is due to 
a non-zero value of $\widehat{a}_7=\sqrt{-VW_7\widetilde{a}^2}$. For
 $\widehat{a}_7 =0 $ (lowest curves) the average number of additional real modes behaves like $\widehat{a}_8^{2\nu+2}$, see Ref.~\cite{Kieburg:2011uf}.}
\end{figure}

The average number of the real modes does not depend on the low energy constant 
$W_6=-\widehat{a}_6^2/(\widetilde{a}^2 V)$ because 
this constant induces overall fluctuations of
the Dirac spectrum parallel to the $\widehat{x}$-axis. 

The asymptotics of $N_{\rm add}$ at small and large lattice spacing is given by
\begin{eqnarray}\label{4.1.5}
 N_{\rm add}&=&\left\{\begin{array}{cl} \displaystyle\sum\limits_{j=0}^{\lfloor\nu/2\rfloor}
\frac{\left(\widehat{a}_8^2-2\widehat{a}_7^2\right)^{2j}(\widehat{a}_8^2+2\widehat{a}_7^2)^{\nu-2j+1}}{2^{2j-1}\Gamma(\nu-2j+2)(j!)^2}
\propto\ha^{2\nu+2}, & \widehat{a}\ll1,\\ \displaystyle\sqrt{\frac{64\widehat{a}_7^2}{\pi^3}}E\left(\sqrt{1-\frac{\widehat{a}_8^2}{2\widehat{a}_7^2}}\right)\propto \ha, & \widehat{a}\gg1,\end{array}\right.
\end{eqnarray}
see Appendix~\ref{app4.2} for a derivation. The function $E$ is the elliptic integral of the second kind, i.e
\begin{eqnarray}\label{4.1.5b}
 E(x)&=&\int\limits_{0}^{\pi/2}\sqrt{1-x^2\sin^2\varphi}d\varphi.
\end{eqnarray}
In Ref.~\cite{Kieburg:2011uf} this result was derived 
for $\ha_6 =\ha_7=0$. Notice that for large lattice spacings the number of additional real 
modes increases linearly with $\ha$ and is independent of $\nu$.

\begin{figure}
\center {\includegraphics[width=1\textwidth]{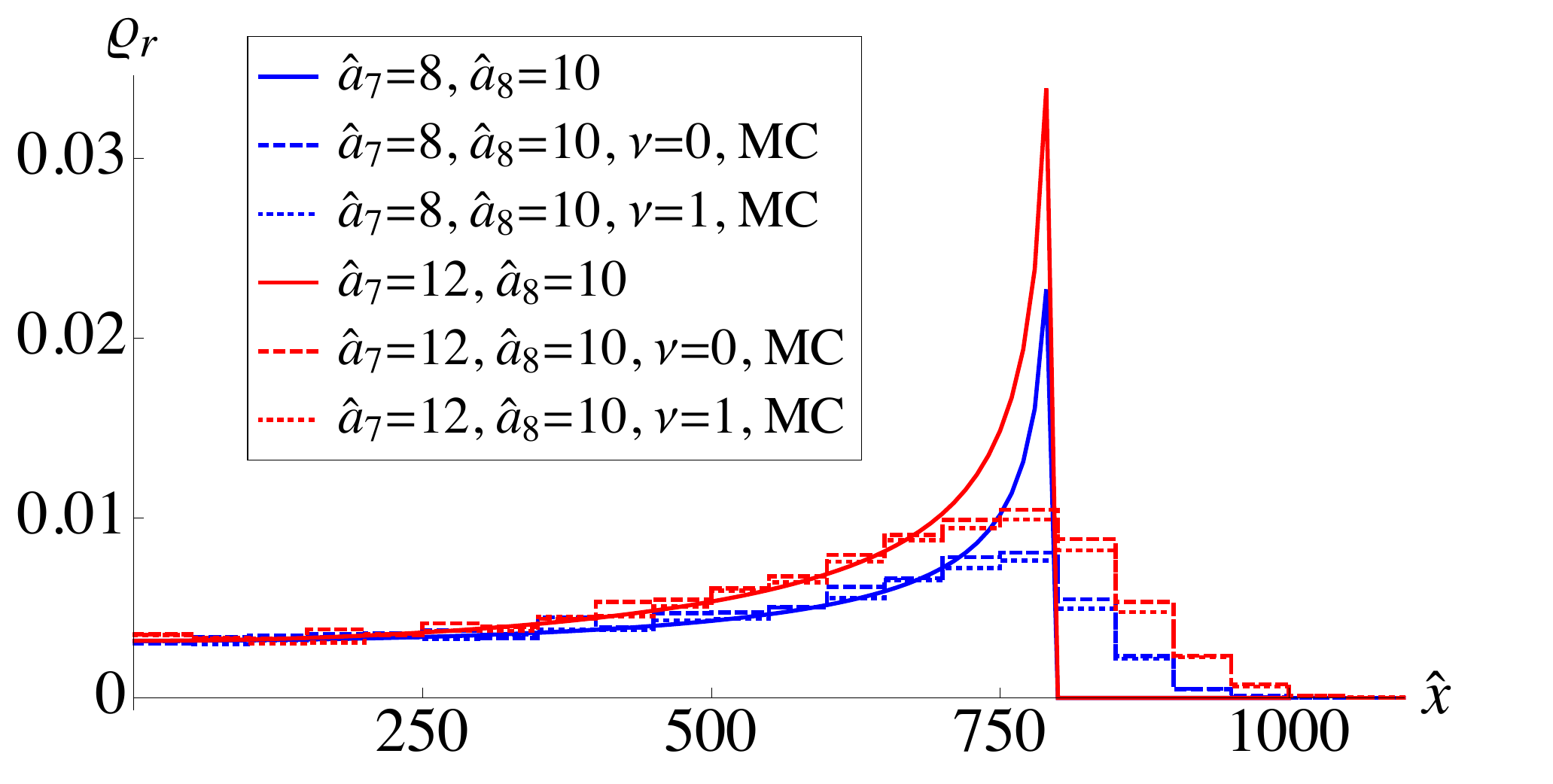}}
 \caption{\label{fig4} At large lattice spacing the density 
of additional real modes develops square root singularities at 
the boundaries. The analytical results at $\ha\to\infty$ 
(solid curves) are compared to Monte Carlo simulations at non-zero, 
but large lattice spacing (histogram [MC], with bin size 50,
 $\widehat{a}_6=\sqrt{-VW_6\widetilde{a}^2}=0.01$ and 
$n=2000$ for an ensemble of 1000 matrices). 
Due to the finite matrix size and the finite lattice spacing,
 $\rho_\rt$ has a tail which drops off much faster than the size 
of the support. The low energy constant 
$\widehat{a}_8=\sqrt{VW_8\widetilde{a}^2}$ is chosen equal to 10. 
Therefore the boundary is at $\hx = 800$ which is  confirmed 
by the Monte Carlo simulations. The dependence on $W_6$ and $\nu$ is completely lost.}
\end{figure}

The average number  of additional real modes can be used to  fix the low energy constants 
from lattice simulations. For $\nu=0$, a sufficient number of eigenvalues~\cite{fnote1} 
can be generated to keep the statistical error small. 
For $\nu = 0 $ and $\nu = 1$ the average number of additional real 
modes  is given by
\begin{eqnarray}
 N_{\rm add}^{\nu=0}&\overset{\widehat{a}\ll1}{=}&2(\widehat{a}_8^2+2\widehat{a}_7^2)=2V\widetilde{a}^2(W_8-2W_7),\label{4.1.5c}\\
 N_{\rm add}^{\nu=1}&\overset{\widehat{a}\ll1}{=}&(\widehat{a}_8^2+2\widehat{a}_7^2)^2+\frac{1}{2}(\widehat{a}_8^2-2\widehat{a}_7^2)^2=V^2\widetilde{a}^4\left[(W_8-2W_7)^2+\frac{1}{2}(W_8+2W_7)^2\right].\label{4.1.5d}
\end{eqnarray}
These simple relations can be used to  fit lattice data at small 
lattice spacing.
In Fig.~\ref{fig2} we illustrate the behavior of $N_{\rm add}$ by a log-log plot.

The density $\rho_{\rm r}$ takes a much simpler form at large lattice spacing. Then, the integrals can be evaluated by a
saddle point approximation resulting in
the expression (see Appendix~\ref{app4.1})
\begin{eqnarray}\label{4.1.6}
\rho_{\rm r}(\widehat{x})&\overset{\widehat{a}\gg1}{=}&\left\{\begin{array}{cl} \displaystyle\frac{1}{8\pi^2\widehat{a}_7\widehat{a}_6}\int\limits_{0}^\infty d\widetilde{x}\cosh\left(\frac{\widetilde{x}\widehat{x}}{8\widehat{a}_6^2}\right)K_0\left(\frac{\widetilde{x}^2}{32\widehat{a}_7^2}\right)\widetilde{x}\exp\left[-\frac{\widetilde{x}^2}{32\widehat{a}_7^2}-\frac{\widetilde{x}^2+\widehat{x}^2}{16\widehat{a}_6^2}\right], & \widehat{a}_8=0, \\ & \\
\displaystyle\frac{\Theta(8\widehat{a}_8^2-|\widehat{x}|)}{2(2\pi)^{3/2}\widehat{a}_8^2}\sqrt{\widehat{a}_8^2+2\widehat{a}_7^2\frac{\widehat{x}^2}{(8\widehat{a}_8^2)^2-\widehat{x}^2}}, & \widehat{a}_8\neq0.\end{array}\right.
\end{eqnarray}
Notice that we have square root singularities at the two edges of the support if both $\widehat{a}_{7}\neq0$ 
and $\widehat{a}_{8}\neq0$,  cf. Fig.~\ref{fig4}. 
So the effect of the  low energy constant $W_7$ 
is different than what we would have expected naively.

\subsubsection{Density of the complex eigenvalues}\label{sec4.2}

The expression for the density of the complex eigenvalues  
given in Eq.~\eqref{rhoc} can be simplified by performing the integral
of $\widehat{m}_6$ and $\widehat{\lambda}_7$ resulting in
\begin{eqnarray}
  \rho_{\rm c}(\widehat{z})&=&\frac{|\widehat{y}|}{ 2(2\pi)^{5/2}\sqrt{\widehat{a}_8^2+2\widehat{a}_6^2}}\int\limits_{[0,2\pi]^2}d\varphi_1d\varphi_2\sin^2\left[\frac{\varphi_1-\varphi_2}{2}\right]\cos[\nu(\varphi_1+\varphi_2)]{\rm sinc}\left[\widehat{y}(\cos\varphi_1-\cos\varphi_2)\right]\label{4.2.1}\\
 &&\hspace*{-1.0cm}\times\exp\left[-4\widehat{a}_8^2\left(\left(\cos\varphi_1-\frac{\widehat{x}}{8\widehat{a}_8^2}\right)^2+\left(\cos\varphi_2-\frac{\widehat{x}}{8\widehat{a}_8^2}\right)^2\right)+\frac{4\widehat{a}_6^2 \widehat{a}_8^2}{\widehat{a}_8^2+2\widehat{a}_6^2}\left(\cos\varphi_1+\cos\varphi_2-\frac{\widehat{x}}{4\widehat{a}_8^2}\right)^2-4\widehat{a}_7^2(\sin\varphi_1+\sin\varphi_2)^2\right].\nonumber
\end{eqnarray}
The function ${\rm sinc}(x)=\sin x/x$ is the \textit{sinus cardinalis}.
This result reduces to the expressions obtained in Ref.~\cite{Kieburg:2011uf} for $\widehat{a}_6=\widehat{a}_7=0$. 

\begin{figure}
\center {\includegraphics[width=1\textwidth]{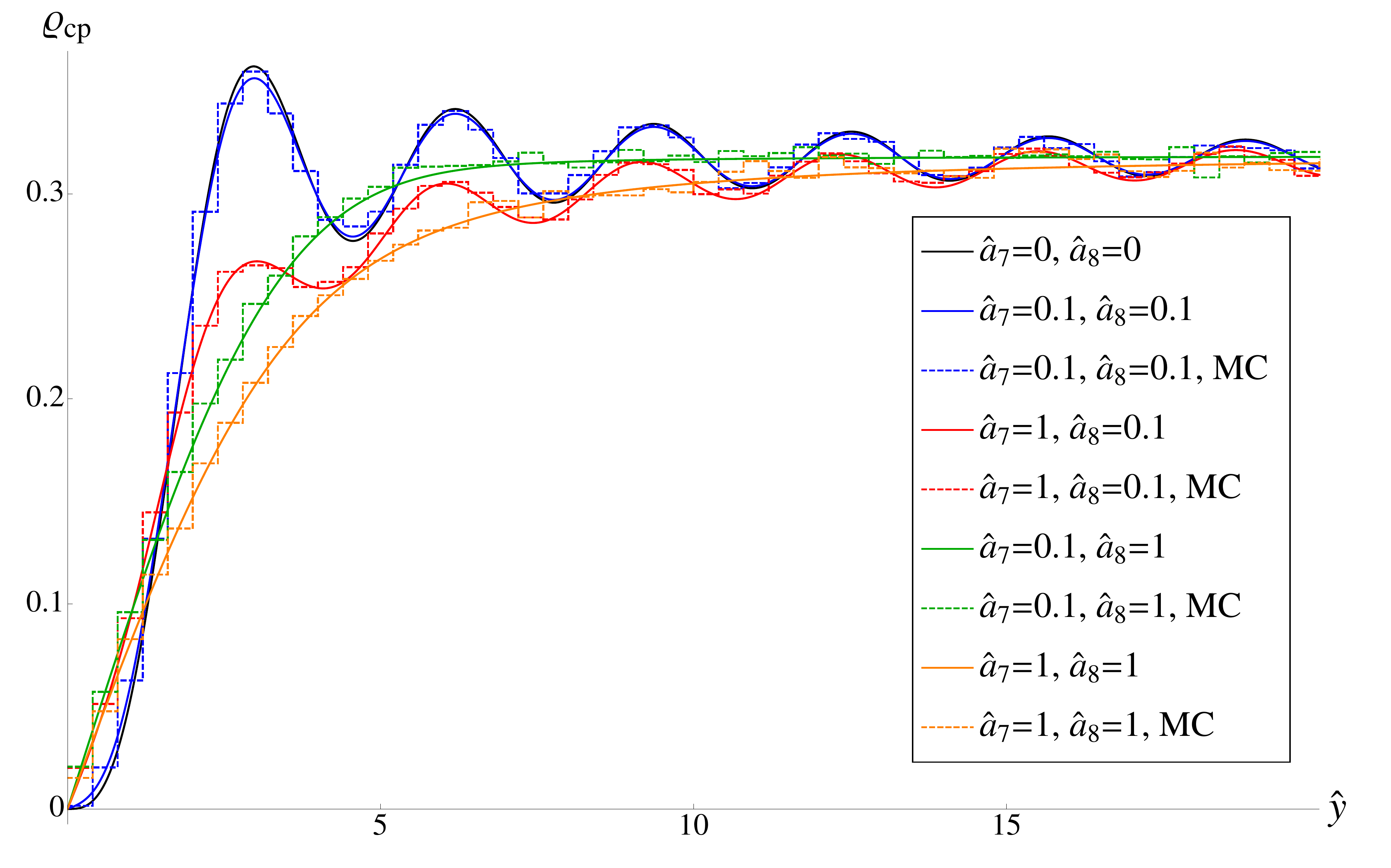}}
 \caption{\label{fig5}Comparison of the analytical result (solid curves) 
and Monte Carlo simulations of the 
Random Matrix Theory (histogram [MC] with bin size equal to 0.4 and  varying ensemble size and matrix size such that the statistical error is about 1-5\%) 
for the density of the complex eigenvalues projected onto the imaginary axis. The index of the Wilson Dirac operator is 
$\nu=1$ for all curves. Notice that $\widehat{a}_6=\sqrt{-VW_6}\widetilde{a}$ does not affect this density. 
The comparison of $\widehat{a}_7=\widehat{a}_8=0.1$ with the continuum result (black, thick curve) shows that
 $\rho_{\rm cp}$ is still a good quantity to extract the chiral condensate $\Sigma$ at small lattice spacing.}
\end{figure}

To compare to numerical  simulations it is useful to consider the projection of the complex modes onto 
the imaginary axis. The result for the projected eigenvalue density 
can be simplified to
\begin{eqnarray}\label{4.2.2}
 \rho_{\rm cp}(\widehat{y})&=&\int\limits_{-\infty}^\infty \rho_{\rm c}(\widehat{x}+\imath\widehat{y})d\widehat{x}\\
 &=&\frac{|\widehat{y}|}{ (2\pi)^{2}}\int\limits_{[0,2\pi]^2}d\varphi_1d\varphi_2\sin^2\left[\frac{\varphi_1-\varphi_2}{2}\right]{\rm sinc}\left[\widehat{y}(\cos\varphi_1-\cos\varphi_2)\right]\cos[\nu(\varphi_1+\varphi_2)]\nonumber\\
 &&\times\exp\left[-2\widehat{a}_8^2(\cos\varphi_1-\cos\varphi_2)^2-4\widehat{a}_7^2(\sin\varphi_1+\sin\varphi_2)^2\right].\nonumber
\end{eqnarray}
Again this function is independent of $W_6$ as was the case for  $N_{\rm add}$. 
The reason is that the Gaussian broadening with respect to the mass $\widehat{m}_6$ is absorbed by the integral  over the real axis. 
 At small lattice spacing $\rho_{\rm cp}$ approaches
the continuum result $\rho_{\rm NZ}$ given in Eq.~\eqref{4.0.2} (see Fig.~\ref{fig5}).
 Therefore it is still a good quantity to determine 
the chiral condensate $\Sigma$ from  lattice simulations.
In Fig. 5, we compare the projected spectral density (solid curves) 
with numerical results from an 
ensemble of random matrices (histograms). The spectral density at
 a couple of lattice spacings away from the origin can be used to 
determine the chiral condensate according to the Banks-Casher formula.

At small lattice spacing, $\rho_{\rm c}$ factorizes into a
Gaussian distribution of the real part of the eigenvalues
and of the level density of the continuum limit,
\begin{eqnarray}
 \rho_{\rm c}(\widehat{z})&\overset{\widehat{a}\ll1}{=}&\frac{|\widehat{y}|}{ 2(2\pi)^{5/2}\sqrt{\widehat{a}_8^2+2\widehat{a}_6^2}}\exp\left[-\frac{\widehat{x}^2}{8(\widehat{a}_8^2+2\widehat{a}_6^2)}\right]\int\limits_{[0,2\pi]^2}d\varphi_1d\varphi_2\sin^2\left[\frac{\varphi_1-\varphi_2}{2}\right]\cos[\nu(\varphi_1+\varphi_2)]{\rm sinc}\left[\widehat{y}(\cos\varphi_1-\cos\varphi_2)\right]\nonumber\\
 &=&\frac{1}{ \sqrt{8\pi(\widehat{a}_8^2+2\widehat{a}_6^2)}}\exp\left[-\frac{\widehat{x}^2}{8(\widehat{a}_8^2+2\widehat{a}_6^2)}\right]\rho_{\rm NZ}(\widehat{y}).\label{4.2.3}
\end{eqnarray}
Therefore the support of $\rho_{\rm c}$ along the real axis is on the scale $\widehat{a}$ while 
it is of order $1$ along the imaginary axis. It also follows from perturbation
theory in the non-Hermitian part of the Dirac operator that 
the first order correction to the continuum 
result is a Gaussian broadening perpendicular to the imaginary axis.
The width of the Gaussian  can be used to determine 
the combination $\widehat{a}_8^2+2\widehat{a}_6^2=V\widetilde{a}^2( W_8-2W_6)$
from fitting the results to  lattice simulations. 
Since most of the eigenvalues of $D_\W$ occur in complex conjugate pairs
at small lattice spacing,
it  is expected to have a relatively small statistical error 
in this limit. A further reduction of the statistical error can
be achieved by integrating the spectral density over $\widehat{y}$ up 
to the Thouless energy  (see Ref.~\cite{Osborn:1998nm} for a definition 
of the Thouless energy in QCD).

The behavior drastically changes in the limit of large lattice spacing. Then the 
density reads (see  Appendix~\ref{app4.3})
\begin{eqnarray}
 \rho_{\rm c}(\widehat{z})&\overset{\widehat{a}\gg1}{=}&\left\{\begin{array}{cl} \displaystyle\frac{\Theta(8\widehat{a}_8^2-|\widehat{x}|)}{16\pi\widehat{a}_8^2}{\rm erf}\left[\frac{|\widehat{y}|}{\sqrt{8\widehat{a}_8^2}}\sqrt{\frac{(8\widehat{a}_8^2)^2-\widehat{x}^2}{(8\widehat{a}_8^2)^2-(1-2\widehat{a}_7^2/\widehat{a}_8^2)\widehat{x}^2}}\right], & \widehat{a}_8>0, \\ & \\ \displaystyle\frac{|\widehat{y}|}{ 16 \pi^2|\widehat{a}_6\widehat{a}_7|}\exp\left[-\frac{\widehat{x}^2}{16\widehat{a}_6^2}+\frac{\widehat{y}^2}{32\widehat{a}_7^2}\right]K_0\left(\frac{\widehat{y}^2}{32\widehat{a}_7^2}\right), & \widehat{a}_8=0. \end{array}\right.\label{4.2.4}
\end{eqnarray}
 There is no dependence on $\nu$, and in the case
of $\widehat{a}_8>0$, the result does not depend on $\widehat{a}_6$ as well
and becomes a strip of width $16\widehat{a}_8^2$ along 
the imaginary axis. To have any structure, the imaginary part 
of the eigenvalues has to be of order $\widehat{a}$. 
In the mean field limit, where $|\hy|/\sqrt{8 \ha_8^2} \gg 1$,
$\rho_{\rm c}$ is equal to 
$1/(16\pi\widehat{a}_8^2)$  on a strip of width
$16\widehat{a}_8^2$.
 Hence, the low energy constants $W_{6/7}$, do not alter the mean field limit of
$\rho_{\rm c}$,
 cf. Ref.~\cite{Kieburg:2011uf}. This was already observed in Ref.~\cite{Kieburg:2012fw}.

The effect of $\ha_6$ is an overall Gaussian fluctuation perpendicular
to the strip of the eigenvalues, and for $\ha_8=0$, when there is no 
strip, only the Gaussian fluctuations remain.
 The second case of Eq.~\eqref{4.2.4} can also be obtained from Eq.~\eqref{4.0.4} since for large $\widehat{y}$,
$\rho_{\rm NZ}$ is equal to  $1/\pi$.

\subsection{The distribution of chirality over the real eigenvalues}\label{sec4.3}

The distribution of chirality over the real eigenvalues  given in Eq.~\eqref{rhochi} is
an expression in terms of the graded partition function $Z_{1/1}^\nu$ and the partition function of two fermionic flavors, $Z_{2/0}^\nu$, which is
evaluated in Appendix~\ref{app3}. Including the
  integrals over $\widehat{m}_6$ and $\widehat{\lambda}_7$ we obtain from Eq.~\eqref{zchi}
\begin{eqnarray}\label{4.3.1}
 \rho_\chi(\widehat{x})&=&\frac{(-1)^{\nu}}{(16\pi)^{3/2}\widehat{a}^2_8|\widehat{a}_7|}\int\limits_{-\infty}^\infty d\widehat{\lambda}_7\int\limits_{\mathbb{R}^2} \frac{ds_1ds_2}{s_1-\imath s_2}(\imath s_2+\widehat{\lambda}_7)^\nu(s_1-\widehat{\lambda}_7)^\nu\\
 &&\hspace*{-0cm}\times\exp\left[-\frac{1}{16\widehat{a}^2_8}\left((s_1-\widehat{x})^2+(s_2+\imath \widehat{x})^2\right)+\frac{\widehat{a}_6^2}{16\widehat{a}_8^4}(s_1-\imath s_2)^2-\frac{\widehat{\lambda}_7^2}{16\widehat{a}_7^2}\right]\nonumber\\
 &&\hspace*{-0cm}\times \left[\frac{\delta^{(\nu-1)}(s_1+\widehat{\lambda}_7)}{(\nu-1)!(s_1-\widehat{\lambda}_7)^\nu}\left(\frac{s_1^2-\widehat{\lambda}_7^2}{s_2^2+\widehat{\lambda}_7^2}\right)^{\nu/2}Z_{1/1}^\nu\left(\sqrt{s_1^2-\widehat{\lambda}_7^2},\imath\sqrt{s_2^2+\widehat{\lambda}_7^2};\widehat{a}=0\right)\right.
\nonumber\\
 &&\hspace*{-0cm}\left.-{\rm sign}(\widehat{\lambda}_7)\Theta(|\widehat{\lambda}_7|-|s_1|) \left(s_1^2+s_2^2\right)\frac{Z_{2/0}^\nu\left(\sqrt{s_1^2-\widehat{\lambda}_7^2},\imath\sqrt{s_2^2+\widehat{\lambda}_7^2};\widehat{a}=0\right)}{[(s_1^2-\widehat{\lambda}_7^2)(s_2^2+\widehat{\lambda}_7^2)]^{\nu/2}}\right].\nonumber
\end{eqnarray}
We recognize the two terms that were obtained in Eqs.~\eqref{rhochinu1} and \eqref{rhochinu2} 
from the expansion in the first column of the determinant in the joint probability density.

\begin{figure}
 \center{\includegraphics[width=1\textwidth]{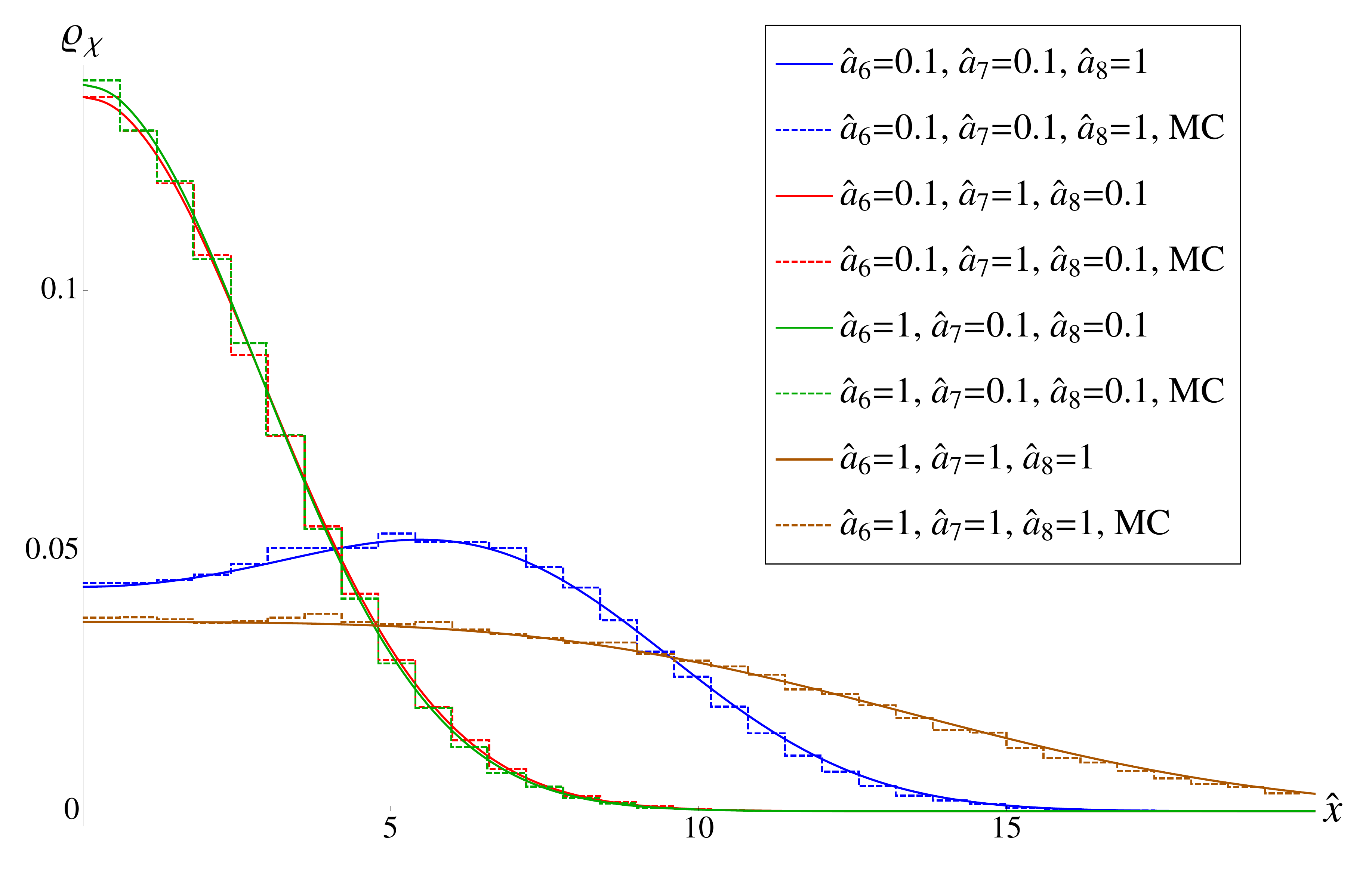}}
 \caption{\label{fig6}The analytical result (solid curves) for 
$\rho_\chi$ is compared to Monte Carlo simulations of RMT (histogram [MC] with bin size 0.6 and varying 
ensemble and matrix size such that the statistical error is about 1-5\%) for $\nu=1$. 
We plotted only the positive real axis since the distribution is 
symmetric around the origin. At small 
$\widehat{a}_8=\sqrt{VW_8\widetilde{a}^2}$ the distributions 
for $(\widehat{a}_6,\widehat{a}_7)=
(\sqrt{-VW_6\widetilde{a}^2},\sqrt{-VW_7\widetilde{a}^2})=(1,0.1), (0.1,1)$ 
are almost the same Gaussian as the analytical result predicts. 
At large $\widehat{a}_8$ the maximum reflects the predicted square root singularity which starts to build up. We have not included the case $\widehat{a}_{6/7/8}=0.1$ since it exceeds the other curves by a factor of $10$ to $100$.}
\end{figure}

Equation~\eqref{4.3.1} is a complicated expression which is quite hard to numerically 
 evaluate. However, it is possible to derive an alternative expression in terms of an integral over the supersymmetric coset manifold $U\in{\rm Gl}(1/1)/\U(1/1)$.
We start from  the equality
\be
&&\int_{-\infty}^\infty \exp\left[-\frac{\hla^2}{16\hala^2} -\frac{\imath\hla}{2} {\rm Str}(U+ U^{-1})\right]d\widehat{\lambda}_7\\
&=&4\sqrt{\pi}\widehat{a}_7\exp\left[-\widehat{a}_7^2\Str^2(U+U^{-1})\right]\nn \\
&=&\exp\left[4\hala^2({\rm Sdet} U + {\rm Sdet} U^{-1}-2)\right] \int_{-\infty}^\infty \exp\left[-\frac{\hla^2}{16\hala^2} -\frac{\imath\hla}{2} {\rm Str}(U- U^{-1})\right]d\widehat{\lambda}_7\nn\\ 
&=& \sum_{j=-\infty}^\infty I_j(8\hala^2) {\rm Sdet}^{j}U e^{-8\hala^2} \int_{-\infty}^\infty \exp\left[-\frac{\hla^2}{16\hala^2} -\frac{\imath\hla}{2} {\rm Str}(U- U^{-1})\right]d\widehat{\lambda}_7, \nn
\ee
based on an identity  for the ${\rm Gl}(1/1)/\U(1/1)$ graded unitary matrices,
\be
{\rm Str}^2(U+U^{-1}) = 8-4( {\rm Sdet} U +{\rm Sdet} U^{-1})+{\rm Str}^2(U-U^{-1}),
\ee
and the expansion of the generating function for the modified Bessel functions of the first kind, $I_j$,
\be
\exp \left [ x \left ( t +\frac 1t \right ) \right ] 
= \sum_{j=-\infty}^\infty I_j(2x) t^j.
\ee
This allows us to absorb $\hma$ and $\hla$ by a shift of 
the eigenvalues of the auxiliary supermatrix $\sigma$ introduced to
linearize the terms quadratic in $U$. The integral over $U$ can now be identified as a graded $1/1$ partition function 
at $\ha = 0$ and we obtain the result
\begin{eqnarray}
 \rho_\chi(\widehat{x})&=&\frac{\exp(-8\widehat{a}_7^2)}{16\pi \widehat{a}^2_8}\sum\limits_{j=1}^\infty\left(I_{j-\nu}(8\widehat{a}_7^2)-I_{j+\nu}(8\widehat{a}_7^2)\right)\int\limits_{\mathbb{R}^2} \exp\left[-\frac{1}{16\widehat{a}^2_8}\left((s_1-\widehat{x})^2+(s_2+\imath \widehat{x})^2\right)+\frac{\widehat{a}_6^2+\widehat{a}_7^2}{16\widehat{a}_8^4}(s_1-\imath s_2)^2\right]\nonumber\\
 &&\times \frac{(-|s_1|)^j\delta^{(j-1)}(s_1)}{(j-1)!}Z_{1/1}^j\left(|s_1|,\imath s_2;\widehat{a}=0\right)\frac{ds_1ds_2}{s_1-\imath s_2}.\label{4.3.5}
\end{eqnarray}
Notice that the $j= 0$ term does not contribute to the distribution of chirality over the real modes because of the symmetry of the modified Bessel function $I_\nu=I_{-\nu}$. 
The derivatives of Dirac delta-function originate
from the ${\rm Im}[ 1/(s_1-\imath\epsilon)^j]$-term. 

\begin{figure}
 \center{\includegraphics[width=1\textwidth]{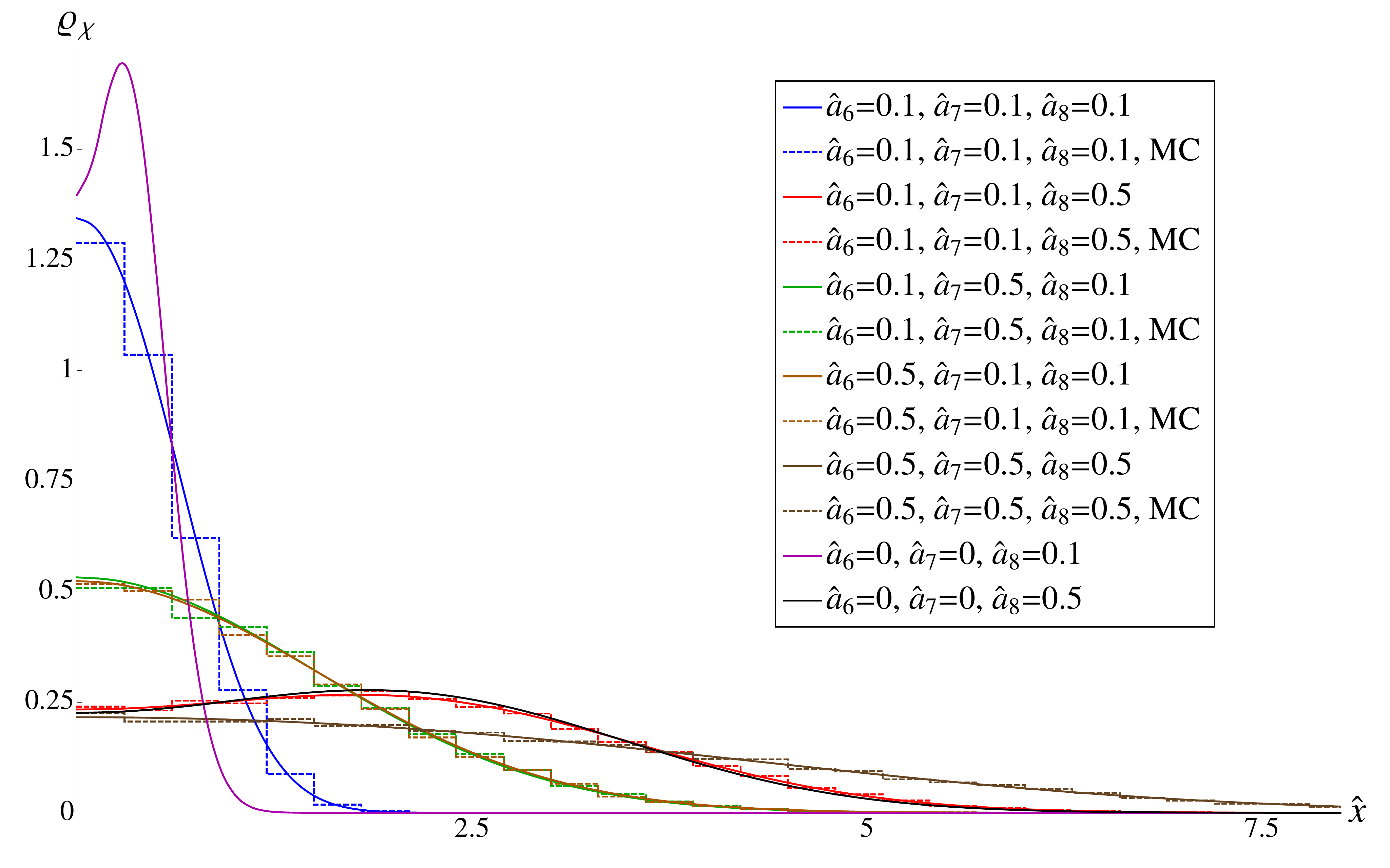}}
 \caption{\label{fig7}We compare the analytical result of $\rho_\chi$ (solid curves) 
with Monte Carlo simulations of RMT (histogram [MC] 
with bin size 0.6 and with varying ensemble and matrix size such that the statistical error is about 1-5\%) for $\nu=2$. 
Again we only plotted the positive real-axis because $\rho_\chi$ 
is symmetric in the quenched theory. Two curves with 
$W_{6/7}=0$ and $W_8=0.1,0.5$ ( highest, purple and thick, black curve) are 
added to emphasize that the two peaks ($\rho_\chi$ has to be reflected at the origin) can be strongly suppressed by 
non-zero $W_{6/7}$ although they are only of the same order as $W_8$. 
Recall that the two peaks are relics of a $2\times 2$
GUE which is formed by $W_8$.}
\end{figure}

The representation~\eqref{4.3.5} is effectively a one-dimensional integral due to the Dirac delta-function. 
Please notice that Eq.~\eqref{4.3.5} reduces to Eq.~\eqref{4.0.3} 
for $\widehat{a}_8=0$. Two plots, Fig.~\ref{fig6} ($\nu=1$) and Fig.~\ref{fig7}, ($\nu=2$) illustrate the effect of  each low-energy constant 
$\widehat{a}_{6/7/8}$ on the distribution $\rho_\chi$.

For $\ha_7=0$ and $\nu=1$ one can derive a more compact result in a straightforward way starting from the expression~\eqref{rhochinu1}. In this case the two-point weight for two real eigenvalues $g_\rt(x_1,x_2)$ is anti-symmetric in its two arguments, see Eq.~\eqref{2.13}. Then the integral in Eq.~\eqref{rhochinu1} involving $Z_{2/0}^1$  is absent.    
Employing the representation of the one-flavor partition function as a unitary integral, see Eq.~\eqref{2.2a}, we perform the integral over $\widehat{m}_6$. Thus, $\rho_\chi(\hx)|_{\nu=1}$ can be expressed as
\be 
\rho_\chi(\hx)|_{\nu=1} &=&\frac 1{\sqrt{16\pi(\ha_8^2+ \ha_6^2)}}
\int_{-\pi}^\pi \frac {d\theta}{2\pi} 
\exp\left [ \frac{(\hx+8\ha^2_8 \sin\theta)^2}{\ha^2_8+\ha_6^2}   \right ].
\label{rhochisimple}
\ee

Let us come back to the general result~\eqref{4.3.5}. At small lattice spacing, $0<\widehat{a}\ll1$, the distribution $\rho_\chi$ 
as well as the integration variables $s_{1/2}$ are of order 
$\widehat{a}$. Since $I_j(8\widehat{a}_7^2)\propto \widehat{a}_7^{2j}$, the 
leading order term is given by $j = \nu$ in the
 sum over $j$.  Thus we have
\begin{eqnarray}
 \rho_\chi(\widehat{x}) &\overset{\widehat{a}\ll1}{=}&\frac{1}{16\pi \widehat{a}^2_8}\int\limits_{\mathbb{R}^2}\exp\left[-\frac{1}{16\widehat{a}^2_8}\left((s_1-\widehat{x})^2+(s_2+\imath \widehat{x})^2\right)+\frac{\widehat{a}_6^2+\widehat{a}_7^2}{16\widehat{a}_8^4}(s_1-\imath s_2)^2\right]\nonumber\\
 &&\times\frac{(-|s_1|)^\nu\delta^{(\nu-1)}(s_1)}{(\nu-1)!}
Z_{1/1}^\nu\left(s_1,\imath s_2;\widehat{a}=0\right)
\frac{ds_1ds_2}{s_1-\imath s_2}. \label{4.3.7}
\end{eqnarray}
In the small $\widehat a$ limit we can replace 
$Z_{1/1}^\nu\left(s_1,\imath s_2;\widehat{a}=0\right) \to (\imath s_2/|s_1| )^\nu$. 
The result becomes a polynomial in $\hx^2$ times a Gaussian 
of width $\sqrt{32(\widehat{a}_8^2+\widehat{a}_6^2+\widehat{a}_7^2)}$.
 Notice that the polynomial is not  the one of a GUE anymore as in the case
of $\widehat{a}_6=\widehat{a}_7 =0$ \cite{Akemann:2010zp}.  
For $\nu=1$, $\rho_\chi$ is a pure Gaussian,
\begin{eqnarray}
 \rho_\chi^{\nu=1}(\widehat{x})&\overset{\widehat{a}\ll1}{=}&\frac{1}{\sqrt{16\pi (\widehat{a}^2_8+\widehat{a}^2_6+\widehat{a}^2_7)}}\exp\left[-\frac{\widehat{x}^2}{16 (\widehat{a}^2_8+\widehat{a}^2_6+\widehat{a}^2_7)}\right], \label{4.3.7b}
\end{eqnarray}
and for $\nu=2$ it is given by
\begin{eqnarray}
 \rho_\chi^{\nu=2}(\widehat{x})&\overset{\widehat{a}\ll1}{=}
&\frac{1}{\sqrt{16\pi (\widehat{a}^2_8+\widehat{a}^2_6+\widehat{a}^2_7)^3}}
\left[\widehat{a}_8^2+2(\widehat{a}_6^2+\widehat{a}_7^2)
+ \frac{\widehat{a}_8^2}{8(\widehat{a}_8^2+\widehat{a}_6^2+\widehat{a}_7^2)}\widehat{x}^2\right]\exp\left[-\frac{\widehat{x}^2}{16 (\widehat{a}^2_8+\widehat{a}^2_6+\widehat{a}^2_7)}\right]. \label{4.3.7c}
\end{eqnarray}
At small lattice spacing, $\rho_\chi$ only depends on the combinations
 $\widehat{a}_8^2$ and $(\widehat{a}_6^2+\widehat{a}_7^2)$.
Therefore it is in principle possible to determine the two following combinations of low energy constants, $W_8$ and $W_6+W_7$,
by fitting $\rho_\chi$ to lattice results. 
For example the second moment (variance) of $\rho_\chi$  given by
\begin{eqnarray}
 \frac{1}{\nu}\int\limits_{-\infty}^\infty\rho_\chi(\widehat{x})\widehat{x}^2d\widehat{x}\overset{\widehat{a}\ll1}{=}8(\nu\widehat{a}_8^2+\widehat{a}_6^2+\widehat{a}_7^2)=8V\widetilde{a}^2(\nu W_8-W_6-W_7),\ \nu>0, \label{4.3.7d}
\end{eqnarray}
at small lattice spacing can be used to fit the combinations $\nu W_8-W_6-W_7$.
The statistical error in this quantity scales with the inverse
square root of the  number of
configurations with the index $\nu$. The ensemble of configurations generated in
 Ref.~\cite{Damgaard:2012gy} 
 yields a statistical error of 
about two to three percent.  The statistics can be drastically increased by 
performing a fit of the variance of $\rho_\chi$ to a linear function 
in the index $\nu$, cf. Eq.~\eqref{4.3.7d}. The slope is then determined by $W_8$ and the off-set by $W_6+W_7$ yielding two important quantities.

In Appendix~\ref{app4.4} we calculate $\rho_\chi$ in the limit of large lattice spacing. Then the distribution of chirality over the real eigenvalues has a support on the scale of $\widehat{a}^2$. The function $\rho_\chi$ reads
\begin{eqnarray}
 \rho_\chi(\widehat{x})&\overset{\widehat{a}\gg1}{=}&\left\{\begin{array}{cl} \displaystyle\frac{\nu}{\pi}\frac{\Theta(8\widehat{a}_8^2-|\widehat{x}|)}{\sqrt{(8\widehat{a}_8^2)^2-\widehat{x}^2}}, & \widehat{a}_8>0, \\ \displaystyle\frac{\nu}{\sqrt{16\pi(\widehat{a}_6^2+\widehat{a}_7^2)}}\exp\left[-\frac{\widehat{x}^2}{16(\widehat{a}_6^2+\widehat{a}_7^2)}\right], & \widehat{a}_8=0. \end{array}\right.\label{4.3.8}
\end{eqnarray}
Interestingly, the low energy constants $W_{6/7}$ have no effect on the behavior of $\rho_\chi$ in this limit if $\widehat{a}_8\neq0$ which is completely different in comparison to $\rho_{\rm r}$ and $\rho_{\rm c}$. The square root singularities at the boundary of the support are unexpected and were already mentioned in Ref.~\cite{Kieburg:2011uf}.

\section{Conclusions}\label{sec5}

Starting from RMT for the Wilson Dirac operator,  
we have derived the microscopic limit of the spectral density and the distribution of the
chiralities over the Dirac spectrum. We have focused on the quenched theory, but all arguments
can be simply  extended to dynamical Wilson fermions. Wilson RMT is equivalent
to the $\epsilon$-limit of the Wilson chiral Lagrangian and  describes the Wilson QCD partition
function and Dirac spectra in this limit. 
The starting point of our analytical calculations is the joint probability density of the random matrix 
ensemble for the non-Hermitian Wilson-Dirac operator $D_\W$. This density was first obtained
in Ref.~\cite{Kieburg:2011uf}, but a detailed derivation is given in this paper, see Appendix~\ref{app1}.

More importantly,  we studied in detail the effect of the  three low energy constants, $W_{6/7/8}$, on 
the quenched microscopic level density of the complex eigenvalues,
the additional real eigenvalues and the distribution of chirality over the real eigenvalues.
In terms of the effect on the spectrum of $D_\W$, the low energy constants $W_6$ and $W_7$ are structurally 
different from $W_8$. The first two can be interpreted in terms of ``collective" fluctuations of
the eigenvalues, whereas a non-zero $W_8$ induces stochastic interactions between all modes, particularly those with different chiralities.
Therefore, the effect of a non-zero $W_6$ and $W_7$ at $W_8=0$ is just a Gaussian broadening of
the Dirac spectrum on the scale of $\ha$. When $a^2 V W_8 \gg 1$ the interactions between the modes result in
a strip of Dirac eigenvalues in the complex plane with real part inside
the interval $[-8VW_8\widetilde{a}^2,8VW_8\widetilde{a}^2]$.
 The structure along the imaginary axis is on the scale $\widehat{a}$. 
As was already discussed in Ref.~\cite{Kieburg:2012fw},
in the mean field limit, the lattice spacing $\widetilde{a}^2 V$ and the eigenvalues $V\widetilde{z}$ fixed,  this structure becomes a box-like strip with 
hard edges at the boundary of the support and with height $1/(16\pi VW_8\widetilde{a}^2)$. 

We also discussed the limit of small lattice spacing, i.e. the limit 
$|VW_{6/7/8}|\widetilde{a}^2$ $\ll1 $. In  practice, this limit is already reached when
$|VW_{6/7/8}|\widetilde{a}^2 \le0.1 $.
Such values can be indeed achieved via clover improvement as discussed 
in Ref.~\cite{Damgaard:2012gy}. In the small $\ha $ limit
we have identified several 
quantities that
are suitable to fit the four low energy constants,  
$W_{6/7/8}$ and $\Sigma$,  to lattice simulations
and our analytical results.  

Several  promising quantities  are (applicable \textbf{only} at small lattice spacing):
\begin{itemize}

\item According to the Banks-Casher formula we have
\be\label{5.0}
\Delta\overset{\widehat{a}\ll1}{=} \frac \pi{\Sigma V}.
\ee
          for the average spacing $\Delta$ of the imaginary part of the eigenvalues several eigenvalue spacings
from the origin.
 \item  The average number of the additional real modes for $\nu = 0$:
			\begin{eqnarray}
 				N_{\rm add}^{\nu=0}&\overset{\widehat{a}\ll1}{=}&2V\widetilde{a}^2(W_8-2W_7).\label{5.1}
           \end{eqnarray}
 \item The width of the Gaussian shaped strip of complex eigenvalues:
			\begin{eqnarray} 
\frac{\sigma^2}{\Delta^2}
&\overset{\widehat{a}\ll1}{=}&\frac 4{\pi^2} 
\widetilde{a}^2 {V(W_8-2W_6)}.\label{5.3}
           \end{eqnarray}
 \item The variance of the distribution of chirality over the real eigenvalues:			
 			\begin{eqnarray}
			  \frac{\langle \widetilde{x}^2\rangle_{\rho_\chi}}{\Delta^2}
&\overset{\widehat{a}\ll1}{=}&\frac 8 {\pi^2}V\widetilde{a}^2(\nu W_8-W_6-W_7),\ \nu>0.\label{5.4}
		    \end{eqnarray}
\end{itemize}
These quantities are easily accessible in lattice simulations. We believe they will lead to an 
improvement of the fits performed in Refs.~\cite{DWW,DHS,Damgaard:2012gy}. 
Note that $\rho_\chi$ is close to the density of the real eigenvalues in the limit of small 
lattice spacing (again we mean by this $|VW_{6/7/8}|\widetilde{a}\approx0.1$ and smaller). 
This statement is not true in the limit of large lattice spacing where the density 
of the additional real modes dominates the density of the real eigenvalues.

The relations~(\ref{5.0}-\ref{5.4}) are an over-determined set for the low energy constants 
$W_{6/7/8}$ and $\Sigma^2$ and are only consistent if we have relations between these quantities.
This can be seen by writing the relations as
\begin{eqnarray}\label{5.5}
\widetilde a^2 V \left[\begin{array}{ccc}  0 & -2 & 1 \\ 
-2 & 0 & 1 \\ 
 -1 & -1 & 1 \\ 
 -1 & -1 & 2 \end{array}\right] \left[\begin{array}{c} W_6 \\ W_7 \\ W_8  \end{array}\right]
= \frac {\pi^2}8
\left[ \begin{array}{c }
\displaystyle{4 N_{\rm add}^{\nu =0}}/{\pi^2}\\
2\displaystyle\sigma^2 / \Delta^2\\
\displaystyle\langle\widetilde{x}^2\rangle_{\rho_\chi}^{\nu=1} / \Delta^2\\
\displaystyle\langle\widetilde{x}^2\rangle_{\rho_\chi}^{\nu=2} / \Delta^2\\
\end{array} \right ].
\ee
The first three relations are linearly dependent, but none of the other triplets are. We thus
have the consistency relation
\be
\frac { \langle \widetilde{x}^2\rangle_{\rho_\chi}^{\nu=1} } { \Delta^2}
=\frac { \sigma^2 } { \Delta^2}
+\frac 2 {\pi^2} N_{\rm add}^{\nu =0}.
\ee
There are more relations like Eqs.~(\ref{5.0}-\ref{5.4}) which can be derived
from our analytical results. The only assumption is a sufficiently small lattice spacing. 

The value of $W_8$ follows immediately from the $\nu$ 
dependence of $\langle \widetilde{x}^2\rangle_{\rho_\chi}$. If there are additional real modes, it cannot be that $W_7$ and $W_8$ are both equal to zero.  
In Ref.~\cite{Damgaard:2012gy} it was  found $W_8=0$ (with clover 
improvement) and results were fitted  as a function of $W_6$ with $W_7=0$. Our prediction is that
the number of additional real modes is zero and it would be interesting if the authors of 
Ref.~\cite{Damgaard:2012gy} could confirm that.

The non-trivial effect of $W_7$ on the quenched spectrum was a surprise for us. 
In Ref.~\cite{Kieburg:2012fw} it was argued  that $W_7$ does not affect the phase structure of the
Dirac spectrum. Indeed, we found that the complex eigenvalue density only shows a weak dependence on $W_7$, and
actually becomes $W_7$ independent in the small $\widetilde a^2V$-limit.  Such a dependence on $W_7$ can be found in the large $\widetilde a^2V$-limit but vanishes again in the thermodynamic limit. Since in the thermodynamic limit the number of real eigenvalues is suppressed as
$1/\sqrt V$ with respect to the number of complex eigenvalues, $W_7$ will not affect the phase
structure of the partition function.
However, a non-zero value of $W_7$ significantly changes the density
of the real eigenvalues.
In particular, in the large $\widetilde a^2V$-limit, we 
find a square root singularity  at the boundary of the support of
the additional real eigenvalues if $W_7\neq0$, while it is a uniform density for $W_7=0$,
see Ref.~\cite{Kieburg:2011uf}.  Nevertheless, we expect in the case of dynamical fermions
that the 
discussion of Ref.~\cite{Kieburg:2012fw} also applies to the real spectrum of $D_\W$. 

\section*{Acknowledgements}

MK acknowledges financial support by the Alexander-von-Humboldt Foundation. JV and SZ acknowledge support by U.S. DOE Grant No. DE-FG-88ER40388. We thank Gernot Akemann, Poul Damgaard, Urs Heller and Kim Splittorff for fruitful discussions.

\appendix

\section{Derivation of the Joint Probability Density}\label{app1}

In this appendix, we derive the joint probability density in three steps.  
In Appendix~\ref{app1.1}, following the derivation for the joint probability density of
the Hermitian Dirac operator \cite{AN}
we introduce an auxiliary Gaussian integral 
such that we obtain  
a Harish-Chandra-Itzykson-Zuber like integral that mixes two different types of variables.
In Appendix~\ref{app1.2} this problem is reduced to a Harish-Chandra-Itzykson-Zuber like integral considered in a bigger framework. We derive an educated guess which fulfills a set of differential equations and a boundary value problem. The asymptotics of the integral for large arguments serves as the boundary. In Appendix~\ref{s2} we perform a stationary phase approximation which already yields the full solution implying that the semi-classical approach is exact and the Duistermaat-Heckman localization theorem~\cite{DuistHeck} applies. In the last step we plug the result of Appendix~\ref{app1.2} into the original 
problem, see Appendix~\ref{app1.3}, and integrate over the remaining variables to arrive at the result for
the joint probability density given in the main text. 

\subsection{Introducing auxiliary Gaussian integrals}\label{app1.1}

We consider the functional $I[f]$, see Eq.~\eqref{2.7}, with an integrable test-function $f$ invariant 
under $\U(n,n+\nu)$. The idea is to rewrite the exponent of the probability 
density $P(D_\W)$  as the sum of a $\U(n,n+\nu)$ invariant term
${\rm Tr} D_\W^2$  and a symmetry breaking term which 
is linear in $D_\W$. 
This is achieved by introducing  two Gaussian distributed Hermitian matrices 
$S_\rt$ and $S_\lt$ with dimensions $n\times n$ and 
$(n+\nu)\times(n+\nu)$, respectively, i.e.
\begin{eqnarray}
 I[f]&=&(2\pi\sqrt{1+a^2})^{-n^2-(n+\nu)^2}\left(-\frac{n}{2\pi}\right)^{n(n+\nu)}\exp\left[-\frac{a^2}{2}\left(\mu_\rt^2+\frac{n+\nu}{n}\mu_\lt^2\right)\right]\int d[D_\W] f(D_\W)\int d[S_\rt,S_\lt]\nonumber\\
 &&\hspace*{-0cm}\times\exp\left[\frac{n}{2}\tr D_\W^2+\imath\tr D_\W\diag(S_\rt,S_\lt)-\frac{a^2}{2n(1+a^2)}\left(\tr(S_\rt+\imath\mu_\rt\eins_n)^2+\tr(S_\lt+\imath\mu_\lt\eins_{n+\nu})^2\right)\right].\label{a1.1.1}
\end{eqnarray}
The matrix ${\rm diag}(S_\rt, S_\lt)$ is a block-diagonal matrix with $S_\rt$ and $S_\lt$
on the diagonal blocks.
The measure for $S_{\rt/\lt}$ is
\begin{eqnarray}
 d[S_\rt,S_\lt]&=&\prod\limits_{j=1}^{n}dS^{(\rt)}_{jj}\prod\limits_{1\leq i<j\leq n}2\,d\,\RE S^{(\rt)}_{ij}d\,\IM S^{(\rt)}_{ij}\prod\limits_{j=1}^{n+\nu}dS^{(\lt)}_{jj}\prod\limits_{1\leq i<j\leq n+\nu}2\,d\,\RE S^{(\lt)}_{ij}d\,\IM S^{(\lt)}_{ij}.\label{a1.1.2}
\end{eqnarray}
Then the non-compact unitary matrix diagonalizing $D_\W$ only appears 
quadratically in the exponent.
Notice that we have to integrate first over the Hermitian matrices $S_{\rt/\lt}$ 
and have to be careful when interchanging integrals 
with integrals over $D_\W$. Obviously the integrations over the eigenvalues of 
$D_\W$ are divergent without performing the $S_{\rt/\lt}$ integrals  
first and cannot be interchanged
with these integrals. Also the coset integrals over $\mathbb{G}_l=\U(n,n+\nu)/[\U^{2n+\nu-l}(1)\times\Or^{l}(1,1)]$, cf. Eq.~\eqref{2.7}, 
are not absolutely convergent. However we can understand them in a weak way and, below, 
we will find Dirac delta functions resulting from the non-compact integrals.

Diagonalizing the matrices $D_l=UZ_lU^{-1}$ and $S_{\rt/\lt}=V_{\rt/\lt}s_{\rt/\lt}V_{\rt/\lt}^\dagger$ with $s_{\rt}=\diag(s_{1}^{(\rt)},\ldots,s_{n}^{(\rt)})$ and $s_{\lt}=\diag(s_{1}^{(\lt)},\ldots,s_{n+\nu}^{(\lt)})$ we can absorb the integrals over $V_\rt$ and $V_\lt$ in the $U\in\mathbb{G}_l$ integral. Then we end up with the integral
\begin{eqnarray}
I[f]&=&\frac{C}{n!(n+\nu)!}\sum\limits_{l=0}^{n}\frac{1}{2^l(n-l)!l!(n+\nu-l)!}\int\limits_{\mathbb{R}^{\nu+2(n-l)}\times\mathbb{C}^l}d[Z_l]|\Delta_{2n+\nu}(Z_l)|^2\int\limits_{\mathbb{R}^{2n+\nu}}d[s_\rt,s_\lt]\Delta_n^2(s_\rt)\Delta_{n+\nu}^2(s_\lt) f(Z_l)\qquad\label{a1.1.3}\\
 &&\hspace*{-1cm}\times\exp\left[\frac{n}{2}\tr Z_l^2-\frac{a^2}{2n(1+a^2)}\left(\tr(s_\rt+\imath\mu_\rt\eins_n)^2+\tr(s_\lt+\imath\mu_\lt\eins_{n+\nu})^2\right)\right]\int\limits_{\mathbb{G}_l}\exp\left[\imath\tr UZ_lU^{-1}\diag(s_\rt,s_\lt)\right]d\mu_{\mathbb{G}_l}(U)\nonumber
\end{eqnarray}
and the normalization constant
\begin{eqnarray}
 C&=&\left(-\frac{n}{2\pi}\right)^{n(n+\nu)}\prod\limits_{j=0}^{n-1}\frac{(2\pi)^j\exp\left[-a^2\mu_\rt^2/2n\right]}{j!(2\pi\sqrt{1+a^2})^{2j+1}}\prod\limits_{j=0}^{n+\nu-1}\frac{(2\pi)^j\exp\left[-a^2\mu_\lt^2/2n\right]}{j!(2\pi\sqrt{1+a^2})^{2j+1}}.\label{a1.1.4}
\end{eqnarray}
See Sec.~\ref{sec2.2} for a discussion of the prefactors in the sum. 

\subsection{The Harish-Chandra-Itzykson-Zuber integral over the non-compact coset $\mathbb{G}_l$}\label{app1.2}

In the next step we calculate the integral
\begin{eqnarray}\label{a1.2.1}
\mathcal{I}_l(Z_l,s)=\int\limits_{\mathbb{G}_l}\exp\left[\imath\tr UZ_lU^{-1}s\right]d\mu_{\mathbb{G}_l}(U).
\end{eqnarray}
with $s=\diag(s_\rt,s_\lt)$. For $l=0$ this integral was derived in Ref.~\cite{FyoStr02}. 

We calculate this integral by determining a complete set of functions
and expanding the integral for asymptotically large
$s$ in this set. In this limit it can be calculated by a stationary phase
approximation. It turns out that this integral, as  is the case with
the usual Harish-Chandra-Itzykson-Zuber integral, is semi-classically exact.

\subsubsection{Non-compact Harish-Chandra-Itzykson-Zuber Integral}
\label{s1}

Let us consider the non-compact integral
\begin{eqnarray} 
 \mathcal{I}_l(Z_l,Z_{l'}^\prime)&=&\int\limits_{\mathbb{G}_l }
\exp\left[\imath\tr UZ_lU^{-1}Z_{l'}^\prime\right]d\mu_{\mathbb{G}_l}(U)
\label{a1.2.15}
\end{eqnarray}
in a bigger framework where $Z'_{l'}$ is a quasi-diagonal matrix with $l'$ complex conjugate eigenvalue pairs. The integral is invariant under the Weyl group $\mathbf{S}(n-l)\times\mathbf{S}(l)\times\mathbf{S}(n+\nu-l)\times\mathbb{Z}_2^l$ in $Z_l$. To make the integral well-defined we have to assume that $l\geq l'$ otherwise the integral is divergent since the non-compact subgroup ${\rm O}^{l'-l}(1,1)\subset\mathbb{G}_l$ commutes with $Z'_{l'}$.

The integral~\eqref{a1.2.15} should be contrasted with the well-known compact Harish-Chandra-Itzykson-Zuber
integral~\cite{HC,IZ}
\begin{eqnarray} 
 \mathcal{I}^{\rm com}(X, X^\prime)&=&\int\limits_{\U(2n+\nu) }
\exp\left[\imath\tr UXU^{-1}X^\prime\right]d\mu_{\U(2n+\nu)}(U)=\frac{(-2\pi\imath)^{\nu(\nu-1)/2}}{\Delta_{\nu}(x)\Delta_{\nu}(x^\prime)}\det\left[\exp\left(\imath x_ix'_j\right)\right]_{1\leq i,j\leq \nu}\label{a1.2.15c}
\end{eqnarray} 
with Weyl group $\mathbf{S}(2n+\nu)$. Moreover the compact case is symmetric when interchanging $X$ with $X'$. This symmetry is broken in $Z_l$ and $Z'_{l'}$ due to the coset $\mathbb{G}_l$.

For a $\gamma_5$-Hermitian matrix $V$
with eigenvalues $Z_l$, we can rewrite the integral (\ref{a1.2.15})  as
\be
 \mathcal{I}_l(Z_l,Z_{l'}^\prime)&=&\mathcal{I}_l(V,Z_{l'}^\prime)=\int\limits_{\mathbb{G}_l }
\exp\left[\imath\tr UVU^{-1}Z_{l'}^\prime\right]d\mu_{\mathbb{G}_l}(U).\label{seki-W}
\ee
This trivially satisfies the Sekigushi-like differential equation \cite{OkoOls97,KGG09}
\begin{eqnarray}
 \det\left(\frac{\partial}{\partial V_{kl}}+u\eins_{2n+\nu} \right )\mathcal{I}_l(V,Z_{l^\prime}^\prime)=\det\left(\imath Z_{l'}^\prime+u\eins_{2n+\nu}\right)\mathcal{I}_l(V,Z_{l'}^\prime)\ \text{for all}\ u\in\mathbb{C}. \label{a1.2.16}
\end{eqnarray}
This equation is written in terms of the independent matrix elements of $V$ and, hence, is independent of the fact to which sector $l$ the matrix $V$ can be quasi-diagonalized.

We would like to rewrite Eq.~\eqref{a1.2.16} in terms of derivatives with respect
to the eigenvalues~\cite{fnote2}. Because of the coefficients
that enter after applying the chain rule when changing coordinates, the derivatives 
do not commute and a direct
evaluation of the determinant is cumbersome. Therefore we will calculate
$\mathcal{I}_l(Z_l,Z_{l'}^\prime)$ in an indirect way. We will do this by constructing a complete set of 
$\mathbf{S}(n-l)\times\mathbf{S}(l)\times\mathbf{S}(n+\nu-l)\times\mathbb{Z}_2^l$
symmetric functions in the
space of the $\{Z_l\}$ with the $\{Z_{l'}'\}$ as quantum numbers which have to be $\mathbf{S}(n-l^\prime)\times\mathbf{S}(l^\prime)\times\mathbf{S}(n+\nu-l^\prime)\times\mathbb{Z}_2^{l^\prime}$ symmetric. Then we
expand   $\mathcal{I}_l(Z_l,Z_{l^\prime}^\prime)$ in this set of functions and determine
the coefficients for asymptotic large $\{Z_l\}$ where the integral can
be evaluated by a stationary phase approximation.

To determine the complete set of functions,
we start from the usual Harish-Chandra-Itzykson-Zuber integral over the  compact group
$\U(2n+\nu)$. 
This integral is well-known and satisfies the 
 Sekigushi-like differential equation \cite{OkoOls97,KGG09}
with
\begin{eqnarray}
 &&\frac{1}{\Delta_{2n+\nu}(X)}\det\left(\frac{\partial}{\partial X}+u\eins_{2n+\nu}\right)\Delta_{2n+\nu}(X)\mathcal{I}^{\rm com}(X,X')=\det \left (\imath X'+u\eins_{2n+\nu}\right)  \mathcal{I}^{\rm com}(X,X')
\label{seki-compact}
\end{eqnarray}
in terms of the $(2n+\nu)$ real eigenvalues $X=\diag(x_1,\ldots,x_{2n+\nu})$ with
\begin{eqnarray}
 \det\left(\frac{\partial}{\partial X}+u\eins_{2n+\nu}\right)
=\prod\limits_{j=1}^{2n+\nu}\left(\frac{\partial}{\partial x_j}+u\right).
\label{seki-compact.b}
\end{eqnarray}
The expansion in powers of $u$ gives the complete
set of $2n+\nu$ independent Casimir operators on the Cartan subspace 
of $\U(2n+\nu)$, so that the Sekigushi equation  determines a complete
set of functions
$ \mathcal{I}_l(Z_l,Z_{l'}^\prime)$ up to the Weyl group. 
Since the non-compact group
$\U(n+\nu,n)$ shares the same complexified Lie algebra as $\U(2n+\nu)$
the  Casimir operators are the same, i.e. the corresponding operator for $\U(n+\nu,n)$ to the one in Eq.~\eqref{seki-compact} is
\begin{eqnarray}
 D_{Z_l}(u)&=&\frac{1}{\Delta_{2n+\nu}(Z_l)}\det\left(\frac{\partial}{\partial Z_l}+u\eins_{2n+\nu}\right)\Delta_{2n+\nu}(Z_l)
\label{seki-non-compact}
\end{eqnarray}
with
\begin{eqnarray}
 &&\det\left(\frac{\partial}{\partial Z_l}+u\eins_{2n+\nu}\right)=\prod\limits_{j=1}^{n-l}\left(\frac{\partial}{\partial x_j^{(1)}}+u\right)\prod\limits_{j=1}^{l}\left(\frac{\partial}{\partial z_j^{(2)}}+u\right)
\left(\frac{\partial}{\partial z_j^{(2)*}}+u\right)\prod\limits_{j=1}^{n+\nu-l}\left(\frac{\partial}{\partial x_j^{(3)}}+u\right).\label{a1.2.17}
\end{eqnarray}

In the compact case, the Sekigushi-like equation (\ref{seki-compact}) follows from Eq. (\ref{a1.2.16})  by transforming
the equation in terms of the eigenvalues and eigenvectors of $V = U X U^{-1}$, see Ref.~\cite{KGG09}. The only difference in the non-compact case is
that the parameters of $U$ as well as some of the eigenvalues $x$ become complex, but the algebraic manipulations to obtain
the Sekigushi-like differential equation in terms of eigenvalues remain the same.
Let $f$ be an integrable test-function on the Cartan-subset $\mathbb{R}^{2n+\nu-2l'}\times\mathbb{C}^{l'}$. 
Then the non-compact integral (\ref{a1.2.15}) satisfies the weak Sekigushi-like equation
\begin{eqnarray}
&&D_{Z_l}(u) \int\limits_{\mathbb{R}^{2(n-l')+\nu}\times\mathbb{C}^{l'}} d[Z'_{l'}] f(Z'_{l'}) \mathcal{I}_l(Z_l,Z_{l'}^\prime)
 = \int\limits_{\mathbb{R}^{2(n-l')+\nu}\times\mathbb{C}^{l'}} d[Z'_{l'}] f(Z'_{l'}) \det \left (\imath Z'_{l'}+u\eins_{2n+\nu}\right)\mathcal{I}_l(Z_l,Z_{l'}^\prime),
\label{seki-non-compact.b}
\end{eqnarray}
and solutions of this equation yield a complete set of functions for the non-compact case 
as well. 
The only difference is the corresponding Weyl group. The completeness can be seen because we 
can generate any polynomial of order $k\in\mathbb{N}_0$ (the non-negative integers) in $Z'_{l'}$ symmetric under  
$\mathbf{S}(n-l')\times\mathbf{S}(l')\times\mathbf{S}(n+\nu-l')\times\mathbb{Z}_2^{l'}$ 
via the differential operator $\prod_{j=1}^kD_{Z_l}(u_j)$. Since those polynomials are dense 
in the space of $\mathbf{S}(n-l')\times\mathbf{S}(l')\times\mathbf{S}(n+\nu-l')\times\mathbb{Z}_2^{l'}$ 
invariant functions, it immediately follows that if a function is in the kernel 
of $D_{Z_l}(u)$ for all $u$ it is zero, i.e. 
\begin{eqnarray}
D_{Z_l}(u) F(Z_l) = 0\ \forall u\in\mathbb{C}&\Leftrightarrow& F(Z_l)=0.
\label{seki-non-compact.c}
\end{eqnarray}
Therefore if we found a solution for Eq.~\eqref{seki-non-compact.b} for an arbitrary test-function $f$ we found $\mathcal{I}_l(Z_l,Z_{l'}^\prime)$ up to the normalization which can be fixed in the large $\tr Z_lZ_l^\dagger$-limit.

Some important remarks about Eq.~\eqref{seki-non-compact.b} are in order. The Vandermonde determinant $\Delta_{2n+\nu}(Z_l)$ enters in a trivial way in the operator $D_{Z_l}(u)$ and the remaining operator has plane waves as eigenfunctions which indeed build a complete set of functions. Thus a good ansatz of $\mathcal{I}_l(Z_l,Z_{l'}^\prime)$ is
\begin{eqnarray}\label{solgen.1}
 \mathcal{I}_l(Z_l,Z_{l'}^\prime)=\frac{1}{\Delta_{2n+\nu}(Z_l)}\prod\limits_{j=1}^l \frac{y_j^{(2)}}{|y_j^{(2)}|}\sum\limits_{\omega\in\mathbf{S}} c_{\omega}^{(ll')}(Z'_{l'})\exp\left[\imath\tr \Pi_\omega Z_l\Pi_\omega^{-1} \widehat{Z}'_{l'}\right],
\end{eqnarray}
where the coefficients $c_{\omega}^{(ll')}(Z'_{l'})$ have to be determined. The factors $y_j^{(2)}/|y_j^{(2)}|$ guarantee the invariance under complex conjugation of each complex eigenvalue pair of $Z_l$. We sum over the permutation group $\omega$ and $\Pi_\omega$ is its standard representation in terms of  $(2n+\nu)\times(2n+\nu)$ matrices. The $\mathbf{S}(n-l)\times\mathbf{S}(l)\times\mathbf{S}(n+\nu-l)\times\mathbb{Z}_2^l$ invariance in $Z_l$ and the $\mathbf{S}(n-l^\prime)\times\mathbf{S}(l^\prime)\times\mathbf{S}(n+\nu-l^\prime)\times\mathbb{Z}_2^{l^\prime}$ invariance in $Z'_{l'}$ carry over to the coefficients $c_{\omega}^{(ll')}(Z'_{l'})$. Hence, we can reduce all coefficients to coefficients independent of $\omega$,
\begin{eqnarray}
 \mathcal{I}_l(Z_l,Z_{l'}^\prime)&=&\frac{1}{\Delta_{2n+\nu}(Z_l)}\prod\limits_{j=1}^l \frac{y_j^{(2)}}{|y_j^{(2)}|}{\underset{\omega'\in\mathbf{S}(n-l')\times\mathbf{S}(l')\times\mathbf{S}(n+\nu-l')\times\mathbf{Z}_2^{l'}}{\underset{\omega\in\mathbf{S}(n-l)\times\mathbf{S}(l)\times\mathbf{S}(n+\nu-l)\times\mathbf{Z}_2^l}{\sum}}}\hspace*{-1cm}\sign \omega\ c^{(ll')}(Z_{l^\prime\omega^\prime}^\prime) \exp\left[\imath\tr Z_{l\omega} Z_{l^\prime\omega^\prime}^\prime\right]\label{solgen.2}
\end{eqnarray}
where we employ the abbreviation
\begin{eqnarray}\label{Zpermuted}
 Z_{l\omega}=\Pi_\omega Z_{l}\Pi_\omega^{-1}
\quad\text{and}\quad Z_{l^\prime\omega^\prime}^\prime=
{\Pi}_{\omega'} Z_{l^\prime}^\prime\Pi_{\omega'}^{-1}.
\end{eqnarray}
The sign of elements in the group $\mathbb{Z}_2$ generating the complex conjugation of single complex conjugated pairs is always $+1$. Moreover, any element in the permutation group $\mathbf{S}(l)$ is an even permutation since it interchanges a complex conjugate pair with another one and, thus, always yields a positive sign. Hence the sign of the permutation $\omega$ is the product of the sign of the permutations in $\mathbf{S}(n-l)$ and in $\mathbf{S}(n+\nu-l)$.

Solving the weak Sekigushi-like equation~\eqref{seki-non-compact.b} for the general case $l\neq l'$ is quite complicated but  as we  will show below 
for $l=l'$ the ansatz
\begin{eqnarray}
 \mathcal{I}_l(Z_l,Z_{l}^\prime)&=&\frac{(-2\pi\imath)^{(2n+\nu)(2n+\nu-1)/2}}{\Delta_{2n+\nu}(Z_l)\Delta_{2n+\nu}(Z_l^\prime)}\det\left[\exp\left(\imath x_i^{(1)}x_j^{\prime(1)}\right)\right]_{1\leq i,j\leq n-l}\det\left[\exp\left(\imath x_i^{(3)}x_j^{\prime(3)}\right)\right]_{1\leq i,j\leq n+\nu-l}\nonumber\\
 &&\times{\rm perm}\left[\frac{y_i^{(2)}y_j^{\prime(2)}}{|y_i^{(2)}y_j^{\prime(2)}|}
\left(\exp\left[2\imath \RE\,z_i^{(2)} {z'}_j^{(2)}\right]+\exp\left[2\imath \RE\,z_i^{(2)\,*} {z'}_j^{(2)}\right]
\right)\right]_{1\leq i,j\leq l},\label{a1.2.18}
\end{eqnarray}
i.e. $c^{(ll)}(Z_{l\omega^\prime}^\prime)\propto(\prod_{j=1}^{l'}{y'}_j^{(2)}/|{y'}_j^{(2)}|)/\Delta_{2n+\nu}(Z_{l\omega^\prime}^\prime)$, does the job. Note that we have again the symmetry when interchanging $Z_l$ with $Z'_l$ since both matrices are in the Cartan subspace corresponding to $\mathbb{G}_l$.
The constant can be fixed by a stationary phase approximation when taking $\tr Z_lZ_l^\dagger\to\infty$. The function ``${\rm perm}$'' is the permanent which is defined 
analogously  to the determinant but without the sign-function in the sum over the permutations. It arises because the Vandermonde determinants are even under the interchange
of a complex pair with another one, i.e. it is the $\mathbf{S}(l)$-invariance of the corresponding Weyl-group. It can be explicitly shown  that the ansatz~\eqref{a1.2.18} satisfies the completeness relation in the space 
of functions on $\mathbb{R}^{\nu+2(n-l)}\times\mathbb{C}^{l}$
invariant under 
$\mathbf{S}(n-l)\times\mathbf{S}(l)\times\mathbf{S}(n+\nu-l)\times\mathbb{Z}_2^l$ 
and with the measure $|\Delta_{2n+\nu}(Z_{l})|^2d[Z_{l}]$, i.e.
\begin{eqnarray}
 &&\int\limits_{\mathbb{R}^{\nu+2(n-l)}\times\mathbb{C}^{l}}\mathcal{I}_l(Z_l,Z_{l}^\prime)\mathcal{I}_l(Z_l^{\prime\prime},Z_{l})|\Delta_{2n+\nu}(Z_{l})|^2d[Z_{l}]\nonumber\\
 &\propto&\frac{1}{\Delta_{2n+\nu}(Z_l^\prime)\Delta_{2n+\nu}(Z_l^{\prime\prime})}\det\left[\delta\left( x_i^{\prime(1)}-x_j^{\prime\prime(1)}\right)\right]_{1\leq i,j\leq n-l}\det\left[\delta\left( x_i^{\prime(3)}-x_j^{\prime\prime(3)}\right)\right]_{1\leq i,j\leq n+\nu-l}\nonumber\\
 &&\times{\rm perm}\left[\frac{y_i^{\prime(2)}y_j^{\prime\prime(2)}}{|y_i^{\prime(2)}y_j^{\prime\prime(2)}|}\delta\left( |y_i^{\prime(2)}|-|y_j^{\prime\prime(2)}|\right)\delta\left( x_i^{\prime(2)}-x_j^{\prime\prime(2)}\right)\right]_{1\leq i,j\leq l}.\label{a1.2.19}
\end{eqnarray}
Therefore, for given $l'=l$ and $Z_{l'}'$ 
the ansatz~\eqref{a1.2.18} for  $\mathcal{I}_l(Z_l,Z_{l}^\prime)$ is the unique solution of 
the Sekiguchi-like equation \eqref{seki-non-compact.b}. One has only to show that the global prefactor is correct, see \ref{s2}.

What happens in the general case $l\neq l'$? The ansatz~\eqref{solgen.2} can only fulfill 
the Sekigushi-like differential equation~\eqref{seki-non-compact.b} if we assume 
that the coefficient $c^{(ll')}(Z_{l^\prime\omega^\prime}^\prime)$ restricts the 
matrix $Z'_{l'}$ to a matrix in the sector with $l$ complex conjugate eigenvalue pairs (notice that $Z_l$ has 
the representation given in Eq. (\ref{2.5})). 
This is only possible on the boundary of the Cartan subsets $\mathbb{R}^{2(n-l)+\nu}\times\mathbb{C}^l$ and $\mathbb{R}^{2(n-l')+\nu}\times\mathbb{C}^{l'}$, i.e. the coefficient has to be proportional to Dirac delta functions
\begin{equation}\label{coeff.1}
c^{(ll')}(Z_{l^\prime\omega^\prime}^\prime)\propto\prod\limits_{j=1}^{l-l'}\delta\left({x'}_{\omega'(n-l+j)}^{(1)}-{x'}_{\omega'(j)}^{(3)}\right).
\end{equation}
The reason for this originates in the fact that not all complex pairs of $Z_l$ can couple with a complex eigenvalue pair in $Z'_{l'}$ and, hence, $\tr Z_{l\omega} Z_{l^\prime\omega^\prime}^\prime$ does not depend on the combinations ${x'}_{\omega'(n-l+j)}^{(1)}-{x'}_{\omega'(j)}^{(3)}$. Therefore we would miss it in the determinant $\det(Z'_{l'}+u\eins_{2n+\nu})$ generated by the differential operator $D_{Z_l}(u)$. To cure  this we have to understand $\mathcal{I}_l(Z_l,Z'_{l'})$ as a distribution where the Dirac delta functions set these missing terms to zero. In \ref{s2} we show that the promising ansatz 
\begin{eqnarray}
 \mathcal{I}_l(Z_l,Z_{l'}^\prime)&=&\frac{c^{(ll')}}{\Delta_{2n+\nu}(Z_{l})\Delta_{2n+\nu}(Z_{l'}^\prime)}\prod\limits_{j=1}^{l}\frac{y_{j}^{(2)}}{|y_{j}^{(2)}|}\prod\limits_{j=1}^{l'}\frac{{y'}_{j}^{(2)}}{|{y'}_{j}^{(2)}|}\label{ansatz}\\
 &&\times\hspace*{-0.5cm}\underset{\omega'\in\mathbb{S}(n-l')\times\mathbb{S}(l')\times\mathbb{S}(n+\nu-l')\times\mathbb{Z}_2^{l'}}{\underset{\omega\in\mathbb{S}(n-l)\times\mathbb{S}(l)\times\mathbb{S}(n+\nu-l)\times\mathbb{Z}_2^l}{\sum}}\sign\omega\omega' \exp\left(\imath\tr Z_{l\omega} Z_{l^\prime\omega^\prime}^\prime\right)\prod\limits_{j=1}^{|l-l'|}\left({x'}_{\omega'(n-l+j)}^{(1)}-{x'}_{\omega'(j)}^{(3)}\right)\delta\left({x'}_{\omega'(n-l+j)}^{(1)}-{x'}_{\omega'(j)}^{(3)}\right)\nonumber
\end{eqnarray}
is indeed the correct result.

Note that the ansatz~\eqref{ansatz} agrees with the solution~\eqref{a1.2.18} for the case $l=l'$. Furthermore one can easily verify that it also solves the weak Sekiguchi-like differential equation~\eqref{seki-non-compact.b}. Indeed, the ansatz is trivially invariant under the two Weyl groups $\mathbf{S}(n-l)\times\mathbf{S}(l)\times\mathbf{S}(n+\nu-l)\times\mathbb{Z}_2^l$ and $\mathbf{S}(n-l')\times\mathbf{S}(l')\times\mathbf{S}(n+\nu-l')\times\mathbb{Z}_2^{l'}$ due to the sum. The global prefactor $1/\Delta_{2n+\nu}(Z_{l'}^\prime)$ reflects the singularities when an eigenvalue in ${x'}^{(1)}$ agrees with one in ${x'}^{(3)}$ as well as a complex eigenvalue pair in ${x'}^{(2)}$ degenerates with another eigenvalue in $Z'_{l'}$, namely then $Z'_{l'}$ commutes with some non-compact subgroups in $\mathbb{G}_l$. Hereby the eigenvalues which have to degenerate via the Dirac delta functions are excluded.

In the next section we calculate the global coefficients in Eq.~\eqref{ansatz}. For this we consider the stationary phase approximation which fixes this coefficient.

\subsubsection{The stationary phase approximation of $\mathcal{I}_l(Z_l,Z_{l'}^\prime)$}
\label{s2}

Let us introduce a scalar parameter $t$ as a small parameter in the integral $\mathcal{I}_l(t^{-1} Z_l,Z_{l'}^\prime)$ as a bookkeeping device for the expansion around the saddlepoints.
Taking $t\to0$ the  group integral~\eqref{a1.2.15} can be evaluated by a stationary phase approximation.
The saddlepoint equation is given by
\begin{eqnarray}\label{saddle}
 \tr dUU^{-1}[U Z_l U^{-1}, Z_{l^\prime}^\prime]_-=0.
\end{eqnarray}
If $l \ne l'$ this equation cannot be satisfied in all directions. 
 The reason is that the
quasi-diagonal matrix $Z'_{l'}$ will never commute with a 
 $\gamma_5$-Hermitian matrix with exactly $l\neq l'$  complex conjugate eigenvalue pairs since $U Z_l U^{-1}$ can be at most quasi-diagonalized by $\U(n,n+\nu)$ and generically $[Z_l, Z_{l^\prime}^\prime]_-\neq0$. This means that we can only expand the sub-Lie-algebra ${\rm o}^{l-l'}(1/1)$ to the linear order while the remaining massive modes are expanded to the second order. The extrema are given by
\be
 U_0=\Pi'\Phi\Pi\in\mathbb{G}_l
\ee
where the permutations are
\begin{eqnarray}
\Pi&\in&\mathbf{S}(n-l)\times[\mathbf{S}(l)/[\mathbf{S}(l')\times\mathbf{S}(l-l')]]\times\mathbf{S}(n+\nu-l),\nonumber\\
\Pi'&\in&[\mathbf{S}(n-l')/\mathbf{S}(n-l)]\times\mathbf{S}(l')\times[\mathbf{S}(n+\nu-l')/\mathbf{S}(n+\nu-l)]\times\mathbb{Z}^{l'},\label{permutations}
\end{eqnarray}
and a block-diagonal matrix
\begin{eqnarray}
\Phi=\left[\begin{array}{c|c|c|c|c|c} \eins_{n-l}  & 0 & 0 & 0 & 0 & 0 \\ \hline 0 & \exp[\imath \widehat\Phi] & 0 & 0 & 0 & 0  \\ \hline 0 & 0 & \eins_{l'} & 0 & 0 & 0  \\ \hline 0 & 0 & 0 & \eins_{l'} & 0 & 0 \\ \hline 0 & 0 & 0 & 0 & \exp[-\imath\widehat\Phi] & 0 \\ \hline  0 & 0 & 0 & 0 & 0 & \eins_{n+\nu-l} \end{array}\right],
\label{rotations}
\end{eqnarray}
where the diagonal matrix of angles is $\widehat\Phi=\diag(\varphi_1,\ldots,\varphi_{l-l'})\in[0,\pi]^{l-l'}$. The matrix $\Phi$ describes the set $ \U^{l-l'}(1)$ ($l-l'$ unit circles in the complex plane) which commutes with $Z'_{l'}$ and is a subgroup of $\mathbb{G}_{l}$. Note that other rotations commuting with $Z'_{l'}$ are already divided out in $\mathbb{G}_{l}$.
The matrix of phases already comprises the complex conjugation of the 
complex eigenvalues represented by the finite group $\mathbb{Z}_2^{l-l'}$, 
choosing $\varphi_j=\pi/2$ switches the sign of the imaginary part $y'_j$. However we have to introduce the complex conjugation for those complex conjugated pairs in $Z'_{l'}$ which couple with pairs in $Z_l$, cf. the group $\mathbb{Z}^{l'}$ in $\Pi'$.

The expansion of $U$ reads
\begin{eqnarray}\label{taylor}
U = \Pi'\Phi\left(\eins_{2n+\nu}-t H_1-\sqrt{t} H_2+\frac{t}{2}H_2^2\right)\Pi.
\end{eqnarray}
We employ the notation~\eqref{Zpermuted} for the action of
$\omega\in\mathbf{S}(n-l)\times\mathbf{S}(l)\times\mathbf{S}(n+\nu-l)\times\mathbb{Z}^l$ 
and $\omega^\prime\in\mathbf{S}(n-l^\prime)\times\mathbf{S}(l^\prime)\times\mathbf{S}(n+\nu-l^\prime)\times\mathbb{Z}^{l'}$ on the matrices $Z_{l\omega}$ and $Z'_{l'\omega'}$, respectively. Note that the matrix $\Phi$ commutes with $Z'_{l'\omega'}$ for any $\omega'$ and, hence, only yields an overall prefactor $\pi^{l-l'}$.
The matrix $H_1$ spans the Lie algebra ${\rm o}^{l-l^\prime}(1,1)$
and is embedded as
\begin{eqnarray}\label{a1.2.8a}
 H_1=\left(\begin{array}{c|c|c|c|c|c} 0 & 0 & 0 & 0 & 0 &0 \\ \hline 0 & 0 & 0 & 0 & h & 0 
\\ \hline 0 & 0 & 0 & 0 & 0 &0 \\ \hline 0 & 0 & 0 & 0 & 0 &0 \\ \hline 0 & h & 0 & 0 & 0 & 0 \\ \hline 0 & 0 & 0 & 0 & 0 & 0 \end{array}\right)\ \text{with}\  h=\diag(h_1,\ldots,h_{l-l^\prime})\in\mathbb{R}^{l-l^\prime}.\qquad
\end{eqnarray}
 The matrix $H_2$ is in the tangent space of the coset
$\mathbb{G}_l/[\U^{l-l'}(1)\times {\rm O}^{l-l'}(1,1)]=\U(n,n+\nu)/[\U^{2n+\nu-2l+l'}(1)\times{\rm O}^{l'}(1,1)\times\U^{l-l'}(1,1)]$ and has the form
\begin{eqnarray}\label{a1.2.9a}
 H_2=\left(\begin{array}{c||c|cc|c||c} H_{11} & H_{12} & H_{13} & H_{14} & H_{15} & H_{16} \\ \hline\hline -H_{12}^\dagger & H_{22} & H_{23} & H_{24} & H_{25} & H_{26} \\ \hline -H_{13}^\dagger & -H_{23}^\dagger & H_{33} & H_{34} & H_{35} & H_{36} \\ H_{14}^\dagger & H_{24}^\dagger & H_{34}^\dagger & H_{44} & H_{45} & H_{46} \\ \hline H_{15}^\dagger & H_{25}^\dagger & H_{35} ^\dagger & -H_{45}^\dagger & H_{55} & H_{56} \\ \hline\hline H_{16}^\dagger  & H_{26}^\dagger  & H_{36}^\dagger & -H_{46}^\dagger & -H_{56}^\dagger & H_{66} \end{array}\right),
\end{eqnarray}
where $H_{11}$, $H_{22}$, $H_{55}$, and $H_{66}$ are anti-Hermitian matrices 
without diagonal elements since they are divided out in the coset $\mathbb{G}_l$ or are lost to $\Phi$. The two matrices $H_{33}$ and $H_{44}$ are anti-Hermitian matrices whose diagonal elements are the same with opposite sign which is also because of the subgroup we divide out in $\mathbb{G}_l$. The matrices $H_{12}$, $H_{13}$, $H_{14}$, $H_{15}$, $H_{16}$, $H_{23}$, $H_{24}$, $H_{26}$, $H_{35}$, $H_{36}$, $H_{45}$, $H_{46}$, and $H_{56}$ are arbitrary complex matrices. Since we have to remove the
degrees of freedom already included in $H_1$ and in the subgroups quotient out in $\mathbb{G}_l$
the matrix $H_{25}$ is a complex matrix with all $l-l'$ diagonal elements
removed and $H_{34}$ is a complex matrix whose diagonal entries are real. The sizes of the blocks of $H_1$ and $H_2$ correspond to the sizes shown in the diagonal matrix of phases $\Phi$, see Eq.~\eqref{rotations}. The double lines in the matrix~\eqref{a1.2.9a} shall show the decomposition of $Z_{l}$ in its real and complex eigenvalues whereas the single lines represent the decomposition for $Z'_{l'}$.

The exponent in the coset integral~\eqref{a1.2.15} takes the form
\begin{eqnarray}
 \tr U Z_lU^{-1}Z_{l^\prime}^\prime=\tr 
Z_{l\omega}Z_{l^\prime\omega^\prime}^\prime -t\tr [Z_{l\omega},Z_{l^\prime\omega^\prime}^\prime]_- H_1
-\frac{t}{2}\tr[Z_{l\omega},H_2]_-[Z_{l^\prime\omega^\prime}^\prime,H_2]_-.\label{a1.2.10}
\end{eqnarray}
The measure for $H_1$ and $H_2$ is the induced Haar measure, i.e.
\begin{eqnarray}
 \tr[U^{-1}dU,Z_{l\omega}]_-^2
=\tr[\Phi^\dagger d\Phi,Z_{l\omega}]_-^2
+t^2\tr[dH_1,Z_{l\omega}]_-^2
+t\tr[dH_2,Z_{l\omega}]_-^2\label{a1.2.11}
\end{eqnarray}
which gives
\begin{eqnarray}
 d\mu_{\mathbb{G}}(U)&=&t^{(2n+\nu)(2n+\nu-1)/2}d[H_1]d[\Phi]d[H_2]=(-1)^{n(n+\nu)}\left(\frac{2}{\imath}\right)^{l'}\prod\limits_{k=1}^{l-l'}
\frac{4t}{\imath}  d\varphi_kdh_k\prod\limits_{j,i}2t\,d\RE(H_2)_{ij}d\IM(H_2)_{ij}.\label{a1.2.11a}
\end{eqnarray}
The product over the two indices $i$ and $j$ is over all independent matrix elements of $H_2$.

We emphasize again that the integrand in $\mathcal{I}_l(t^{-1}Z_l,Z'_{l'})$ does not depend on  $\Phi$  making this integration
trivial and yielding the prefactor $\pi^{l-l'}$.
The integral over $H_1$ yields the $l-l'$ Dirac delta functions mentioned in Eq.~\eqref{coeff.1}, i.e. it yields
\begin{eqnarray}
(2\pi)^{l-l'}\prod\limits_{j=1}^{l-l'}\delta\left(2y_{\omega(j)}^{(2)}\left[{x'}_{\omega'(n-l+j)}^{(1)}-{x'}_{\omega'(j)}^{(3)}\right]\right)=\prod\limits_{j=1}^{l-l'}\frac{\pi\delta\left({x'}_{\omega'(n-l+j)}^{(1)}-{x'}_{\omega'(j)}^{(3)}\right)}{|y_{\omega(j)}^{(2)}|}.
\end{eqnarray}
Notice that the other term in the expansion of the Dirac delta function does not contribute because of the order of the 
integrations~\cite{fnote3}.

The integrals over $H_2$ are simple Gaussian integrals
resulting in the main result of this section,
\begin{eqnarray}\label{asymp}
 \mathcal{I}_l(t^{-1}Z_l,Z_{l'}^\prime)&=&\frac{(-2\pi\imath)^{l-l'}}{l'!(l-l')!(n-l)!(n+\nu-l)!2^{l}}\frac{(-2\pi\imath)^{(2n+\nu)(2n+\nu-1)/2}}{\Delta_{2n+\nu}(t^{-1}Z_{l})\Delta_{2n+\nu}(Z_{l'}^\prime)}\prod\limits_{j=1}^{l}\frac{y_{j}^{(2)}}{|y_{j}^{(2)}|}\prod\limits_{j=1}^{l'}\frac{{y'}_{j}^{(2)}}{|{y'}_{j}^{(2)}|}\\
 &&\hspace*{-1.5cm}\times\hspace*{-0.5cm}\underset{\omega'\in\mathbb{S}(n-l')\times\mathbb{S}(l')\times\mathbb{S}(n+\nu-l')\times\mathbb{Z}_2^{l'}}{\underset{\omega\in\mathbb{S}(n-l)\times\mathbb{S}(l)\times\mathbb{S}(n+\nu-l)\times\mathbb{Z}_2^l}{\sum}}\sign\omega\omega' \exp\left(\frac{\imath}{t}\tr Z_{l\omega} Z_{l^\prime\omega^\prime}^\prime\right)\prod\limits_{j=1}^{|l-l'|}\left({x'}_{\omega'(n-l+j)}^{(1)}-{x'}_{\omega'(j)}^{(3)}\right)\delta\left({x'}_{\omega'(n-l+j)}^{(1)}-{x'}_{\omega'(j)}^{(3)}\right).\nonumber
\end{eqnarray}
The overall coefficient $c^{(ll')}$ in Eq.~\eqref{ansatz} can be easily read off. Thereby the numerator of the first factor results from the integral over $H_1$ and is related to the $l-l'$ Dirac delta functions. The denominator is the volume of the finite group $\mathbf{S}(n-l)\times\mathbf{S}(l')\times\mathbf{S}(l-l')\times\mathbf{S}(n+\nu-l)\times\mathbb{Z}^l$ which we extend to summing over the full Weyl groups for $Z_l$ and $Z'_{l'}$. We recall that the sum over permutations in $\mathbf{S}(l)$ and $\mathbf{S}(l')$ describe the interchange of complex pairs which are even permutations because we
interchange both $z_k$ and $z^*_k$ with another pair. The numerator of the term with the Vandermonde determinants essentially result from the Gaussian integrals and always 
appears independent of how many complex pairs $Z_l$ and $Z'_{l'}$ have. The factors of $t^{-1}$ appear as prefactors of $Z_l$ and can be omitted again since they have done their job as bookkeeping device.

Let us summarize what we have found. Comparing the result~\eqref{asymp} with the $Z_l$ dependence of the ansatz
$\mathcal{I}(Z_l,Z_{l}^\prime)$ given in Eq.~\eqref{ansatz}, 
we observe that they are exactly the same.
This implies that the asymptotic large $Z_l$ result for the integral~\eqref{a1.2.15} is actually equal to the exact result.
 We conclude that the non-compact Harish-Chandra-Itzykson-Zuber integral
is semi-classically exact and seems to fulfill the conditions of the Duistermaat-Heckman theorem~\cite{DuistHeck}.

Let us consider two particular cases. For $l=l'$ we sum over
all permutations in $\mathbf{S}(l)$ which yields the permanent in Eq.~\eqref{a1.2.19}, whereas the sum
over permutations in $\mathbf{S}(n+\nu-l) $ and $\mathbf{S}(n-l)$ gives determinants and, thus, agrees. The special case $n=0$ yields the original Harish-Chandra-Itzykson-Zuber integral~\cite{HC,IZ}, see Eq.~\eqref{a1.2.15c}.

\subsubsection{The joint probability density}\label{app1.3}

We explicitly write out $Z_{l}$ and apply  Eq.~\eqref{asymp} for $Z'_{l'}=s$. Then, we find for our original non-compact group integral
\be\label{jl-final}
\mathcal{I}_l(Z_l,s)&=&\frac{(-2\pi\imath)^{(2n+\nu)(2n+\nu-1)/2}}{(n-l)!l!(n+\nu-l)!}\frac{(-2\pi\imath)^l}{\Delta_{2n+\nu}(Z_l)\Delta_{2n+\nu}(s)}  
\underset{\omega\in\mathbf{S}(n)\times\mathbf{S}(n+\nu)}
{\underset{\omega^\prime\in\mathbf{S}(n-l)\times\mathbf{S}(l)
\times\mathbf{S}(n+\nu-l)}{\sum}}\sign(\omega \omega^\prime) 
\prod\limits_{j=1}^{n-l}\exp\left(\imath x_{\omega^\prime(j)}^{(1)}s_{\omega(j)}^{({\rm r})}\right)\\
 &&\hspace*{-1.5cm}\times\prod\limits_{j=1}^{n+\nu-l}\exp\left(\imath x_{\omega^\prime(j)}^{(3)}s_{\omega(l+j)}^{({\rm l})}\right)\prod\limits_{j=1}^{l}\frac{y_{\omega^\prime(j)}^{(2)}}{|y_{\omega^\prime(j)}^{(2)}|}\left(s_{\omega(n-l+j)}^{({\rm r})}-s_{\omega(j)}^{({\rm l})}\right)\delta\left(s_{\omega(n-l+j)}^{({\rm r})}-s_{\omega(j)}^{({\rm l})}\right)\exp\left(\imath x_{\omega^\prime(j)}^{(2)}\left(s_{\omega(n-l+j)}^{({\rm r})}+s_{\omega(j)}^{({\rm l})}\right)\right).\nn
\ee
Now we are ready to integrate over $s$.

We plug Eq.~\eqref{jl-final} into the integral~\eqref{a1.1.3}. 
The sum over the permutations can be absorbed by the integral due 
to relabelling resulting in
\begin{eqnarray}
 I[f]&=&C\sum\limits_{l=0}^{n}\frac{(-2\pi\imath)^{(2n+\nu)(2n+\nu-1)/2+l}}{2^l(n-l)!l!(n+\nu-l)!}\int\limits_{\mathbb{R}^{\nu+2(n-l)}\times\mathbb{C}^l}d[Z_l]\Delta_{2n+\nu}(Z_l^*) f(Z_l)\prod\limits_{j=1}^{l}\frac{y_{j}^{(2)}}{|y_{j}^{(2)}|}\int\limits_{\mathbb{R}^{2n+\nu}}d[s_\rt,s_\lt]\frac{\Delta_n^2(s_\rt)\Delta_{n+\nu}^2(s_\lt)}{\Delta_{2n+\nu}(s)} \nonumber\\
 &&\times\prod\limits_{j=1}^{n-l}\exp\left[\frac{n}{2}(x_j^{(1)})^2+\imath x_j^{(1)}s_j^{(\rt)}-\frac{a^2}{2n(1+a^2)}(s_j^{(\rt)}+\imath\mu_\rt)^2\right]\nonumber\\
 &&\times\prod\limits_{j=1}^{n+\nu-l}\exp\left[\frac{n}{2}(x_j^{(3)})^2+\imath x_j^{(3)}s_{l+j}^{(\lt)}-\frac{a^2}{2n(1+a^2)}(s_{l+j}^{(\lt)}+\imath\mu_\lt)^2\right]\nonumber\\
 &&\times\prod\limits_{j=1}^{l}\left(s_{n-l+j}^{(\rt)}-s_{j}^{(\lt)}\right)\delta\left(s_{n-l+j}^{(\rt)}-s_{j}^{(\lt)}\right)\exp\left[\frac{a^2}{4n(1+a^2)}(\mu_\rt-\mu_\lt)^2\right]\nonumber\\
 &&\times\exp\left[n((x_j^{(2)})^2-(y_j^{(2)})^2)+2\imath x_j^{(2)}s_{j}^{(\lt)}-\frac{a^2}{n(1+a^2)}\left(s_{j}^{(\lt)}+\imath\frac{\mu_\rt+\mu_\lt}{2}\right)^2\right].
\label{total}
\end{eqnarray}
The quotient of the Vandermonde determinants is
\begin{eqnarray}\label{a1.3.2}
 \frac{\Delta_n^2(s_\rt)\Delta_{n+\nu}^2(s_\lt)}{\Delta_{2n+\nu}(s)}=(-1)^{n(n-1)/2+\nu(\nu-1)/2}\det\left[\begin{array}{c} \displaystyle\left\{\frac{1}{s_i^{(\rt)}-s_j^{(\lt)}}\right\}\underset{1\leq j\leq n+\nu}{\underset{1\leq i\leq n}{}} \\ \displaystyle\left\{(s_j^{(\lt)})^{i-1}\right\}\underset{1\leq j\leq n+\nu}{\underset{1\leq i\leq \nu}{}} \end{array}\right].
\end{eqnarray}
This determinant also appears in the supersymmetry method of RMT \cite{KGG09,KieGuh09a} and is a square root of a Berezinian (the supersymmetric analogue of the Jacobian).

Expanding the determinant~\eqref{a1.3.2} in the first $l$ columns not all terms will survive. Only those terms which cancel the prefactor of the Dirac delta functions do not vanish. The integration over $\diag(s_{n-l+1}^{(\rt)},\ldots,s_{n}^{(\rt)},s_1^{(\lt)},\ldots,s_l^{(\lt)})$ yields
\begin{eqnarray}
 I[f]&=&C\sum\limits_{l=0}^{n}\frac{(-2\pi\imath)^{(2n+\nu)(2n+\nu-1)/2+l}(-1)^{n(n-1)/2+\nu(\nu-1)/2+(n+l)l}}{2^l(n-l)!l!(n+\nu-l)!} \int\limits_{\mathbb{R}^{\nu+2(n-l)}\times\mathbb{C}^l}d[Z_l]\Delta_{2n+\nu}(Z_l^*)f(Z_l)\int\limits_{\mathbb{R}^{\nu+2(n-l)}}d[s_\rt,s_\lt]\nonumber\\
 &&\times\det\left[\begin{array}{c} \displaystyle\left\{\frac{1}{s_i^{(\rt)}-s_{l+j}^{(\lt)}}\right\}\underset{1\leq j\leq n+\nu-l}{\underset{1\leq i\leq n-l}{}} \\ \displaystyle\left\{(s_{l+j}^{(\lt)})^{i-1}\right\}\underset{1\leq j\leq n+\nu-l}{\underset{1\leq i\leq \nu}{}} \end{array}\right]\prod\limits_{j=1}^{n-l}\exp\left[\frac{n}{2}(x_j^{(1)})^2+\imath x_j^{(1)}s_j^{(\rt)}-\frac{a^2}{2n(1+a^2)}(s_j^{(\rt)}+\imath\mu_\rt)^2\right]\nonumber\\
 &&\times\prod\limits_{j=1}^{n+\nu-l}\exp\left[\frac{n}{2}(x_j^{(3)})^2+\imath x_j^{(3)}s_{l+j}^{(\lt)}-\frac{a^2}{2n(1+a^2)}(s_{l+j}^{(\lt)}+\imath\mu_\lt)^2\right]\prod\limits_{j=1}^{l}\sqrt{\frac{n\pi(1+a^2)}{a^2}}\frac{y_{j}^{(2)}}{|y_{j}^{(2)}|}\exp\left[\frac{a^2}{4n(1+a^2)}(\mu_\rt-\mu_\lt)^2\right]\nonumber\\
 &&\times\exp\left[-\frac{n}{a^2}(x_j^{(2)})^2-n(y_j^{(2)})^2+ x_j^{(2)}(\mu_\rt+\mu_\lt)\right].\label{a1.3.3}
\end{eqnarray}
The other exponential functions as well as the remaining integrations 
over $s_\rt$ and $s_\lt$  can be pulled into the determinant. 
The integrals in the $\nu$ bottom rows yield 
harmonic oscillator wave function. 
These can be reordered into monomials times a 
Gaussian. This results in
\begin{eqnarray}\label{a1.3.4}
 I[f]&=&C\sum\limits_{l=0}^{n}\frac{(-2\pi\imath)^{(2n+\nu)(2n+\nu-1)/2+l}(-1)^{n(n-1)/2+\nu(\nu-1)/2+(n+l)l}}{2^l(n-l)!l!(n+\nu-l)!}(2\pi)^{\nu/2}\imath^{\nu(\nu-1)/2}\left(\frac{n(1+a^2)}{a^2}\right)^{\nu^2/2}\\
 &&\times\int\limits_{\mathbb{R}^{\nu+2(n-l)}\times\mathbb{C}^l}d[Z_l]\Delta_{2n+\nu}(Z_l^*)f(Z_l)\det\left[\begin{array}{c} 
\displaystyle\left\{\widetilde{G}(x_i^{(1)},x_j^{(3)})\right\}\underset{1\leq j\leq n+\nu-l}{\underset{1\leq i\leq n-l}{}} \\ \displaystyle\left\{( x_j^{(3)})^{i-1}\exp\left[-\frac{n}{2a^2}(x_j^{(3)})^2+\mu_\lt x_j^{(3)}\right]\right\}\underset{1\leq j\leq n+\nu-l}{\underset{1\leq i\leq \nu}{}} \end{array}\right] \nonumber\\
 &&\times\prod\limits_{j=1}^{l}\sqrt{\frac{n\pi(1+a^2)}{a^2}}\frac{y_{j}^{(2)}}{|y_{j}^{(2)}|}\exp\left[\frac{a^2}{4n(1+a^2)}(\mu_\rt-\mu_\lt)^2-\frac{n}{a^2}(x_j^{(2)})^2-n(y_j^{(2)})^2+ x_j^{(2)}(\mu_\rt+\mu_\lt)\right].\nonumber
\end{eqnarray}
What remains is to simplify the function
\begin{eqnarray}\label{a1.3.5}
 \widetilde{G}(x_i^{(1)},x_j^{(3)})&=&\int\limits_{\mathbb{R}^2}ds_\rt ds_\lt\frac{\exp\left[x_i^{(1)}\mu_\rt+x_j^{(3)}\mu_\lt\right]}{s_\rt-s_\lt+\imath n(1+a^2)(x_i^{(1)}-x_j^{(3)})/a^2}\\
 &&\times\exp\left[-\frac{n}{2a^2}\left((x_i^{(1)})^2+(x_j^{(3)})^2\right)-\frac{a^2}{2n(1+a^2)}\left((s_{\rt}+\imath\mu_\rt)^2+(s_{\lt}+\imath\mu_\lt)^2\right)\right].\nonumber
\end{eqnarray}
We use the difference $x_i^{(1)}-x_j^{(3)}$ as a regularization of the integral.
 This works because generically this difference is not equal to zero. 
Then we can express the denominator as an exponential function. 
Let $\beta=(x_i^{(1)}-x_j^{(3)})/|x_i^{(1)}-x_j^{(3)}|$ be the sign of this difference. 
The integral~\eqref{a1.3.5} can be written as
\begin{eqnarray}
 \widetilde{G}(x_i^{(1)},x_j^{(3)})&=&\frac{\beta}{\imath}\exp\left[x_i^{(1)}\mu_\rt+x_j^{(3)}\mu_\lt-\frac{n}{2a^2}\left((x_i^{(1)})^2+(x_j^{(3)})^2\right)\right]\int\limits_{0}^\infty dt\int\limits_{\mathbb{R}^2}ds_\rt ds_\lt\nonumber\\
 &&\times\exp\left[-\frac{n(1+a^2)}{a^2}|x_i^{(1)}-x_j^{(3)}|t+\imath\beta(s_\rt-s_\lt)t-\frac{a^2}{2n(1+a^2)}\left((s_{\rt}+\imath\mu_\rt)^2+(s_{\lt}+\imath\mu_\lt)^2\right)\right]\nonumber\\
 &=&\frac{-2\pi\imath n(1+a^2)}{a^2}\beta\exp\left[x_i^{(1)}\mu_\rt+x_j^{(3)}\mu_\lt-\frac{n}{2a^2}\left((x_i^{(1)})^2+(x_j^{(3)})^2\right)\right]\nonumber\\
 &&\times\int\limits_0^\infty\exp\left[-\frac{n(1+a^2)}{a^2}t^2+\left(\beta(\mu_\rt-\mu_\lt)-\frac{n(1+a^2)}{a^2}|x_i^{(1)}-x_j^{(3)}|\right)t\right] dt\nonumber\\
 &=&-\pi\imath \sqrt{\frac{\pi n(1+a^2)}{a^2}}\beta\exp\left[-\frac{n}{4a^2}\left(x_i^{(1)}+x_j^{(3)}\right)^2+\frac{n}{4}\left(x_i^{(1)}-x_j^{(3)}\right)^2+\frac{1}{2}\left(x_i^{(1)}+x_j^{(3)}\right)\left(\mu_\rt+\mu_\lt\right)\right]\nonumber\\
 &&\times\exp\left[\frac{a^2}{4n(1+a^2)}\left(\mu_\rt-\mu_\lt\right)^2\right]{\rm erfc}\left[\sqrt{\frac{n(1+a^2)}{4a^2}}|x_i^{(1)}-x_j^{(3)}|-\beta\sqrt{\frac{a^2}{4n(1+a^2)}}(\mu_\rt-\mu_\lt)\right].\label{a1.3.6}
\end{eqnarray}
Plugging this result into Eq.~\eqref{a1.3.4} we get 
the joint probability density for a fixed number of real eigenvalues
given in Eq. (\ref{2.15}). 
 Moreover one can perform the sum over $l$ to find the 
joint probability density of all eigenvalues given in Eq.~\eqref{2.12}.

\section{Two useful Integral Identities}\label{app2}

In this appendix we evaluate two integrals that have been used to
simplify the expression for $\rho_\rt$ and $\rho_{\rm c}$.

\subsection{Convolution of a Gaussian with an error function}\label{app2.1}

Let $\RE\gamma^2>-1$. We consider the integral
\begin{eqnarray}\label{a2.1.1}
 I(\alpha,\gamma)=\int\limits_{\mathbb{R}}\exp[-(x+\alpha)^2]{\rm erf}(\gamma x)dx.
\end{eqnarray}
The solution can be obtained by constructing an initial value problem. Since the Gaussian is symmetric and the error function anti-symmetric around the origin we have
\begin{eqnarray}\label{a2.1.2}
 I(0,\gamma)=0.
\end{eqnarray}
The derivative is
\begin{eqnarray}\label{a2.1.3}
 \partial_\alpha I(\alpha,\gamma)&=&\int\limits_{\mathbb{R}}{\rm erf}(\gamma x)\partial_x\exp[-(x+\alpha)^2]dx =-\frac{2\gamma}{\sqrt{\pi}}\int\limits_{\mathbb{R}}\exp[-(x+\alpha)^2-\gamma^2 x^2]dx =-\frac{2\gamma}{\sqrt{\gamma^2+1}}\exp\left[-\frac{\gamma^2\alpha^2}{\gamma^2+1}\right].\qquad
\end{eqnarray}
Integrating the derivative from $0$ to $\alpha$ we find the desired result
\begin{eqnarray}\label{a2.1.4}
 \int\limits_{\mathbb{R}}\exp[-(x+\alpha)^2]{\rm erf}(\gamma x)dx=-\sqrt{\pi}\,{\rm erf}\left(\frac{\gamma\alpha}{\sqrt{\gamma^2+1}}\right).
\end{eqnarray}
This integral is needed to simplify the term~\eqref{term1}.

Another integral identity  which is used for the derivation of the 
level density of the real eigenvalues with positive chirality is given by
\begin{eqnarray}\label{a2.1.5}
&&\int\limits_{\mathbb{R}^2}\exp(-\alpha_1x_1^2-\alpha_2x_2^2+\beta_1x_1+\beta_2x_2){\rm erf}\left(\frac{x_1+\delta x_2}{\gamma}+\epsilon\right)dx_1dx_2\\
&=&\frac{\pi}{\sqrt{\alpha_1\alpha_2}}\exp\left[\frac{1}{4}\left(\frac{\beta_1^2}{\alpha_1}+\frac{\beta_2^2}{\alpha_2}\right)\right]{\rm erf}\left(\frac{\alpha_2\gamma\beta_1+\alpha_1\gamma\delta\beta_2+2\alpha_1\alpha_2\gamma^2\epsilon}{2\sqrt{\alpha_1\alpha_2\gamma^2(\alpha_1\alpha_2\gamma^2+\alpha_1\delta^2+\alpha_2)}}\right).\nonumber
\end{eqnarray}
This identity is a direct consequence of the identity~\eqref{a2.1.4}.
The constants $\alpha_i$ (with $\RE\alpha_i>0$), 
$\beta_i$, $\gamma\neq0$, $\delta$ and $\epsilon$ are arbitrary. 

\subsection{Convolution of a Gaussian with a {\it sinus cardinalis}}\label{app2.2}

The second integral enters in the simplification of the asymptotic 
behavior of $\rho_{\rm c}$. It  is the convolution integral
\begin{eqnarray}\label{a2.2.1}
 \widetilde{I}(\alpha,\gamma)=\int\limits_{\mathbb{R}}dx\exp[-(x+\alpha)^2]{\rm sinc}(\gamma x).
\end{eqnarray}
To evaluate this integral we introduce an auxiliary integral 
to obtain a Fourier transform of a Gaussian, i.e.
\begin{eqnarray}\label{a2.2.2}
 \widetilde{I}(\alpha,\gamma)&=&\frac{1}{\gamma}\int\limits_0^\gamma d\widetilde{\gamma}\int\limits_{\mathbb{R}}dx\exp[-(x+\alpha)^2]{\rm cos}(\widetilde{\gamma} x).
\end{eqnarray}
First we integrate over $x$ and then over $\widetilde{\gamma}$ 
to obtain an expression in terms of error functions,
\begin{eqnarray}\label{a2.2.3}
 \widetilde{I}(\alpha,\gamma)&=&\frac{\pi}{\gamma}\exp(-\alpha^2)\RE\, {\rm erf}\left(\frac{\gamma}{2}+\imath \alpha\right).
\end{eqnarray}

\section{The $Z_{1/1}^{\nu}$--Partition Function}\label{app3}

In this appendix we evaluate the partition function $Z_{1/1}^\nu$ which enters
in the expression for the distribution of the chiralities over the real
eigenvalues of $D_\W$. The derivation below is along the lines given in Ref.~\cite{Splittorff:2011bj}.

 We employ the parametrization~\eqref{3.2.7} to evaluate
\begin{eqnarray}\label{3.2.9}
 &&\hspace*{-2cm}\underset{\varepsilon\to0}{\lim}\IM\int\frac{\det(D_\W-\widehat{z}_1/(2n)\eins_{2n+\nu})}{\det(D_\W-\widehat{x}_2/(2n)\eins_{2n+\nu}\mp\imath\varepsilon\gamma_5)}P(D_\W)d[D_\W]\\
 &\overset{n\gg1}{=}&\hspace*{-0.2cm}-\underset{\varepsilon\to0}{\lim}\IM\int \frac{d e^{\imath\varphi}}{2\pi\imath}d e^{\vartheta}d\eta d\eta^* \Sdet^\nu U\exp[-\widehat{a}^2_8\Str(U^2+U^{-2})]\nonumber\\
 &&\times\exp\left[\pm\frac{\imath}{2}\Str\diag(\widehat{x}_2-\widehat{m}_6,\widehat{z}_1-\widehat{m}_6)(U-U^{-1})-\left(\varepsilon\pm\frac{\imath\widehat{\lambda}_7}{2}\right)\Str(U+U^{-1})\right].\nonumber
\end{eqnarray}
We employ the same trick as in Ref.~\cite{Splittorff:2011bj} to linearize the exponent in $U$ and $U^{-1}$ by introducing an auxiliary Gaussian integral over a supermatrix, i.e.
\begin{eqnarray}\label{3.2.10}
 \exp[-\widehat{a}^2_8\Str(U^2+U^{-2})]=\int d[\sigma]\exp\left[-\frac{1}{16\widehat{a}^2_8}\Str\sigma^2\pm\frac{\imath}{2}\Str\sigma(U-U^{-1})\right]
\end{eqnarray}
with
\begin{eqnarray}\label{3.2.11}
 \sigma=\left[\begin{array}{cc} \sigma_1 & \eta_\sigma \\ \eta_\sigma^* & \imath\sigma_2 \end{array}\right]\ {\rm and}\ d[\sigma]=d\sigma_1 d\sigma_2 d\eta_\sigma d\eta_\sigma^*.
\end{eqnarray}
After plugging Eq.~\eqref{3.2.10} in Eq.~\eqref{3.2.9} we diagonalize $\sigma=V\diag(s_1,s_2)V^\dagger$ and integrate over $V\in\U(1/1)$. We obtain
\begin{eqnarray}\label{3.2.12}
 &&\hspace*{-0.3cm}\underset{\varepsilon\to0}{\lim}\IM\int\frac{\det(D_\W-\widehat{z}_1/(2n)\eins_{2n+\nu})}{\det(D_\W-\widehat{x}_2/(2n)\eins_{2n+\nu}\mp\imath\varepsilon\gamma_5)}P(D_\W)d[D_\W]\\
 &\overset{n\gg1}{=}&\frac{1}{16\pi\widehat{a}^2_8}
(\widehat{z}_1-\widehat{x}_2)\underset{\varepsilon\to0}{\lim}\IM
\int  \frac{ds_1ds_2}{s_1-\imath s_2} 
\left(\frac{s_1-\widehat{\lambda}_7\pm\imath\varepsilon}{s_1+\widehat{\lambda}_7\mp\imath\varepsilon}\frac{\imath s_2+\widehat{\lambda}_7\mp\imath\varepsilon}{\imath s_2-\widehat{\lambda}_7\pm\imath\varepsilon}\right)^{\nu/2}\nonumber\\
 &&\times\exp\left[-\frac{1}{16\widehat{a}^2_8}\left((s_1-\widehat{x}_2+\widehat{m}_6)^2-(s_2+\imath \widehat{z}_1-\imath\widehat{m}_6)^2\right)\right] Z_{1/1}^\nu\left(\sqrt{s_1^2-(\widehat{\lambda}_7\mp\imath\varepsilon)^2},\imath\sqrt{s_2^2+(\widehat{\lambda}_7\mp\imath\varepsilon)^2};\widehat{a}=0\right)\nonumber,
\end{eqnarray}
which expresses the partition function at non-zero lattice spacing in terms of an integral over
 the partition function with one bosonic and one fermionic flavor at zero lattice spacing~\eqref{Z11cont}.
 
The resolvent $G_{1/1}$ is given by the derivative with respect to $\widehat{z}_1$, see Eq.~\eqref{Greenfunc}. To obtain a non-zero result we necessarily have to differentiate the prefactor $(\widehat{z}_1-\widehat{x}_2)$. The distribution
of the chiralities of the real eigenvalues of $D_\W$ follows from the imaginary part of the resolvent.
The Efetov-Wegner term \cite{Weg83,Efe83} appearing after  diagonalizing $\sigma$ is the normalization $Z_{1/1}^\nu(1,1)=1$ and vanishes when taking the imaginary part.

Two terms contribute to the imaginary part of the resolvent. First, the imaginary part of
\be
\frac{1}{\pi}\underset{\varepsilon\to0}{\lim}\IM\left [ \frac{1}{(s_1+\widehat{\lambda}_7-\imath\varepsilon)^\nu} \right ]=\frac{(-1)^{\nu-1}}{(\nu-1)!}\delta^{(\nu)}(s_1+\hla)
\ee
is the $\nu$-th derivative of the Dirac delta-function.
Second, when $|s_1|< |\hla|$, the imaginary part arising from 
the logarithmic contribution of $K_\nu(z)$, i.e.
\be
K_\nu(z) = (-1)^{\nu+1}I_\nu(z) \log z + \frac 1{z^{\nu}} \sum_{k=0}^\infty a_k z^k  
\ee
also contributes to the imaginary part of the resolvent.
 The Bessel functions of the imaginary part of $Z_{1/1}^\nu(x_1,x_2,\ha=0)$ 
combine into the two-flavor partition function $Z_{2/0}^\nu(x_1,x_2,\ha=0)$.
Adding both contributions  we arrive at the result
\be
 &&\hspace*{-0.3cm}\underset{\varepsilon\to0}{\lim}\IM\int\frac{\det(D_\W-z_1\eins_{2n+\nu})}{\det(D_\W-x_2\eins_{2n+\nu}\mp\imath\varepsilon\gamma_5)}P(D_\W)d[D_\W]
\label{zchi}\\
 &\overset{n\gg1}{=}&\frac{1}{16\pi\widehat{a}^2_8}
(\widehat{z}_1-\widehat{x}_2)\underset{\varepsilon\to0}{\lim}\IM
\int  \frac{ds_1ds_2}{s_1-\imath s_2} 
(\imath s_2+\widehat{\lambda}_7)^\nu(s_1+\widehat{\lambda}_7)^\nu\exp\left[-\frac{1}{16\widehat{a}^2_8}\left((s_1-\widehat{x}_2+\widehat{m}_6)^2-(s_2+\imath \widehat{z}_1-\imath\widehat{m}_6)^2\right)\right]\nonumber\\
 &&
\hspace*{-0.6cm}\times \left[\frac{\delta^{(\nu-1)}(s_1+\widehat{\lambda}_7)}{(\nu-1)!(s_1-\widehat{\lambda}_7)^\nu}\left(\frac{s_1^2-\widehat{\lambda}_7^2}{s_2^2+\widehat{\lambda}_7^2}\right)^{\nu/2}Z_{1/1}^\nu\left(\sqrt{s_1^2-\widehat{\lambda}_7^2},\imath\sqrt{s_2^2+\widehat{\lambda}_7^2};\widehat{a}=0\right)\right.
\nonumber\\
 &&\hspace*{2cm}\left.-{\rm sign}(\widehat{\lambda}_7)\Theta(|\widehat{\lambda}_7|-|s_1|) \left(s_1^2+s_2^2\right)\frac{Z_{2/0}^\nu\left(\sqrt{s_1^2-\widehat{\lambda}_7^2},\imath\sqrt{s_2^2+\widehat{\lambda}_7^2};\widehat{a}=0\right)}{[(s_1^2-\widehat{\lambda}_7^2)(s_2^2+\widehat{\lambda}_7^2)]^{\nu/2}}\right]\nonumber
\ee
yielding Eq.~\eqref{4.3.1}.

\section{Derivations of the Asymptotic Results given in Sec.~\ref{sec4}}\label{app4}

The derivation of asymptotic limits of the spectral density can be quite non-trivial because of
 cancellations of the leading contributions so that a naive saddle point approximation cannot
be used. 
In the subsections below, we derive asymptotic expressions for the average number of 
additional real modes (Appendix~\ref{app4.2}), the level density of the right handed modes (Appendix~\ref{app4.1})
 and the level density of the complex modes (Appendix~\ref{app4.3}). 
In Appendix~\ref{app4.4} we consider the distribution of chirality over the real modes.

\subsection{The average number of additional real modes}\label{app4.2}

The limit of small lattice spacing is obvious and will not be discussed here. At large lattice spacing 
we rewrite Eq.~\eqref{4.1.4} as
\begin{eqnarray}\label{a4.2.1}
 N_{\rm add}&=&\int\limits_{[0,2\pi]^2}\frac{d\Phi d\varphi}{ 8\pi^2}\cos[2\nu\Phi]\frac{1-\exp\left[-8(\widehat{a}_8^2\sin^2\varphi+2\widehat{a}_7^2\cos^2\varphi) \sin^2\Phi\right]}{\sin^2\Phi}.
\end{eqnarray}
Since $\widehat{a}_{7/8}$ are large we expand the angle $\Phi$ around the origin, in particular
\begin{eqnarray}\label{a4.2.2}
 \Phi=\frac{\delta\Phi}{\sqrt{\widehat{a}_8^2\sin^2\varphi+2\widehat{a}_7^2\cos^2\varphi}}\ll1.
\end{eqnarray}
Note that we have two equivalent saddlepoints at $0$ and at $\pi$. We thus have
\begin{eqnarray}\label{a4.2.3}
N_{\rm add}&=&\int\limits_{\mathbb{R}\times[0,2\pi]}\frac{d\delta\Phi d\varphi}{ 4\pi^2}\frac{1-\exp\left[-8\delta\Phi^2\right]}{\delta\Phi^2}\sqrt{\widehat{a}_8^2\sin^2\varphi+2\widehat{a}_7^2\cos^2\varphi}.
\end{eqnarray}
The integral over $\delta\Phi$ is equal to $\sqrt{32\pi}$, and the  integral 
over $\varphi$ is the elliptic integral of the second kind. Hence we obtain the result~\eqref{4.1.5}.

\subsection{The density of the additional real modes}\label{app4.1}

We have two different cases for the behavior of $\rho_{\rm r}$ at large lattice spacing. 
To derive the large $\ha$ asymptotics in the case $\widehat{a}_8^2=0$ we rewrite Eq.~\eqref{4.1.2}
 as a group integral, i.e.
\begin{eqnarray}
 \rho_{\rm r}(\widehat{x})&=&\frac{1}{2^{15/2}\pi\widehat{a}_7\sqrt{\widehat{a}_8^2+2\widehat{a}_6^2}}\int\limits_{\mathbb{R}^2}d\widehat{\lambda}_7d\widetilde{x}\int\limits_{\U(2)}d\mu(U){\det}^{\nu}U\exp\left[-\frac{\widehat{\lambda}_7^2}{16\widehat{a}_7^2}\right]\left[{\rm sign}(\widetilde{x}-\widehat{x})-{\rm erf}\left(\frac{\widetilde{x}-\widehat{x}+2\widehat{\lambda}_7}{\sqrt{32\widehat{a}_8^2}}\right)\right](\widetilde{x}-\widehat{x})\nonumber\\
 &&\times\exp\left[-\widehat{a}_8^2\tr\left(U+U^{-1}-\frac{\widetilde{x}+\widehat{x}}{8\widehat{a_8}^2}\eins_2\right)^2+\frac{\widehat{a}_6^2\widehat{a}_8^2}{\widehat{a}_8^2+2\widehat{a}_6^2}\tr^2\left(U+U^{-1}-\frac{\widetilde{x}+\widehat{x}}{8\widehat{a_8}^2}\eins_2\right)\right]\nonumber\\
 &&\times\exp\left[\frac{\widetilde{x}-\widehat{x}}{4}\tr\diag(1,-1)(U+U^{-1})+\frac{\widehat{\lambda}_7}{2}\tr(U-U^{-1})\right].\hspace*{0.5cm}\label{a4.1.1}
\end{eqnarray}
For $\widehat{a}_8^2=0$ Eq.~\eqref{a4.1.1} simplifies to
\begin{eqnarray}
 \rho_{\rm r}(\widehat{x})&=&\frac{1}{256\pi\widehat{a}_7\widehat{a}_6}\int\limits_{\mathbb{R}^2}d\widehat{\lambda}_7d\widetilde{x}\int\limits_{\U(2)}d\mu(U){\det}^{\nu}U\exp\left[-\frac{\widehat{\lambda}_7^2}{16\widehat{a}_7^2}\right]\left[{\rm sign}(\widetilde{x}-\widehat{x})-{\rm sign}(\widetilde{x}-\widehat{x}+2\widehat{\lambda}_7)\right](\widetilde{x}-\widehat{x})\nonumber\\
 &&\times\exp\left[-\frac{(\widetilde{x}+\widehat{x})^2}{64\widehat{a}_6^2}+\frac{\widetilde{x}-\widehat{x}}{4}\tr\diag(1,-1)(U+U^{-1})+\frac{\widehat{\lambda}_7}{2}\tr(U-U^{-1})\right]\nonumber\\
 &=&\frac{1}{32\pi\widehat{a}_7\widehat{a}_6}\int\limits_{\mathbb{R}}d\widetilde{x}\left( \Theta(\widetilde{x})\int\limits_{-\infty}^{-\widetilde{x}}-\Theta(-\widetilde{x})\int\limits_{-\widetilde{x}}^{\infty}\right)d\widehat{\lambda}_7\int\limits_{\U(2)}d\mu(U){\det}^{\nu}U\nonumber\\
 &&\times \widetilde{x}\exp\left[-\frac{\widehat{\lambda}_7^2}{16\widehat{a}_7^2}-\frac{(\widetilde{x}+\widehat{x})^2}{16\widehat{a}_6^2}-\frac{\sqrt{\widehat{\lambda}_7^2-\widetilde{x}^2}}{2}\tr(U-U^{-1})\right].\label{a4.1.2}
\end{eqnarray}
For the second equality we substituted $\widetilde{x}\to 2\widetilde{x}+\widehat{x}$ and 
replaced the sign functions by the integration domains of $\widehat{\lambda}_7$. 
Moreover we used the fact that the group integral only depends on 
$\sqrt{\widehat{\lambda}_7^2-\widetilde{x}^2}$.

The saddlepoint equation of the $U$ integral in  Eq.~\eqref{a4.1.2} gives four saddle points,
\begin{eqnarray}\label{a4.1.3}
 U=\pm\imath\eins_2,\ {\rm and}\ U=\pm \imath\diag(1,-1).
\end{eqnarray}
The saddlepoints which are proportional to unity are algebraically suppressed while the
contribution of the other two saddle points is the same.
 We thus find
\begin{eqnarray}
 \rho_{\rm r}(\widehat{x})&=&\frac{1}{8\pi^2\widehat{a}_7\widehat{a}_6}\int\limits_{0}^\infty d\widetilde{x}\int\limits_{\widetilde{x}}^{\infty} d\widehat{\lambda}_7\frac{\widetilde{x}}{\sqrt{\widehat{\lambda}_7^2-\widetilde{x}^2}}\cosh\left(\frac{\widetilde{x}\widehat{x}}{8\widehat{a}_6^2}\right)\exp\left[-\frac{\widehat{\lambda}_7^2}{16\widehat{a}_7^2}-\frac{\widetilde{x}^2+\widehat{x}^2}{16\widehat{a}_6^2}\right].\label{a4.1.4}
\end{eqnarray}
After substituting $\widehat{\lambda}_7\to\widetilde{x}\cosh\vartheta$ the integral 
over $\vartheta$ yields the first case of Eq.~\eqref{4.1.6}.

For $\widehat{a}_8\neq 0$ we  again  start with Eq.~\eqref{4.1.2}.
The integration over the two error functions, 
see Eq.~\eqref{4.1.3}, makes it difficult to evaluate the result directly, 
particularly when $\widehat{a}_7\neq 0$. As long as $\widehat{a}_7$ is finite, 
the second error function does not yield anything apart from giving  a Gaussian cut-off to the integral. 
The imaginary part of the argument of the second error function shows strong oscillations resulting in cancellations.
These oscillations also impede a numerical evaluation of the integrals for large lattice spacing.

Let $\widehat{a}_6=0$ to begin with. A non-zero value of $\ha_6$ can be introduced  later 
 by a convolution with a Gaussian in $\widehat{x}$. To obtain the correct contribution 
from the first term we consider a slight modification of $\rho_{\rm r}$,
\begin{eqnarray}
I(X,\alpha)=\int\limits_{[0,2\pi]^2}d\varphi_1d\varphi_2\sin^2\left[\frac{\varphi_1-\varphi_2}{2}\right]e^{\imath\nu(\varphi_1+\varphi_2)}\frac{\widehat{k}(X,\varphi_1,\varphi_2)-\widehat{k}(X,\varphi_2,\varphi_1)}{\cos\varphi_2-\cos\varphi_1}\label{a4.1.5}
\end{eqnarray}
with
\begin{eqnarray}\label{a4.1.6}
 \widehat{k}(X,\varphi_1,\varphi_2)&=&\exp\left[4\widehat{a}_8^2 \left(\cos\varphi_1-X\right)^2-4\widehat{a}_8^2 \left(\cos\varphi_2-X\right)^2-4\widehat{a}_7^2(\sin\varphi_1+\sin\varphi_2)^2\right]\\
 &&\times\left[{\rm erf}\left[\sqrt{8}\widehat{a}_8(X-\cos\varphi_1)\right]+{\rm erf}\left[2\widehat{a}_8\alpha(\cos\varphi_1-X)\right]\right].\nonumber
\end{eqnarray}
The variable $X$ plays the role of $\widehat{x}/(8\widehat{a}_8^2)$. The error function 
with the constant $\alpha$ replaces the second error function in  Eq.~\eqref{4.1.3} 
and is of order one in the limit $\widehat{a}\to\infty$. It regularizes the integral and its contribution will be removed at the end.  
However it has to fulfill some constraints to guarantee the existence of the saddlepoints
\begin{eqnarray}\label{a4.1.7}
 \varphi_1^{(0)},\varphi_2^{(0)}\in\{\pm\arccos X\}\ {\rm with}\ X\in[-1,1].
\end{eqnarray}
Nevertheless these saddlepoints are independent of $\alpha$.
The saddle point  $\varphi_1^{(0)}=\varphi_2^{(0)}$ is algebraically suppressed in comparison to 
$\varphi_1^{(0)}=-\varphi_2^{(0)}$ due to the $\sin^2$ factor in the measure. Expanding about the saddlepoints yields
\begin{eqnarray}
I(X,\alpha)&\propto&\frac{\Theta(1-|X|)}{\widehat{a}_8}
\int\limits_{\mathbb{R}^2}
\frac{d\delta\varphi_1d\delta\varphi_2}{\delta\varphi_1+\delta\varphi_2}
\exp\left[-\frac{\widehat{a}_7^2}{\widehat{a}_8^2}\gamma^2(\delta\varphi_1+\delta\varphi_2)^2\right] \label{a4.1.8}\\
&&\times\left[\exp(\delta\varphi_1^2-\delta\varphi_2^2)
\left({\rm erf}(\sqrt{2}\delta\varphi_1)-{\rm erf}(\alpha\delta\varphi_1)\right)+\exp(\delta\varphi_2^2-\delta\varphi_1^2)
\left({\rm erf}(\sqrt{2}\delta\varphi_2)-{\rm erf}(\alpha\delta\varphi_2)\right)\right]\nonumber
\end{eqnarray}
with
\begin{eqnarray}
\gamma(X)=\frac{X}{\sqrt{1-X^2}}. \label{a4.1.9}
\end{eqnarray}
In the next step we change the coordinates to center-of-mass-relative coordinates, i.e. $\Phi=\delta\varphi_1+\delta\varphi_2$ and $\Delta\varphi=\delta\varphi_1-\delta\varphi_2$, and find
\begin{eqnarray}
I(X,\alpha)&\propto&\frac{\Theta(1-|X|)}{\widehat{a}_8}\int\limits_{\mathbb{R}^2}\frac{d\Phi d\Delta\varphi}{\Phi}\exp\left[-\frac{\widehat{a}_7^2}{\widehat{a}_8^2}\gamma^2(X)\Phi^2\right] \label{a4.1.10}\\
&&\times\left[\exp(\Phi\Delta\varphi)\left({\rm erf}\left(\frac{\Phi+\Delta\varphi}{\sqrt{2}}\right)-{\rm erf}\left(\frac{\alpha}{2}(\Phi+\Delta\varphi)\right)\right)+\exp(-\Phi\Delta\varphi)\left({\rm erf}\left(\frac{\Phi-\Delta\varphi}{\sqrt{2}}\right)-{\rm erf}\left(\frac{\alpha}{2}(\Phi-\Delta\varphi)\right)\right)\right].\nonumber
\end{eqnarray}
We perform an integration by parts in $\Delta\varphi$ yielding Gaussian integrals in $\Delta\varphi$
which evaluate to
\begin{eqnarray}
I(X,\alpha)&\propto&\frac{\Theta(1-|X|)}{\widehat{a}_8}\int\limits_{\mathbb{R}}\frac{d\Phi}{\Phi^2}\exp\left[-\frac{\widehat{a}_7^2}{\widehat{a}_8^2}\gamma^2(X)\Phi^2\right]\left[\exp\left(-\frac{\Phi^2}{2}\right)-\exp\left(-\frac{1+\alpha^2}{\alpha^2}\Phi^2\right)\right]. \label{a4.1.11}
\end{eqnarray}
The $1/\Phi^2$ term can be exponentiated  by introducing an auxiliary integral and the resulting Gaussian over $\Phi$ can be performed. We obtain
\begin{eqnarray}
I(X,\alpha)&\propto&\Theta(1-|X|)\int\limits_{(1+\alpha^2)/\alpha^2}^{1/2}\frac{dt}{\sqrt{\widehat{a}_7^2\gamma^2(X)+\widehat{a}_8^2t}}\label{a4.1.12}\\
&\propto&\frac{\Theta(1-|X|)}{\widehat{a}_8^2}\left(\sqrt{\widehat{a}_7^2\gamma^2(X)+\frac{\widehat{a}_8^2}{2}}-\sqrt{\widehat{a}_7^2\gamma^2(X)+\frac{1+\alpha^2}{\alpha^2}\widehat{a}_8^2}\right).\nonumber
\end{eqnarray}
The contribution of the artificial term depending on $\alpha$ can be readily read off, but it fixes
the integral only up to an additive constant. 
This constant can be determined by integrating the result over $\widehat{x}$ which has to
agree with the large $\ha$ limit of  $N_{\rm add}$, cf. Eq.~\eqref{4.1.5}. It turns out that this constant
is equal to zero. The overall constant is also obtained by comparing to $N_{\rm add}$.

The convolution with the Gaussian distribution generating $\widehat{a}_6$ does not give
something new in the limit of large lattice spacing. The width of this  Gaussian scales with $ \ha$ 
while the density $\rho_\rt$ has support on $\ha^2$, so that it  becomes 
a Dirac delta-function in the large $\ha$ limit.

\subsection{The density of the complex eigenvalues}\label{app4.3}

Let $\widehat{a}_8>0$ and $\ha \gg 1$. Then we perform a saddlepoint approximation of Eq.~\eqref{4.2.1} in the integration variables $\varphi_{1/2}$. The saddlepoints are given by
\begin{equation}\label{a4.3.1}
 \varphi_{1}^{(0)}=- \varphi_{2}^{(0)}=\pm{\rm arccos}\left(\frac{\widehat{x}}{8\widehat{a}_8^2}\right)\quad{\rm with}\quad \widehat{x}\in[-8\widehat{a}_8^2,8\widehat{a}_8^2].
\end{equation}
We have also the saddlepoints $\varphi_1^{(0)}=\varphi_2^{(0)}$ if $\widehat{a}_7=0$. 
However they are algebraically suppressed due to the Haar measure. 
Notice the two saddlepoints in Eq.~\eqref{a4.3.1} 
 yield the same contribution. After the integration over the massive modes about the saddlepoint 
we find the first case of Eq.~\eqref{4.2.4}. In the calculation we used the convolution  integral derived 
 in Appendix~\ref{app2.2}.

Let us now look at the case with $\widehat{a}_8=0$. Then we have
\begin{eqnarray}\label{a4.3.2}
 \rho_{\rm c}(\widehat{z})&\overset{\widehat{a}\gg1}{=}&\frac{|\widehat{y}|}{4\pi^2\sqrt{16\pi\widehat{a}_6^2}\sqrt{16\pi\widehat{a}_7^2}}\exp\left[-\frac{\widehat{x}^2}{16\widehat{a}_6^2}\right]\int\limits_{\mathbb{R}}d\widehat{\lambda}_7\exp\left[-\frac{\widehat{\lambda}_7^2}{16\widehat{a}_7^2}\right]\\
 &&\times\int\limits_{[0,2\pi]^2}d\varphi_1 d\varphi_2\sin^2\left[\frac{\varphi_1-\varphi_2}{2}\right]\cos[\nu(\varphi_1+\varphi_2)]{\rm sinc}\left[\widehat{y}(\cos\varphi_1-\cos\varphi_2)\right]\exp\left[\imath\widehat{\lambda}_7(\sin\varphi_1+\sin\varphi_2)\right].\nonumber
\end{eqnarray}
The integrals over the angles can be rewritten as a group integral over $\U(2)$, 
\begin{eqnarray}
&&\int\limits_{[0,2\pi]^2}d\varphi_1 d\varphi_2\sin^2\left[\frac{\varphi_1-\varphi_2}{2}\right]e^{\nu\imath(\varphi_1+\varphi_2)}{\rm sinc}\left[\widehat{y}(\cos\varphi_1-\cos\varphi_2)\right]\exp\left[\imath\widehat{\lambda}_7(\sin\varphi_1+\sin\varphi_2)\right]\nonumber\\
&=&2\pi^2\int\limits_{\U(2)}d\mu(U){\det}^\nu U\exp\left[\frac{1}{2}\tr (\Lambda U-\Lambda^* U^\dagger)\right]\label{a4.3.3}
\end{eqnarray}
with $\Lambda=\diag(\widehat{\lambda}_7+\imath\widehat{y},\widehat{\lambda}_7-\imath\widehat{y})$. This integral only depends on the quantity $\sqrt{\widehat{\lambda}_7^2+\widehat{y}^2}$ because the angle of the combined complex variable $\widehat{\lambda}+\imath\widehat{y}$ can be absorbed into $U$, i.e.
\begin{eqnarray}\label{a4.3.4}
&&\int\limits_{\U(2)}d\mu(U){\det}^\nu U\exp\left[\frac{1}{2}\tr (\Lambda U-\Lambda^*U^\dagger)\right]=\int\limits_{\U(2)}d\mu(U){\det}^\nu U\exp\left[\frac{\sqrt{\widehat{\lambda}_7^2+\widehat{y}^2}}{2}\tr ( U- U^\dagger)\right].
\end{eqnarray}
The variable $\widehat{y}$ as well as the integration variable $\widehat{\lambda}_7$ are of the order $\ha$. Therefore we can perform a saddlepoint approximation and  end up with
\begin{eqnarray}\label{a4.3.5}
 \rho_{\rm c}(\widehat{z})&\overset{\widehat{a}\gg1}{=}&\frac{|\widehat{y}|}{\pi\sqrt{16\pi\widehat{a}_6^2}\sqrt{16\pi\widehat{a}_7^2}}\exp\left[-\frac{\widehat{x}^2}{16\widehat{a}_6^2}\right]\int\limits_{\mathbb{R}}d\widehat{\lambda}_7\frac{\exp\left[-\widehat{\lambda}_7^2/ (16\widehat{a}_7^2)\right]}{\sqrt{\widehat{\lambda}_7^2+\widehat{y}^2}}.
\end{eqnarray}
resulting in 
 the second case of Eq.~\eqref{4.2.4}.

\subsection{The distribution of chirality over the real eigenvalues}\label{app4.4}
 
In this Appendix we derive the large $\ha$ limit of $\rho_\chi(\widehat{x})$ for $\ha_8>0$
given
in Eq.~\eqref{4.3.8}.  The case $\ha_8=0$ reduces $\rho_\chi(\widehat{x})$ to the result~\eqref{4.0.3} 
and will not be discussed in this section. 
We set $\widehat{a}_{6/7}=0$ to begin with and introduce them later on.

The best way to obtain the asymptotics for large lattice spacing is to start with Eq.~\eqref{3.2.9} with $\widehat{m}_6=\widehat{\lambda}_7=\varepsilon=0$. The integral does not need a regularization since the $\widehat{a}_8$-term guarantees the convergence. We also omit the sign in front of the linear trace terms in the Lagrangian because we can change $U\to-U$.

In the first step we substitute $\eta\rightarrow e^{\imath\varphi}\eta$ and $\eta^*\rightarrow e^\vartheta\eta^*$. Then the measure is $d\varphi d\vartheta d\eta d\eta^*$ and the parametrization of $U$ is given by
\begin{eqnarray}\label{a4.4.1}
 U=\left[\begin{array}{cc} e^\vartheta & 0 \\ 0 & e^{\imath\varphi} \end{array}\right]\left[\begin{array}{cc} 1 & \eta^* \\ \eta & 1 \end{array}\right],\  U^{-1}=\left[\begin{array}{cc} 1+\eta^*\eta & -\eta^* \\ -\eta & 1-\eta^*\eta \end{array}\right]\left[\begin{array}{cc} e^{-\vartheta} & 0 \\ 0 & e^{-\imath\varphi} \end{array}\right].
\end{eqnarray}
There are two saddlepoints in the variables $\vartheta$ and $\varphi$, i.e.
\begin{eqnarray}\label{a4.4.2}
 e^{\vartheta_0}=-\frac{\imath\widehat{x}_2}{8\widehat{a}_8^2}+\sqrt{1-\left(\frac{\widehat{x}_2}{8\widehat{a}_8^2}\right)^2},\quad e^{\imath\varphi_0}=-\frac{\imath\widehat{z}_1}{8\widehat{a}_8^2}+L\sqrt{1-\left(\frac{\widehat{z}_1}{8\widehat{a}_8^2}\right)^2}
\end{eqnarray}
with $L={\pm1}$. Moreover, the variables 
$\widehat{z}_1,\widehat{x}_2$ have to be in the interval 
$[-8\widehat{a}_8^2,8\widehat{a}_8^2]$ else 
the contributions will be exponentially suppressed. 
We have no second saddlepoint for the variable $\vartheta$ since the 
real part of the exponential has to be positive definite. 
Other saddlepoints which can be reached by shifting $\varphi$ and $\vartheta$  by $2\pi\imath$ independently are forbidden since they are not accessible in the limit $\widehat{a}_8\to\infty$ . Notice that the saddlepoint solutions~\eqref{a4.4.2} are phases, i.e. $|e^{\vartheta_0}|=|e^{\imath\varphi_0}|=1$.

In the second step we expand the integration variables
\begin{eqnarray}\label{a4.4.3}
 e^{\vartheta}= e^{\vartheta_0}\left(1+\frac{\delta\vartheta}{\sqrt{(8\widehat{a}_8^2)^2-\widehat{x}_2^2}}\right),\quad e^{\imath\varphi}=e^{\imath\varphi_0}\left(1+\frac{\imath\delta\varphi}{\sqrt{(8\widehat{a}_8^2)^2-\widehat{z}_1^2}}\right).
\end{eqnarray}
All terms in front of the exponential as well as of the Grassmann variables are replaced by the saddlepoint solutions $\vartheta_0$ and $\varphi_0$. 
The resulting Gaussian integrals over the variables $\delta\vartheta$ and $\delta\varphi$ yield
\begin{eqnarray}
 &&\hspace*{-1cm}\IM \int d\mu(U) \Sdet^\nu U \exp\left[-\widehat{a}_8^2\Str(U-U^{-1})^2+\frac{\imath}{2}\Str\diag(\widehat{x}_2,\widehat{z}_1)(U-U^{-1})\right]\nonumber\\
 &\propto&\sum\limits_{L\in\{\pm1\}}\IM \frac{\exp[\nu(\vartheta_0-\imath\varphi_0)]}{\sqrt{(8\widehat{a}_8^2)^2-\widehat{x}_2^2}\sqrt{(8\widehat{a}_8^2)^2-\widehat{z}_1^2}}\exp\left[-\frac{\widehat{x}_2^2- \widehat{z}_1^2}{16\widehat{a}_8^2}\right]\nonumber\\
 &&\times\int d\eta d\eta^*(1-\eta^*\eta)^\nu\exp[-2\widehat{a}_8^2(e^{\vartheta_0}+e^{-\imath\varphi_0})(e^{-\vartheta_0}+e^{\imath\varphi_0})\eta^*\eta].\label{a4.4.4}
\end{eqnarray}
After the integration over the Grassmann variables we have two terms, one is of order 
one, and the other one of order $\widehat{a}_8^2$ which exceeds the first term for 
$\ha_8 \gg 1$. Hence we end up with
\begin{eqnarray}
 &&\hspace*{-1cm}\IM \int d\mu(U) \Sdet^\nu U \exp\left[-\widehat{a}_8^2\Str(U-U^{-1})^2+\frac{\imath}{2}\Str\diag(\widehat{x}_2,\widehat{z}_1)(U-U^{-1})\right]\nonumber\\
 &\propto&\sum\limits_{L\in\{\pm1\}}\IM \frac{1}{\sqrt{(8\widehat{a}_8^2)^2-\widehat{x}_2^2}\sqrt{(8\widehat{a}_8^2)^2-\widehat{z}_1^2}} \left(\frac{-\imath \widehat{x}_2+\sqrt{(8\widehat{a}_8^2)^2-\widehat{x}_2^2}}{-\imath\widehat{z}_1+L\sqrt{(8\widehat{a}_8^2)^2-\widehat{z}_1^2}}\right)^\nu\label{a4.4.5}\\
 &&\times\left[\left(\frac{\widehat{z}_1-\widehat{x}_2}{8\widehat{a}_8^2}\right)^2+\left(\sqrt{1-\left(\frac{\widehat{x}_2}{8\widehat{a}_8^2}\right)^2}+L\sqrt{1-\left(\frac{\widehat{z}_1}{8\widehat{a}_8^2}\right)^2}\right)^2\right]\exp\left[-\frac{\widehat{x}_2^2- \widehat{z}_1^2}{16\widehat{a}_8^2}\right].\nonumber
\end{eqnarray}
Notice that both saddlepoints, $L\in\{\pm1\}$, give a contribution 
for independent variables $\widehat{z}_1$ and $\widehat{x}_2$.
To obtain the resolvent we differentiate this expression with respect to $\widehat{z}_1$ and
put $\widehat{z}_1=\widehat{x}_1$ afterwards. The first term between the large brackets and the 
second term for $L=-1$ are quadratic in $\widehat{z}_1-\widehat{x}_2$ and do not contribute to the resolvent.
 For $L=+1$ we obtain
\begin{eqnarray}
 \IM\partial_{\widehat{z}_1}|_{\widehat{z}_1=\widehat{x}_2}\int d\mu(U) \Sdet^\nu U \exp\left[-\widehat{a}_8^2\Str(U-U^{-1})^2+\frac{\imath}{2}\Str\diag(\widehat{x}_2,\widehat{z}_1)(U-U^{-1})\right]\propto\nu \frac{\Theta(8 \ha^2_8-|\widehat{x}_2|)}{2\ha^2_8\sqrt{64\ha_8^4-\widehat{x}_2^2}}
\label{a4.4.6} 
\ee
This limit yields the square root singularity. The normalization of $\rho_\chi$ to $\nu$
yields an overall normalization constant of $2\ha_8^2/\pi$.

The effect of $\widehat{a}_6$ is introduced by the integral
\begin{eqnarray}
\frac 1{4\ha_6 \sqrt \pi}
\int\limits_{\mathbb{R}}\exp\left[-\frac{\widehat{m}_6^2}{16\widehat{a}_6^2}\right]
\rho_\chi(\widehat{x}-\widehat{m}_6)|_{\widehat{a}_6=0} d\widehat{m}_6\label{D26}&=&
\frac 1{4\ha_6 \sqrt \pi}
\int\limits_{\mathbb{R}}\exp\left[-\frac{\widehat{m}_6^2}{16\widehat{a}_6^2}\right]
\frac{\Theta(8\widehat{a}_8^2-|\widehat{x}-\widehat{m}_6|)}
{\pi\sqrt{(8\widehat{a}_8^2)^2-(\widehat{x}-\widehat{m}_6)^2}} d\widehat{m}_6\\
&=&\frac {1}{4\ha_6 \pi^{3/2}}\int\limits_0^{\pi}\exp\left[-\frac{4\widehat{a}_8^4}{\widehat{a}_6^2}\left(\cos\varphi+\frac{\widehat{x}}{8\widehat{a}_8^2}\right)^2\right]d\varphi .\nonumber
\ee
In the large $\ha$ limit this evaluates to
\be
\frac 1{4\ha_6 \sqrt \pi}
\int\limits_{\mathbb{R}}\exp\left[-\frac{\widehat{m}_6^2}{16\widehat{a}_6^2}\right]
\frac{\Theta(8\widehat{a}_8^2-|\widehat{x}-\widehat{m}_6|)}
{\pi\sqrt{(8\widehat{a}_8^2)^2-(\widehat{x}-\widehat{m}_6)^2}} d\widehat{m}_6\overset{\widehat{a}\gg1}{=}\nu \frac{\Theta(8 \ha^2_8-|\widehat{x}|)}{\pi\sqrt{64\ha_8^4-\widehat{x}^2}},\label{D28}
\ee
which is exactly the same Heaviside distribution with the square root singularities in the interval $[-8\widehat{a}_8^2,8\widehat{a}_8^2]$ of Eq.~\eqref{a4.4.6}. 
The introduction of $\widehat{a}_7$ follows from Eq.~\eqref{4.3.5}. We have to 
replace $\ha_6^2 \to \ha_6^2+\ha_7^2$ and sum the result over the index
$j$ with the prefactor $\exp(-8\ha_7^2)[I_{j-\nu}(8\ha_7^2)-I_{j-\nu}(8\ha_7^2)]$.
The intermediate result~\eqref{D28} is independent of $\ha_7$ and  linear in the index, in the sum this index is $j$. The sum over $j$ can be performed according to
\begin{equation}\label{a4.4.8}
 \sum\limits_{j=1}^\infty j\left(I_{j-\nu}(8\widehat{a}_7^2)-I_{j+\nu}(8\widehat{a}_7^2)\right)=\nu\exp(8\widehat{a}_7^2)
\end{equation}
resulting in the asymptotic result~\eqref{4.3.8}.


\begin{thebibliography}{11}
\bibitem{footnote} Notice that in this subsection the terms proportional to $\widehat{m}_6$ and $\widehat{\lambda}_7$ are explicitly included in the Dirac operator rather than in the probability distribution as in the earlier sections. Moreover the operators $D$ and $D_{\rm W}$ are already multiplied with $\Sigma V$ such that we consider the dimensionless, rescaled spectrum of the Dirac operator.
\bibitem{fnote2}
Note that $\mathcal{I}_l(Z_l,Z_{l'}^\prime)$ does not depend
on the unitary transformation that diagonalizes $V$.
\bibitem{fnote3}
 We integrate first over $Z'_{l'}$ and then over $Z_l$.
\end{thebibliography}
\end{document}